\documentclass[a4paper,11pt]{article}
\pdfoutput=1
\usepackage{jheppub} 

\usepackage[T1]{fontenc}

\usepackage{mathtools}
\usepackage{makecell}

\usepackage{verbatim}
\usepackage{graphics}
\usepackage{epsfig} 
\usepackage{color}
\usepackage{amssymb}
\usepackage{amsmath}
\usepackage{doi}
\usepackage{dsfont}
\usepackage{setspace}
\usepackage{float}
\usepackage{caption}
\usepackage{subcaption}
\usepackage{physics}
\usepackage{slashed}
\usepackage{booktabs}
\usepackage{multirow}
\usepackage{soul}

\usepackage{xspace}

\def\LOthree{${\rm LO}_3$\xspace}

\def\LOfive{${\rm LO}_5$\xspace}
\def\LOfull{${\rm LO}$\xspace}
\def\Bornone{${\rm Born}_1$\xspace}
\def\Borntwo{${\rm Born}_2$\xspace}
\def\Bornthree{${\rm Born}_3$\xspace}
\def\Bornfour{${\rm Born}_4$\xspace}
\def\Bornfive{${\rm Born}_5$\xspace}
\def\Bornfull{${\rm Born}$\xspace}
\def\NLOone{${\rm NLO}_1$\xspace}
\def\NLOtwo{${\rm NLO}_2$\xspace}
\def\NLOthree{${\rm NLO}_3$\xspace}
\def\NLOfour{${\rm NLO}_4$\xspace}
\def\NLOfive{${\rm NLO}_5$\xspace}
\def\NLOsix{${\rm NLO}_6$\xspace}
\def\NLOqcd{${\rm NLO}_{\rm QCD}$\xspace}
\def\NLOfull{${\rm NLO}$\xspace}

\def\asLOone{$\mathcal{O}(\alpha_s^4\alpha^2)$\xspace}
\def\asLOtwo{$\mathcal{O}(\alpha_s^3\alpha^3)$\xspace}
\def\asLOthree{$\mathcal{O}(\alpha_s^2\alpha^4)$\xspace}
\def\asLOfour{$\mathcal{O}(\alpha_s^1\alpha^5)$\xspace}
\def\asLOfive{$\mathcal{O}(\alpha_s^0\alpha^6)$\xspace}

\def\asNLOone{$\mathcal{O}(\alpha_s^5\alpha^2)$\xspace}
\def\asNLOtwo{$\mathcal{O}(\alpha_s^4\alpha^3)$\xspace}
\def\asNLOthree{$\mathcal{O}(\alpha_s^3\alpha^4)$\xspace}
\def\asNLOfour{$\mathcal{O}(\alpha_s^2\alpha^5)$\xspace}
\def\asNLOfive{$\mathcal{O}(\alpha_s^1\alpha^6)$\xspace}
\def\asNLOsix{$\mathcal{O}(\alpha_s^0\alpha^7)$\xspace}

\def\lepjet{$\ell+j$\xspace}

\title{Complete NLO corrections to off-shell $\boldsymbol{t
\bar{t}}$ production in the $\boldsymbol{\ell+j}$ decay channel}

\author[\,a]{Leon Mans,}
\author[\,b]{Daniel Stremmer}
\author[\,a]{and Malgorzata Worek}

\affiliation[a]{Institute for Theoretical Particle Physics
and Cosmology, RWTH Aachen University, \\D-52056 Aachen, Germany}
\affiliation[b]{Institute for Theoretical Particle Physics,
Karlsruhe Institute of Technology, \\D-76128 Karlsruhe, Germany}

\emailAdd{leon.mans@rwth-aachen.de}
\emailAdd{daniel.stremmer@kit.edu}
\emailAdd{worek@physik.rwth-aachen.de}

\abstract{We present the calculation of the complete NLO corrections to the off-shell top-quark pair production in the \lepjet decay channel, denoted as $pp \to \ell^- \bar{\nu}_\ell\, j_b j_b \,jj + X$, where $\ell^- = e^-,\, \mu^-$. The calculation consistently preserves the finite-width effects of the top quarks and massive gauge bosons, as well as takes into account all doubly-, singly-, and non-resonant contributions along with their interference effects. All Born-level contributions, at the perturbative orders from ${\cal O}(\alpha_s^4\alpha^2)$ to ${\cal O}(\alpha_s^0\alpha^6)$, are included and corrected by both NLO QCD and NLO EW effects. Furthermore, all possible partonic initial states are taken into account. Particular attention is paid to the infrared safety in the presence of photons and jets. This requires the use of the so-called parton-to-photon fragmentation function and the photon-to-jet conversion function, which makes the democratic photon–parton clustering and the $\gamma\to q\bar q$ splittings finite. We present our findings at the integrated and differential fiducial cross-section levels for the LHC Run III centre-of-mass energy of $\sqrt{s}= 13.6$ TeV. In addition, we quantify the impact of subleading NLO effects, in particular, electroweak Sudakov logarithms and non-resonant QCD backgrounds. Two analysis strategies are employed and compared, namely with and without the  resonance-enhancing requirement on the invariant mass of the two light jets, $|M_{jj}-m_W|<{\cal Q}_{\text{cut}} = 20$ GeV, illustrating the relationship between QCD background suppression, off-shell effects, interferences, and complete NLO corrections.}

\dedicated{\rm TTK-25-47, TTP25-056, P3H-25-111}


\begin{document} 

\maketitle
\flushbottom

\section{Introduction}
\label{sec:intro}

The top quark, the heaviest particle in the Standard Model (SM), is copiously produced at the CERN Large Hadron Collider (LHC) and continues to play a central role in precision tests of the SM and in indirect and direct searches for physics beyond it. At the LHC, the $t\bar{t}$ process is the dominant way top quarks are produced in $pp$ collisions.  Among the different $t\bar t$ final states, the \lepjet channel offers a particularly attractive precision–purity compromise: it combines statistics with good background rejection, allows a full kinematic reconstruction of the $t\bar t$ system thanks to the presence of only a single neutrino, and retains excellent sensitivity to top-quark spin correlations, charge asymmetries, and mass-sensitive observables. These features make the \lepjet mode a workhorse for precision SM studies and a powerful channel for new-physics searches targeting boosted and high-mass regimes. Examples of beyond the SM (BSM) scenarios include, among others, vector-like quarks \cite{Aguilar-Saavedra:2013qpa,ATLAS:2018ziw}, heavy charged and neutral resonances ($W',Z'$) \cite{Harris:2011ez,ATLAS:2013nki,CMS:2012zja}, and contact-interaction scenarios encoded in four-fermion operators \cite{Aguilar-Saavedra:2010uur,CMS:2023xyc}, all of which can populate the TeV tails of various distributions based on $t\bar t$ kinematics or mimic characteristic boosted-jet topologies. With the High-Luminosity LHC (HL-LHC) on the horizon, these phase-space regions will be probed with unprecedented precision and reach, and the corresponding theoretical predictions must be commensurately robust. However, these are precisely the regions of phase space where full off-shell effects are needed to obtain accurate results, since high-$p_T$ tails of dimensionful observables are especially sensitive to singly- and non-resonant contributions. Moreover, they are crucial for describing observables that probe the mass of potentially resonant particles in the process, distinguished by kinematic edges, which are important both for SM template fits and for BSM resonance searches. These include, for example, the reconstructed invariant mass of the top quark, $M(t)$, the minimal invariant mass of the $b$-jet and the charged lepton, $M(\ell^+b)_{min}$, the stransverse mass of the top quark, $M_{T2}(t)$, and the stransverse mass of the $W$ gauge boson, $M_{T2}(W)$ \cite{Lester:1999tx}.

From the experimental side, $t\bar t$ production was first observed at the Tevatron by the CDF and D{\O} collaborations \cite{CDF:1995wbb,D0:1995jca}. Since then, ATLAS and CMS have measured fiducial cross sections of $t\bar t$ production in the \lepjet decay channel in $pp$ collisions at centre-of-mass energies of $\sqrt{s}=5.02, \, 7,  \,8, \,13$ and $13.6$ TeV \cite{ATLAS:2010zaw,CMS:2016csa,CMS:2015rld,ATLAS:2017wvi,CMS:2016oae,ATLAS:2017cez,CMS:2023qyl,CMS:2024ghc,ATLAS:2022jbj}. Precision differential measurements now extend to multi-TeV scales and include top-quark spin correlations \cite{CMS:2024zkc} and ever more refined constraints on the top-quark mass \cite{ATLAS:2025bpp}. The experimental selections in the \lepjet channel typically require one isolated charged lepton, often missing transverse momentum, at least two $b$-tagged jets and two additional light-flavour jets, often with further kinematic requirements tailored to enhance the $t\bar t$ signal over irreducible backgrounds. Matching theory predictions with these realistic fiducial definitions is, therefore, essential.

On the theory side, tremendous progress has been achieved for $t\bar t$ production and top-quark decays. In the dilepton channel, full off-shell next-to-leading order (NLO) QCD predictions \cite{Denner:2010jp,Bevilacqua:2010qb,Denner:2012yc,Frederix:2013gra,Cascioli:2013wga,Heinrich:2013qaa} and NLO electroweak (EW) corrections \cite{Denner:2016jyo} are available, and parton-shower (PS) matched descriptions exist both in the narrow-width approximation (NWA) \cite{Campbell:2014kua} and in frameworks including full off-shell and non-resonant effects for the  $bb4\ell$ final state \cite{Jezo:2016ujg}. Inclusive $t\bar t$ production is known at next-to-next-to-leading order (NNLO) QCD \cite{Czakon:2013goa,Czakon:2015owf,Czakon:2016dgf,Catani:2019iny,Catani:2019hip} as well as with next-to-next-to-leading-logarithmic
(NNLL) threshold resummation \cite{Czakon:2018nun}. Furthermore, approximate NNLO QCD descriptions with exact NNLO QCD decays are available in the literature \cite{Gao:2017goi}. NNLO QCD predictions in the NWA for differential observables \cite{Behring:2019iiv,Czakon:2020qbd}, together with NNLO QCD predictions matched to parton-shower simulations via the \textsc{MiNNLO} method \cite{Mazzitelli:2020jio,Mazzitelli:2021mmm},  have played a key role in further increasing the precision of theoretical modelling for this process. The combination of NNLO QCD and NLO EW corrections, including large electroweak Sudakov logarithms, has been scrutinized for inclusive and differential observables \cite{Czakon:2017wor,Huss:2025nlt}. Finally, a new path towards NNLO QCD predictions for the off-shell $t\bar{t}$ production has recently been established for leptonic decays and massive bottom quarks \cite{Buonocore:2025fqs}. In this paper, all tree-level and one-loop amplitudes have been evaluated including non-resonant and off-shell effects related to the top quarks and the leptonic decays of the $W^\pm$ bosons, while the missing two-loop virtual contribution has been estimated using the so-called double-pole approximation \cite{Denner:2019vbn}.

By contrast, the \lepjet  decay channel has received comparatively less attention. Indeed, NLO QCD studies exist both at the fixed-order level \cite{Denner:2017kzu} and are also matched to parton-shower programs \cite{Jezo:2023rht}.  Moreover, the NNLO QCD parton-shower matched predictions are available for this decay channel  as well \cite{Mazzitelli:2021mmm}. However, a systematic and complete treatment of the full off-shell $t\bar{t}$ process in this decay channel, including subleading LO and NLO contributions, remains an open problem. There are both physical and technical reasons for this imbalance. Compared to the dilepton case, the \lepjet final state suffers from a larger irreducible QCD-mediated $W+jets$ background (e.g. $W(\to \ell \nu_\ell)+ b\bar b +jj$) that contributes at \asLOone and exhibits more complex colour flows and radiation patterns due to the lack of the hadronically decaying $W$ gauge boson. When calculating complete NLO corrections, the \lepjet decay channel exhibits non-trivial infrared (IR) safety issues due to the clustering of photons and jets in the same final state, but also due to the $\gamma \to q\bar{q}$ splittings that are identified as light jets in the jet algorithm. Moreover, a complete NLO calculation with full off-shell effects must take into account not only all doubly-, singly-, and non-resonant Feynman diagrams and their interferences, but also all leading and subleading Born-level contributions and their corresponding NLO QCD and NLO EW corrections. One-loop electroweak Sudakov logarithms of the type $-\alpha\log^n (\frac{s}{M_W^2})$ where $n=1,2$,  can induce suppressions of several tens of percent  in multi-TeV kinematic tails \cite{Czakon:2017wor}, precisely the phase-space regions that the HL-LHC will start to be sensitive to. Therefore, it is necessary to have the QCD and EW effects under excellent control in order to describe the entire range of interesting observables, or even better, to study them within one, fully coherent framework, as recommended for example in Refs. \cite{Stremmer:2024ecl,Huss:2025nlt}.

The purpose of the  article is to mitigate the current situation and to calculate for the first time complete NLO predictions for the full off-shell $pp \to \ell^-\bar{\nu}_\ell\,j_b j_b \,jj +X$  process, where $\ell^-$ stands for  $e^-$ and $\mu^-$. The calculation, which is performed within one common framework, comprises all Born-level contributions at the perturbative orders from  ${\cal O}(\alpha_s^4\alpha^2)$ to ${\cal O}(\alpha_s^0\alpha^6)$ together with the corresponding NLO QCD and NLO EW corrections. Consequently, the complete NLO contribution involves the following six different orders ${\cal O}(\alpha_s^5\alpha^2)$, ${\cal O}(\alpha_s^4\alpha^3)$, ${\cal O}(\alpha_s^3\alpha^4)$, ${\cal O}(\alpha_s^2\alpha^5)$, ${\cal O}(\alpha_s\alpha^6)$  and ${\cal O}(\alpha_s^0\alpha^7)$. Some of these NLO contributions apply higher-order corrections to more than one LO process. Therefore, it is not possible to unambiguously assign a given correction type to a given underlying Born-level process as different production mechanisms, in particular the EW and QCD-induced production modes, are naturally mixed with each other.  Our calculation (i) takes into account finite-width effects of the top quark and $W^\pm/Z$ gauge bosons, (ii) accounts for all doubly-, singly- and non-resonant contributions and their interference effects in a gauge-invariant manner, and (iii) considers all possible production mechanisms and all possible final-state particles, both at LO and NLO. 

In particular, we will provide complete NLO predictions for the $t\bar{t}$ production process in the \lepjet decay channel both at the integrated and differential cross-section level, and study two realistic analysis strategies. We expand the inclusive choice of  phase-space cuts with an additional resonance-enhancing requirement on the invariant mass of the two light jets $|M_{jj}-m_W|< {\cal Q}_{cut} = 20$ GeV, which  is designed to further suppress the irreducible QCD $W+jets$ background. In our study, we will quantify the impact of complete higher-order effects in the \lepjet decay channel. To avoid any problems with the IR-safety, we implement additional, dedicated techniques, namely: the parton-to-photon fragmentation functions \cite{Glover:1993xc,ALEPH:1995zdi} and the photon-to-jet conversion functions \cite{Denner:2019zfp}. We assess the size and kinematic dependence of NLO QCD and EW corrections,  highlighting the role of the photon-initiated subprocesses and demonstrating how the one-loop electroweak Sudakov effects change the shape for various dimensionful observables in the high-$p_T$ tails. Finally, we carefully examine the size of the theoretical uncertainties related to missing higher-order terms in QCD calculations, which are estimated  in the complete \NLOfull and \NLOqcd  predictions via  a seven-point scale variation, and the impact of the ${\cal Q}_{cut}$ cut on the suppression of the QCD background. 

The remainder of this article is organised as follows. In Section~\ref{sec:process}, we  comprehensively define all necessary contributions to the process under consideration, both at the LO and NLO level. The computational framework is described in detail in Section \ref{sec:framework}, where also the implementation of the parton-to-photon fragmentation functions and the photon-to-jet conversion functions into our framework are discussed. In Section \ref{sec:parameters} we list our input parameters, fiducial phase-space cuts, PDF information and the scale  settings we use. The integrated fiducial cross-section predictions are presented in Section \ref{sec:results-int}, including a comparison between results with and without the ${\cal Q}_{cut}$ cut. In Section \ref{sec:results-diff} we analyse several  differential cross-section distributions. To assess the robustness of the presented results, in Section \ref{sec:scale} we examine the impact of an alternative dynamical scale choice on our complete NLO predictions. We conclude in Section~\ref{sec:conclusions} with a summary and an outlook towards possible future studies.

\section{Process definition and setup} 
\label{sec:process}

\begin{figure}[t!]
    \centering

    \begin{minipage}{0.32\linewidth}
        \centering
        \includegraphics[width=\linewidth]{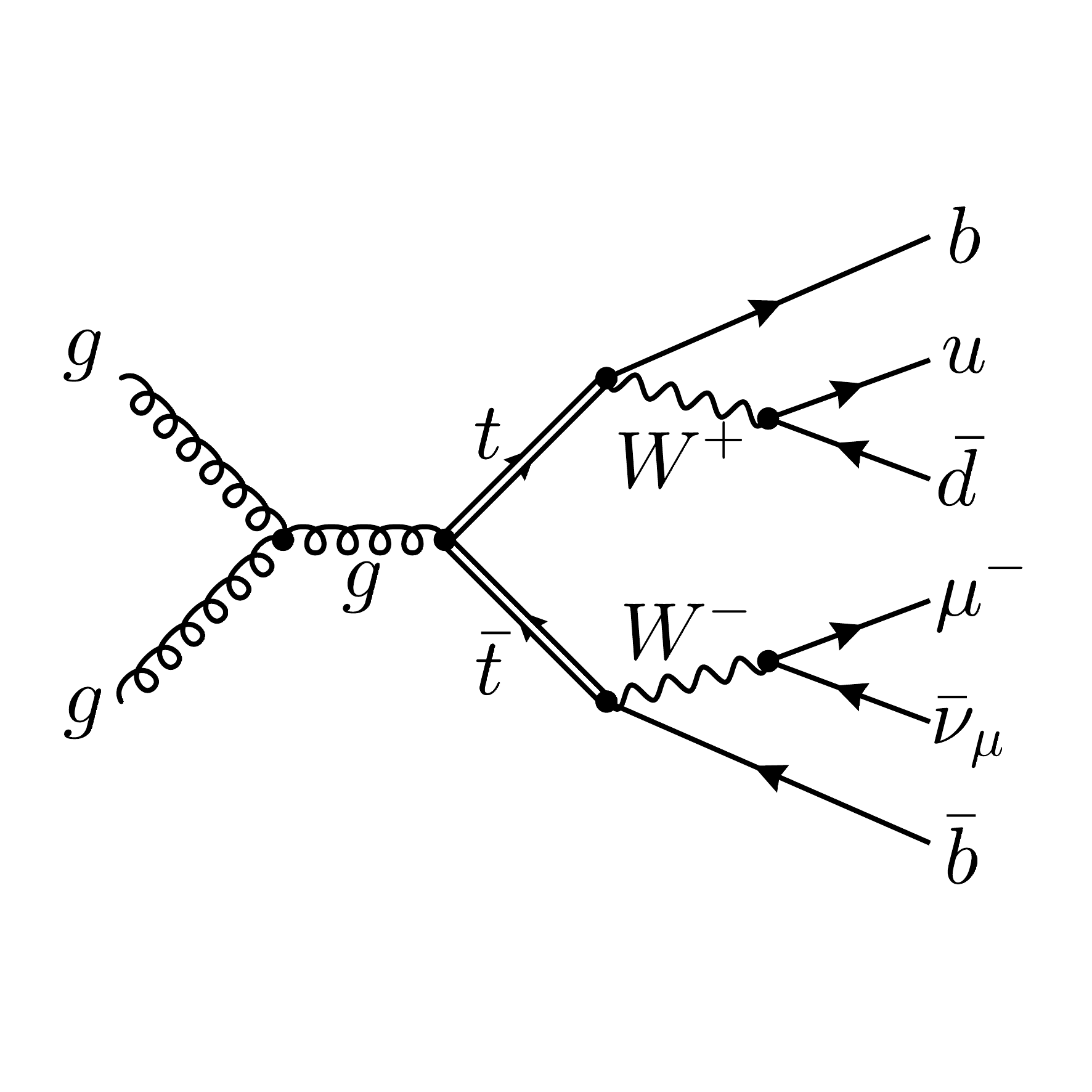}
    \end{minipage}
    \hfill
    \begin{minipage}{0.32\linewidth}
        \centering
        \vspace{0.9cm}
        \includegraphics[width=\linewidth]{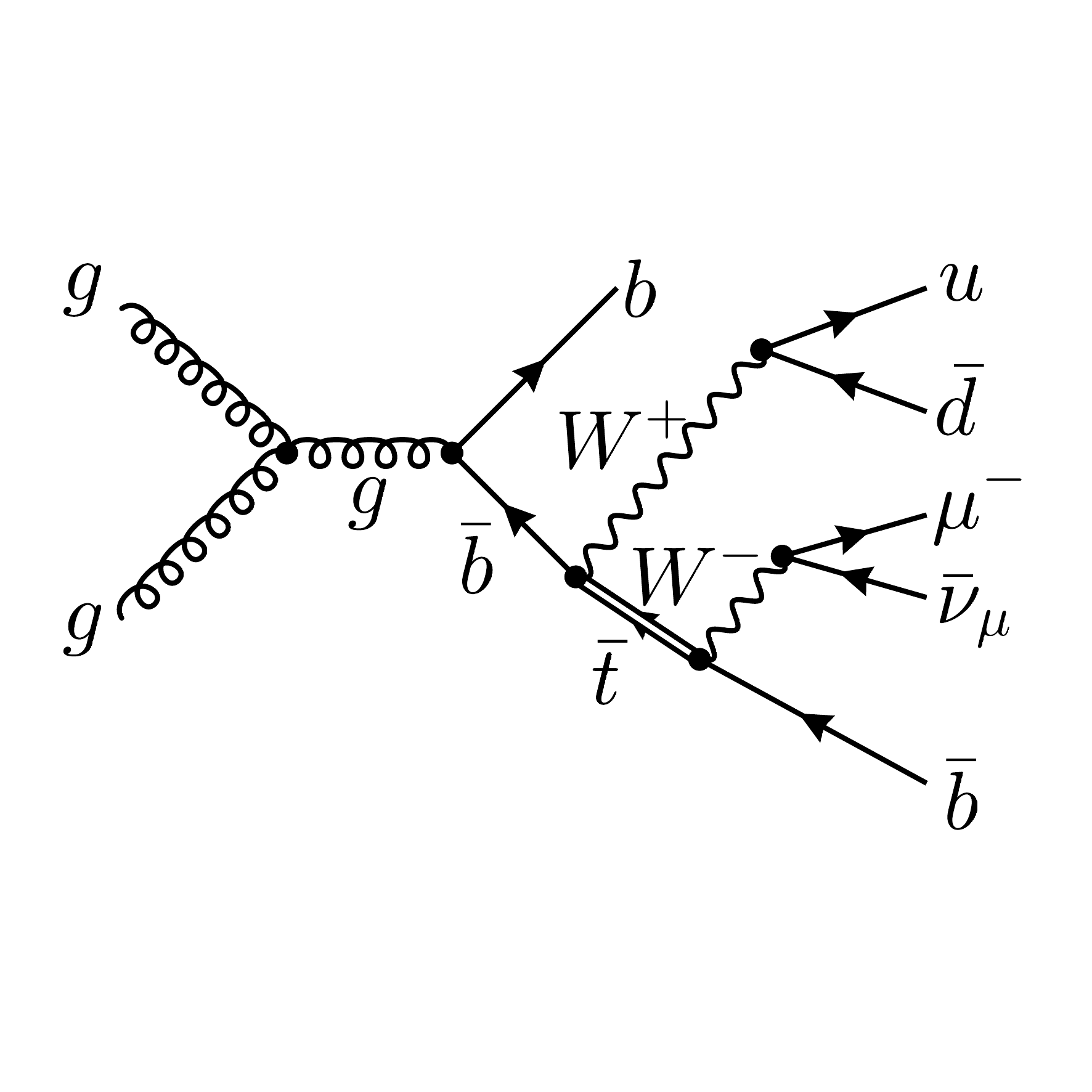}
    \end{minipage}
    \hfill
    \begin{minipage}{0.32\linewidth}
        \centering
        \includegraphics[width=\linewidth]{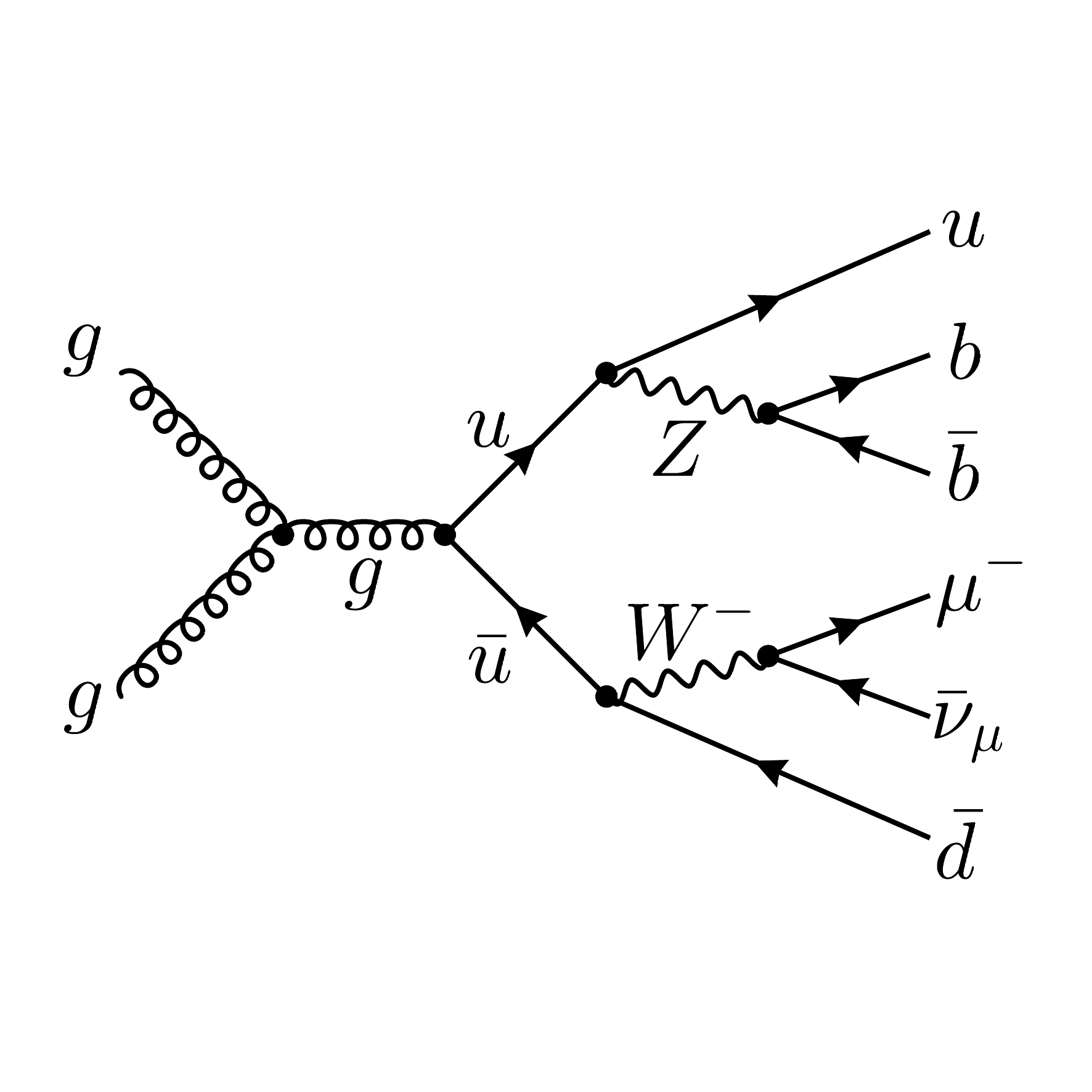}
    \end{minipage}
    \vspace{-1cm}
    \caption{\it Examples of Feynman diagrams with two (left), one (middle) and no top-quark resonances (right) contributing to the $pp \to \ell^-\bar{\nu}_\ell\,j_b j_b\, jj +X$ process at $\text{LO}_3$.}
    \label{fig:born-resonance}
\end{figure}

We consider full off-shell top-quark pair production in the \lepjet  decay channel,  denoted as $pp \to \ell^-\bar{\nu}_\ell\,j_bj_b\,jj +X$ with $\ell^-=e^-,\mu^-$. The full set of leading and subleading Born-level contributions and higher-order corrections is calculated based on the $\ell^-\bar{\nu}_\ell\,j_bj_b\,jj$ matrix element, including all resonant and non-resonant Feynman diagrams, as well as interference effects. Finite-width effects of the top quark, $W^\pm/Z$ gauge bosons and the Higgs boson are incorporated using Breit-Wigner propagators. To this end, the complex-mass scheme is employed \cite{Denner:2005fg,Denner:2006ic,Denner:2019vbn}. The calculation is performed in the $5$-flavour scheme, treating the bottom quarks as massless particles. We consistently include in the initial state all PDF suppressed channels  from bottom quarks as well as photons. Throughout the analysis, the  Cabibbo-Kobayashi-Maskawa (CKM) matrix is kept diagonal.
The LO contributions are classified as follows, where $g_s$ denotes the strong coupling and $g$ the EW gauge coupling; accordingly $\alpha_s = g_s^2/(4\pi)$ and $\alpha = g^2/(4\pi)$:

\begin{itemize}
    \item \textbf{LO$_1$} ($\mathcal{O}(\alpha_s^4 \alpha^2)$): The QCD-induced contribution, characterised by the highest power of  $\alpha_s$. Nevertheless, this contribution is suppressed because Feynman diagrams do not include resonant top quarks and contain at most one resonant $W$ boson. This contribution is considered to be the irreducible QCD background for the production of top-quark pairs.

    \item \textbf{LO$_2$} ($\mathcal{O}(\alpha_s^3 \alpha^3)$): The order that comprises interference terms between the matrix elements at $\mathcal{O}(g_s^4 g^2)$ and $\mathcal{O}(g_s^2 g^4)$, where the latter contribution contains Feynman diagrams with two resonant top quarks. In addition, we encounter photon-initiated subprocesses, which are suppressed by the photon PDF.

    \item \textbf{LO$_3$} ($\mathcal{O}(\alpha_s^2 \alpha^4)$): The dominant contribution that contains Feynman diagrams with two resonant top quarks and $W$ bosons. Additionally, this order receives contributions from interference terms between the matrix elements at  $\mathcal{O}(g_s^4 g^2)$ and $\mathcal{O}(g_s^0 g^6)$ as well as photon-initiated amplitudes at  $\mathcal{O}(g_s^3 g^3)$ and $\mathcal{O}(g_s g^5)$. Different resonance structures for this perturbative order are shown in Figure \ref{fig:born-resonance}. We note here that all Feynman diagrams in this paper are produced with the help of \textsc{FeynGame} \cite{Harlander:2020cyh}.

    \item \textbf{LO$_4$} ($\mathcal{O}(\alpha_s \alpha^5)$): Similarly to LO$_2$, this contribution requires photons in the initial state and is further suppressed by the relative power of $\alpha$ compared to $\alpha_s$. Despite this suppression, Feynman diagrams in this order include two resonant top quarks and $W$ bosons. This order also receives contributions from the interference between the matrix elements at $\mathcal{O}(g_s^2 g^4)$ and $\mathcal{O}(g_s^0g^6)$. Both contributions contain Feynman diagrams with two resonant top quarks that are produced by QCD and EW interactions, respectively. 

    \item \textbf{LO$_5$} ($\mathcal{O}(\alpha_s^0\alpha^6)$): The purely EW induced  contribution. It  incorporates Feynman diagrams with two resonant top quarks and $W$ gauge bosons.
\end{itemize}
\begingroup
\setlength{\extrarowheight}{0.1cm}
\begin{table}[t!]
\centering
\begin{tabular}{|p{4.5cm}|c|c|c|c|c|}
\hline
$\textbf{LO$_i$}$ & $\textbf{LO$_1$}$ & $\textbf{LO$_2$}$ & $\textbf{LO$_3$}$ & $\textbf{LO$_4$}$ & $\textbf{LO$_5$}$ 
\\ \hline
\textbf{Perturbative Order} & $\mathcal{O}(\alpha_s^4 \alpha^2)$ & $\mathcal{O}(\alpha_s^3 \alpha^3)$ & $\mathcal{O}(\alpha_s^2 \alpha^4)$ & $\mathcal{O}(\alpha_s^1 \alpha^5)$ & $\mathcal{O}(\alpha_s^0 \alpha^6)$ 
\\ \hline
\begin{tabular}{@{}ll@{}}
$g\,g \to \ell^- \bar{\nu}_{\ell}\, b\,\bar{b}\, q\,\bar{q}^{\,\prime}$\\
$\bar{q}\,g \to \ell^- \bar{\nu}_{\ell}\, b\,\bar{b}\, \bar{q}^{\,\prime}\,g$\\
$q\,\bar{q}^{\,\prime} \to \ell^- \bar{\nu}_{\ell}\, b\,\bar{b}\, g\,g$\\
$q\,g \to \ell^- \bar{\nu}_{\ell}\, b\,\bar{b}\, q^{\,\prime}\,g$\\
\end{tabular}
 & $ |\mathcal{M}_{4}|^2$ 
 & $ 2\Re(\mathcal{M}_{4}\mathcal{M}_{2}^*)$ 
 & $ |\mathcal{M}_{2}|^2$ 
 &  &  \\ \hline
\begin{tabular}{@{}ll@{}}
$q\,q \to \ell^- \bar{\nu}_{\ell}\, b\,\bar{b}\, q\,q^{\,\prime}$\\
$q\,q \to \ell^- \bar{\nu}_{\ell}\, b\,\bar{b}\, {\bar{q}}^{\,\prime}\,\bar q$\\
$q\,q^{\,\prime}\to \ell^- \bar{\nu}_{\ell}\, b\,\bar{b}\, q\,q$\\
$q\,\bar q^{\,\prime} \to \ell^- \bar{\nu}_{\ell}\, b\,\bar{b}\, q^{\,\prime}\,\bar q^{\,\prime}$\\
$\bar{q}\,\bar{q}^{\,\prime} \to \ell^- \bar{\nu}_{\ell}\, b\,\bar{b}\, \bar{q}\,\bar{q}$\\
$q\,\bar{q} \to \ell^- \bar{\nu}_{\ell}\, b\,\bar{b}\, q\,\bar{q}^{\,\prime}$\\
$q\,\bar{q}^{\,\prime} \to \ell^- \bar{\nu}_{\ell}\, b\,\bar{b}\, q\,\bar{q} $\\
$\bar{q}\,\bar{q} \to \ell^- \bar{\nu}_{\ell}\, b\,\bar{b}\, \bar{q}\,\bar{q}^{\,\prime}$\\
\end{tabular}
 & $ |\mathcal{M}_{4}|^2$ 
 & $ 2\Re(\mathcal{M}_{4}\mathcal{M}_{2}^*)$ 
 & \makecell[c]{%
  $|\mathcal{M}_2|^2$ \\ 
  $2\Re(\mathcal{M}_{4}\mathcal{M}_{0}^*)$
}
 & $ 2\Re(\mathcal{M}_{2}\mathcal{M}_{0}^*)$ 
 & $ |\mathcal{M}_{0}|^2$ \\ \hline
\begin{tabular}{@{}ll@{}}
$q\,Q \to \ell^- \bar{\nu}_{\ell}\, b\,\bar{b}\, q^{\,\prime}\,Q$\\
$\bar q\,Q \to \ell^- \bar{\nu}_{\ell}\, b\,\bar{b}\, \bar q^{\,\prime}\,Q$\\
$q\,\bar Q \to \ell^- \bar{\nu}_{\ell}\, b\,\bar{b}\, q^{\,\prime}\,\bar Q$\\
$\bar q\,\bar Q \to \ell^- \bar{\nu}_{\ell}\, b\,\bar{b}\, \bar q^{\,\prime}\,\bar Q$\\
$q\,Q \to \ell^- \bar{\nu}_{\ell}\, b\,\bar{b}\, q\,Q^{\,\prime}$\\
$\bar q\,Q \to \ell^- \bar{\nu}_{\ell}\, b\,\bar{b}\, \bar q\,Q^{\,\prime}$\\
$q\,\bar Q \to \ell^- \bar{\nu}_{\ell}\, b\,\bar{b}\, q\,\bar Q^{\,\prime}$\\
$\bar q\,\bar Q \to \ell^- \bar{\nu}_{\ell}\, b\,\bar{b}\, \bar q\,\bar Q^{\,\prime}$\\
$q\,\bar{q}^{\,\prime} \to \ell^- \bar{\nu}_{\ell}\, b\,\bar{b}\, Q\,\bar{Q}$\\
$q\,\bar{q} \to \ell^- \bar{\nu}_{\ell}\, b\,\bar{b}\, Q\,\bar{Q}^{\,\prime}$\\
\end{tabular}
 & $ |\mathcal{M}_{4}|^2$ 
 & $ 2\Re(\mathcal{M}_{4}\mathcal{M}_{2}^*)$ 
 & \makecell[c]{%
  $|\mathcal{M}_2|^2$ \\ 
}
 &  
 & $ |\mathcal{M}_{0}|^2$ \\ \hline
\begin{tabular}{@{}ll@{}}
$b\,\bar{b} \to \ell^- \bar{\nu}_{\ell}\, b\,\bar{b}\, q\,\bar{q}^{\,\prime}$\\
$b\,b \to \ell^- \bar{\nu}_{\ell}\, b\,b\, q\,\bar{q}^{\,\prime}$\\
$\bar{b}\,\bar{b} \to \ell^- \bar{\nu}_{\ell}\, \bar{b}\,\bar{b}\, q\,\bar{q}^{\,\prime}$\\
\end{tabular}
 & $ |\mathcal{M}_{4}|^2$ 
 & $ 2\Re(\mathcal{M}_{4}\mathcal{M}_{2}^*)$ 
 & \makecell[c]{%
  $|\mathcal{M}_2|^2$ \\ 
  $2\Re(\mathcal{M}_{4}\mathcal{M}_{0}^*)$
}
 & $ 2\Re(\mathcal{M}_{2}\mathcal{M}_{0}^*)$ 
 & $ |\mathcal{M}_{0}|^2$ \\ \hline
\begin{tabular}{@{}ll@{}}
$q\,\gamma \to \ell^- \bar{\nu}_{\ell}\, b\,\bar{b}\, q^{\,\prime}\,g$\\
$\bar{q}\,\gamma \to \ell^- \bar{\nu}_{\ell}\, b\,\bar{b}\, \bar{q}^{\,\prime}\,g$\\
$g\,\gamma \to \ell^- \bar{\nu}_{\ell}\, b\,\bar{b}\, q\,\bar{q}^{\,\prime}$\\
\end{tabular}
 &  & $ |\mathcal{M}_{3}|^2$ 
 & $ 2\Re(\mathcal{M}_{3}\mathcal{M}_{1}^*)$ 
 & $ |\mathcal{M}_{1}|^2$ 
 &  \\ \hline

\begin{tabular}{@{}c@{}}
$\gamma\,\gamma \to \ell^- \bar{\nu}_{\ell}\, b\,\bar{b}\, q\,\bar{q}^{\,\prime}$\\
\end{tabular}
 &  &  & $ |\mathcal{M}_{2}|^2$ 
 &  & $ |\mathcal{M}_{0}|^2$ \\ \hline
\end{tabular}
\caption{\it The structure of all perturbative orders shown at the level of the partonic channels contributing to the \( pp \to \ell^-\bar{\nu}_\ell\,j_b j_b jj +X \) process at LO. $\mathcal{M}_{i}$ are labelled according to the order of $\alpha_s$. Both $q$ and $q^{\,\prime}$ can span all quarks and anti-quarks. In a few cases we include $q$ and $Q$ to differentiate between first- and second-generation quarks. The empty boxes correspond to contributions that cannot be generated for a given power of $\alpha_s$ and $\alpha$ or to interference terms that vanish due to the colour structures.} 
\label{tab:LO orders}
\end{table}
\endgroup

\begin{figure}[t!]
    \centering
    \begin{minipage}{0.49\linewidth}
        \centering
        \includegraphics[width=\linewidth, trim=1.5cm 0cm 1.5cm 0cm, clip]{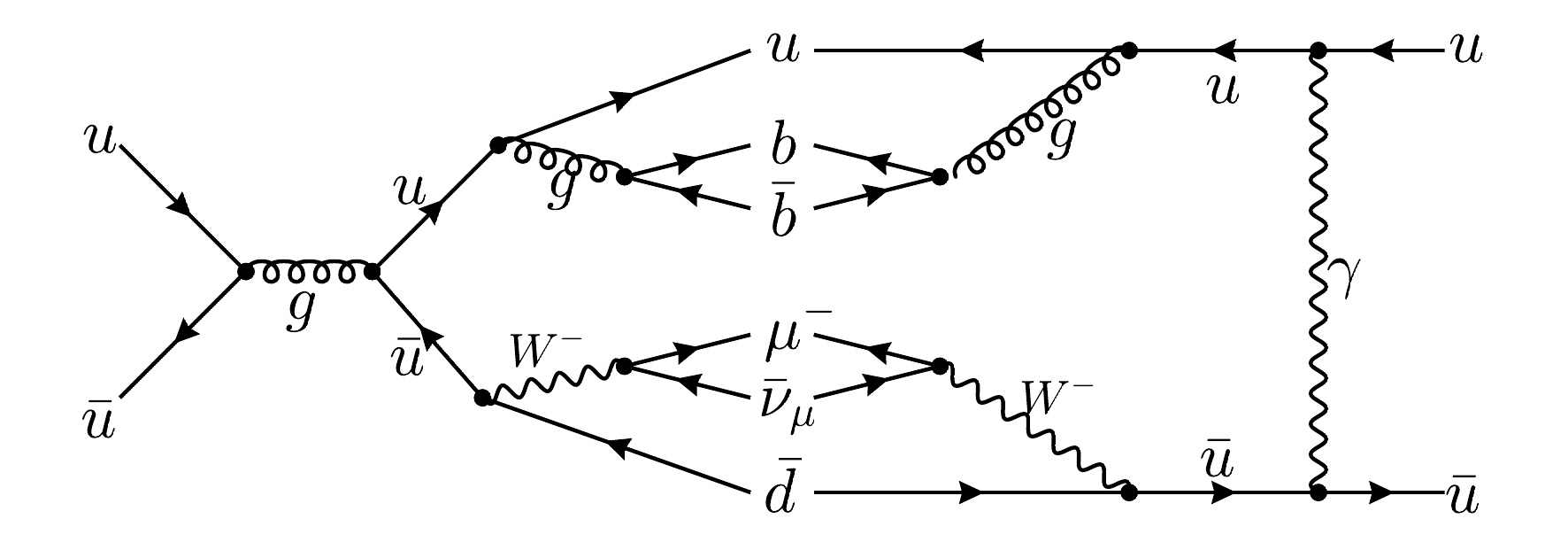}
    \end{minipage}
    \hfill
    \begin{minipage}{0.49\linewidth}
        \centering
        \includegraphics[width=\linewidth, trim=1.5cm 0cm 1.5cm 0cm, clip]{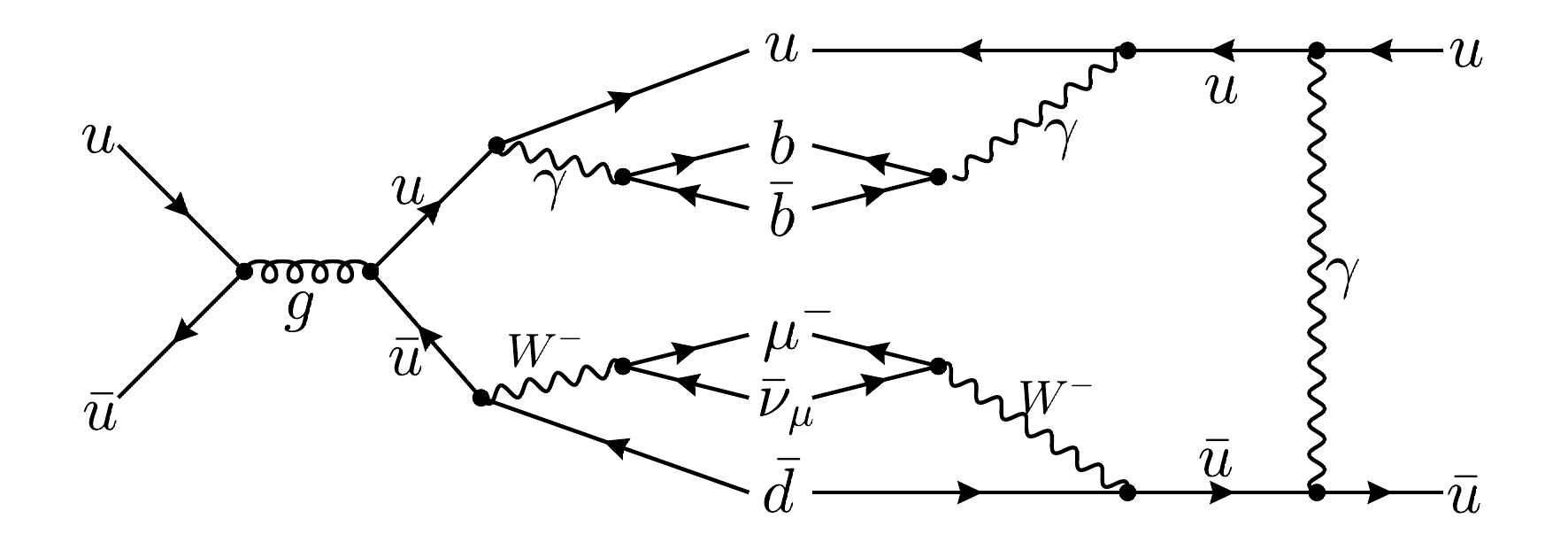}
    \end{minipage}
    \caption{ \it Examples of Feynman diagrams contributing to the $pp \to \ell^-\bar{\nu}_\ell\,j_b j_b jj +X$ process at $\text{LO}_2$ (left) and $\text{LO}_4$ (right) via interference effects. Left: interference effects between  $\text{LO$_1$}$  and $\text{LO$_3$}$. Right: interference effects between $\text{LO$_3$}$  and $\text{LO$_5$}$.}
    \label{fig:born-interference}
\end{figure}

Examples of Feynman diagrams that contribute at $\text{LO$_i$}$, where $i=1,\dots,5$, are shown in Figure~\ref{fig:born-interference}, together with some LO contributions that are possible via interference effects only.
Furthermore, the structure of all perturbation orders shown at the level of the partonic subprocesses that are needed at LO is displayed in Table~\ref{tab:LO orders}. Both $q$ and $q^\prime$ can span all quarks and anti-quarks. In a few cases, however, we include $q$ and $Q$ to distinguish first- and second-generation quarks. Because the charge of the bottom jets is not tagged, we must also include the following partonic channels in our calculations: $b\, b \to \ell^-\, \bar{\nu}_\ell\, b\, b\, u\, \bar{d}$ and $\bar{b}\, \bar{b} \to \ell^-\, \bar{\nu}_\ell\, \bar{b}\, \bar{b}\, u\, \bar{d}$, which contain at most single-resonant top-quark contributions. 

\begin{figure}[t!]
    \centering
    \includegraphics[width=\linewidth]{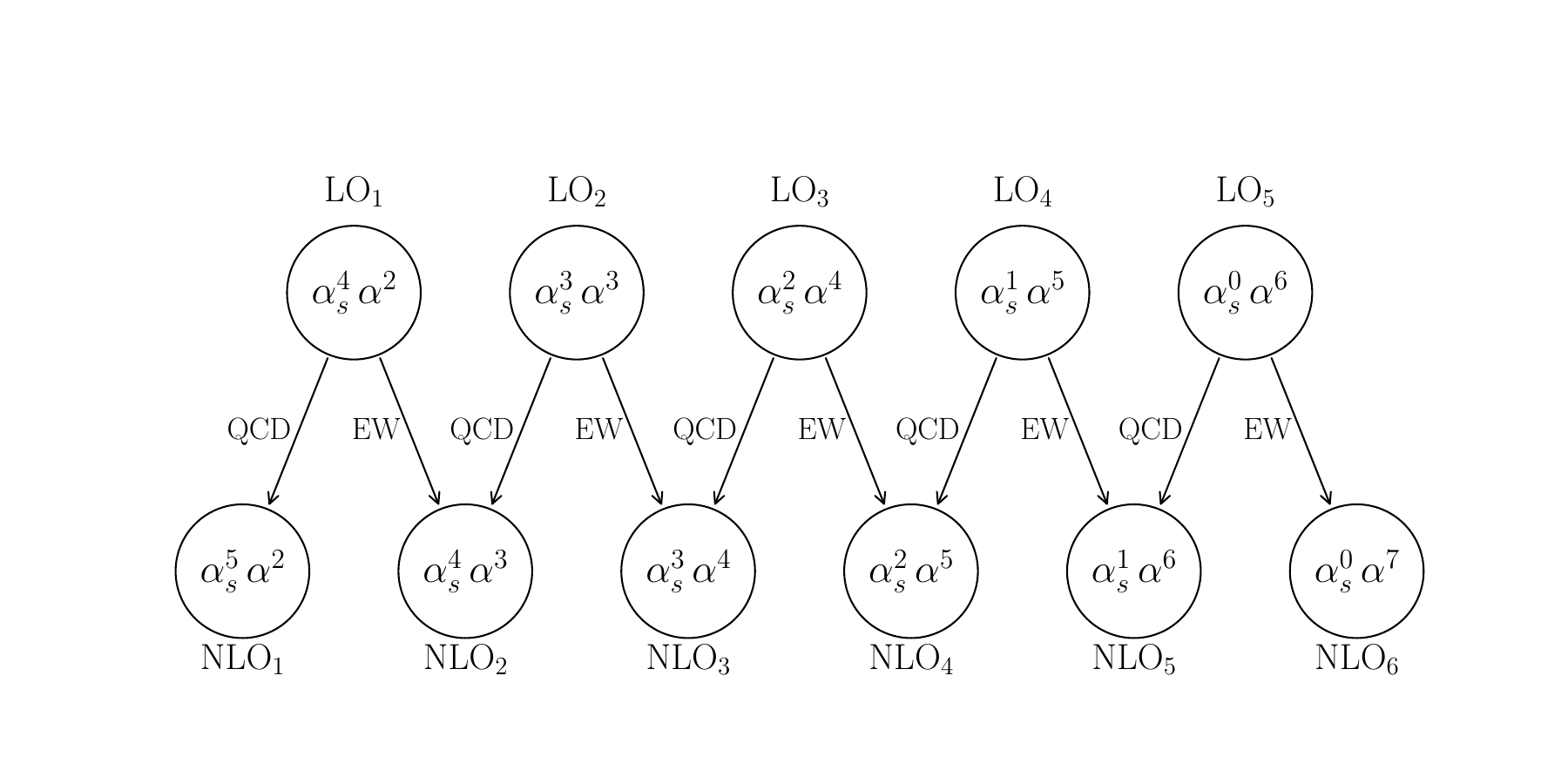}
    \caption{\it Graphical representation of all LO contributions and their NLO corrections for the  $pp \to \ell^-\bar{\nu}_\ell\,j_b j_b jj +X$ process.}
    \label{fig:orders}
\end{figure}

\begin{figure}[t!]
    \centering
    \begin{minipage}{0.32\linewidth}
        \centering
        \includegraphics[width=\linewidth]{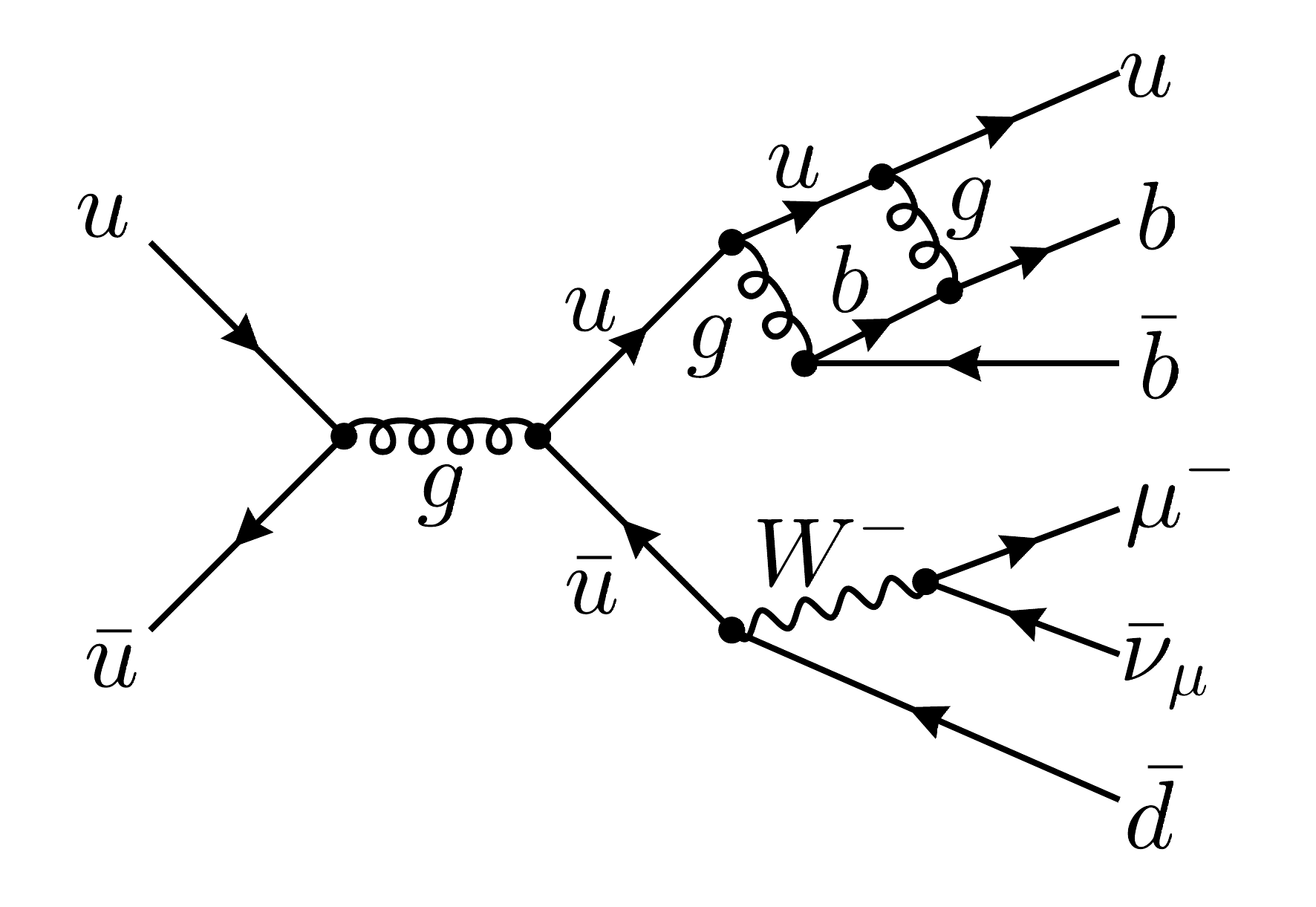}
    \end{minipage}
    \hfill
    \begin{minipage}{0.32\linewidth}
        \centering
        \includegraphics[width=\linewidth]{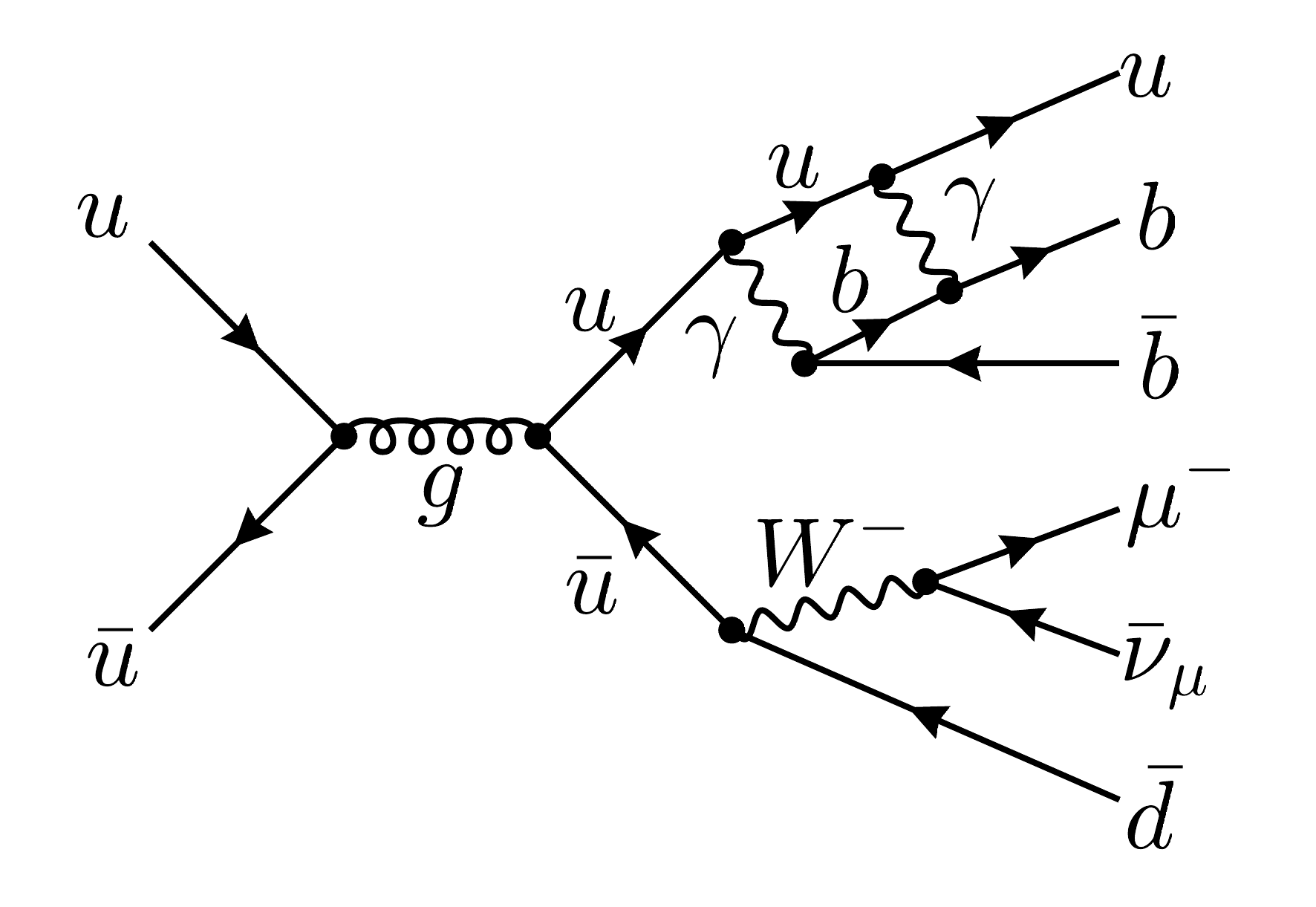}
    \end{minipage}
    \hfill
    \begin{minipage}{0.32\linewidth}
        \centering
        \includegraphics[width=\linewidth]{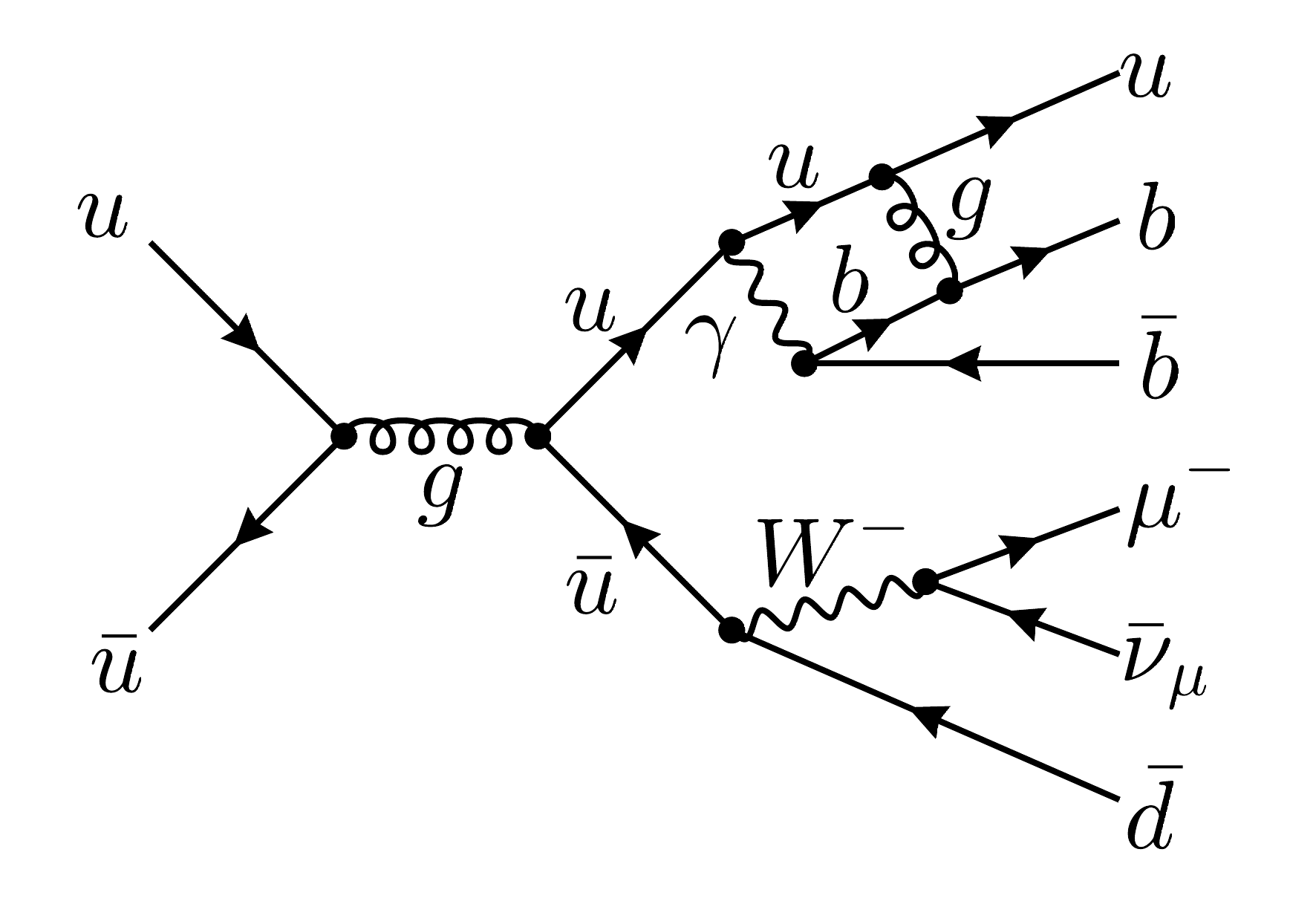}
    \end{minipage}
    \caption{\it Examples of one-loop Feynman diagrams contributing to the $pp \to \ell^-\bar{\nu}_\ell\,j_b j_b jj+X$ process. The diagram on the left contributes at $\mathcal{O}(g_s^6 g^2)$, the one in the middle at $\mathcal{O}(g_s^2 g^6)$ and the one on the right at $\mathcal{O}(g_s^4 g^4)$.}
    \label{fig:loop-orders}
\end{figure}

\begin{figure}[t!]
    \centering
    \begin{minipage}{0.32\linewidth}
        \centering
        \includegraphics[width=\linewidth]{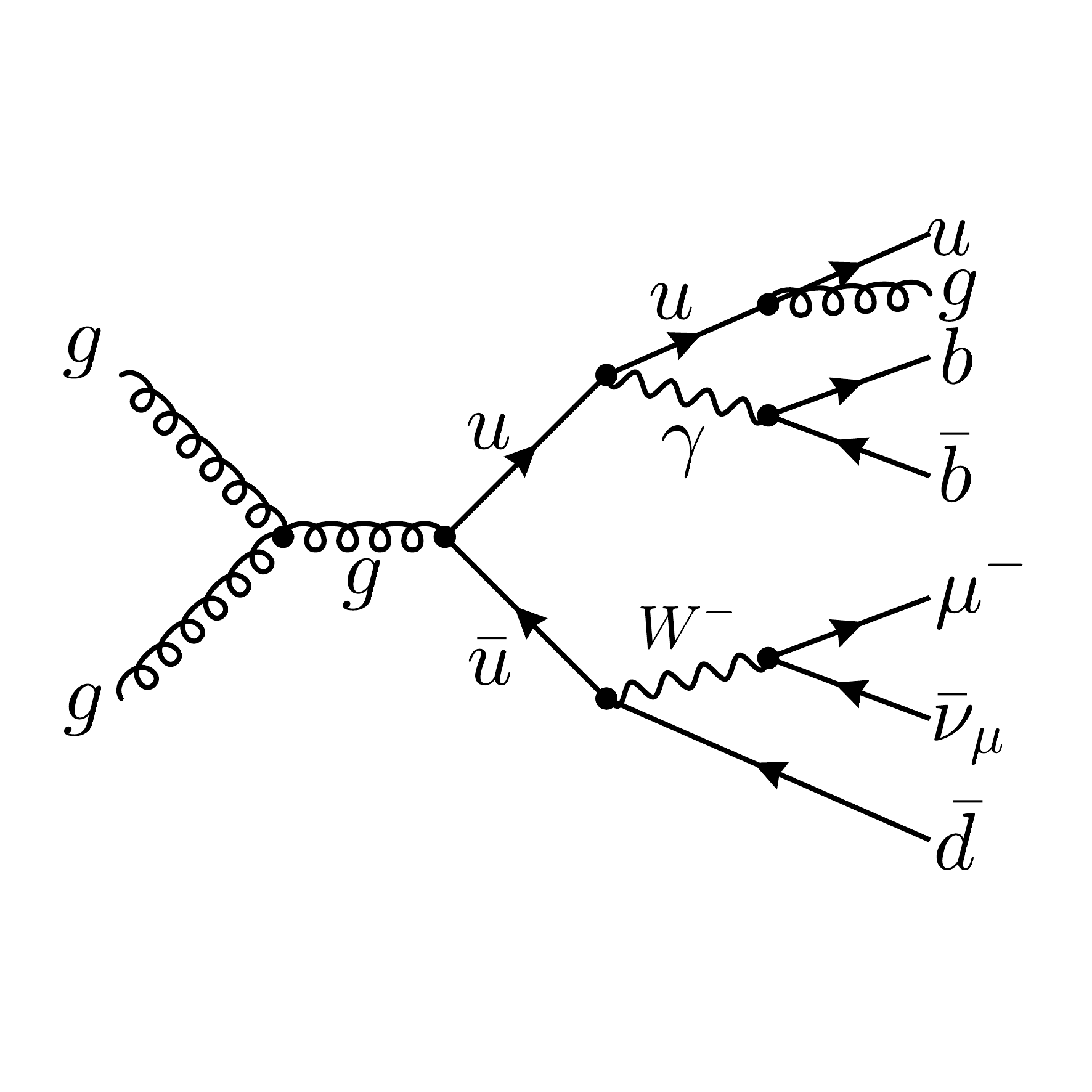}
    \end{minipage}
    \hfill
    \begin{minipage}{0.6\linewidth}
        \centering
        \includegraphics[width=\linewidth]{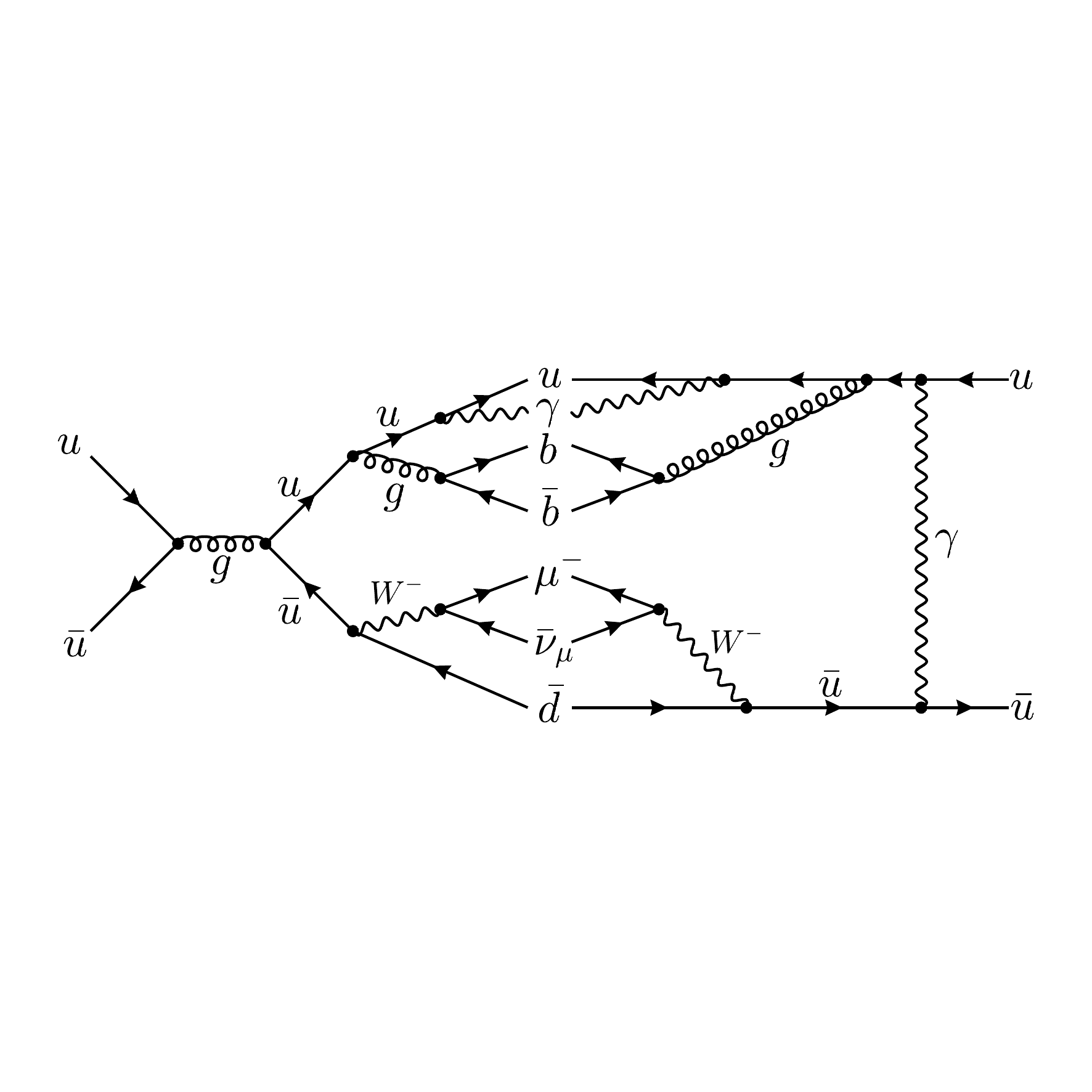}
    \end{minipage}
    \vspace{-3cm}
    \caption{\it Examples of real-emission Feynman  diagrams contributing to the $pp \to \ell^-\bar{\nu}_\ell\,j_b j_b jj+X$ process at $\text{NLO$_3$}$.}
    \label{fig:cancellationdiagramsnlo3}
\end{figure}

\begin{figure}[htbp]
  \centering

  \makebox[\linewidth]{
    \begin{subfigure}[b]{0.3\linewidth}
      \includegraphics[width=\linewidth]{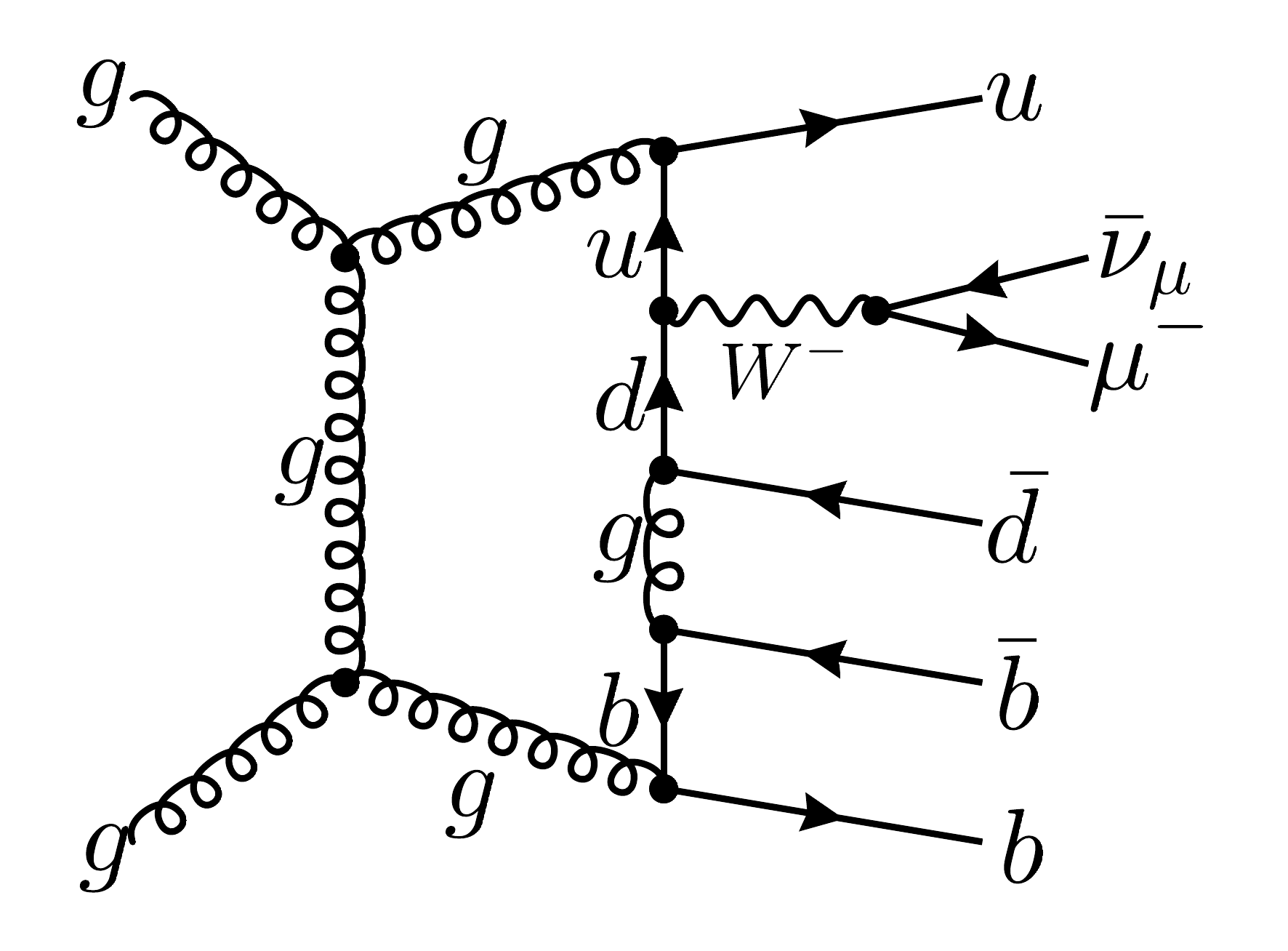}
      \vspace{0.1cm}
      \caption{}
      \label{fig:sub_a}
    \end{subfigure}\hspace{0.04\linewidth}
    \begin{subfigure}[b]{0.3\linewidth}
      \includegraphics[width=\linewidth]{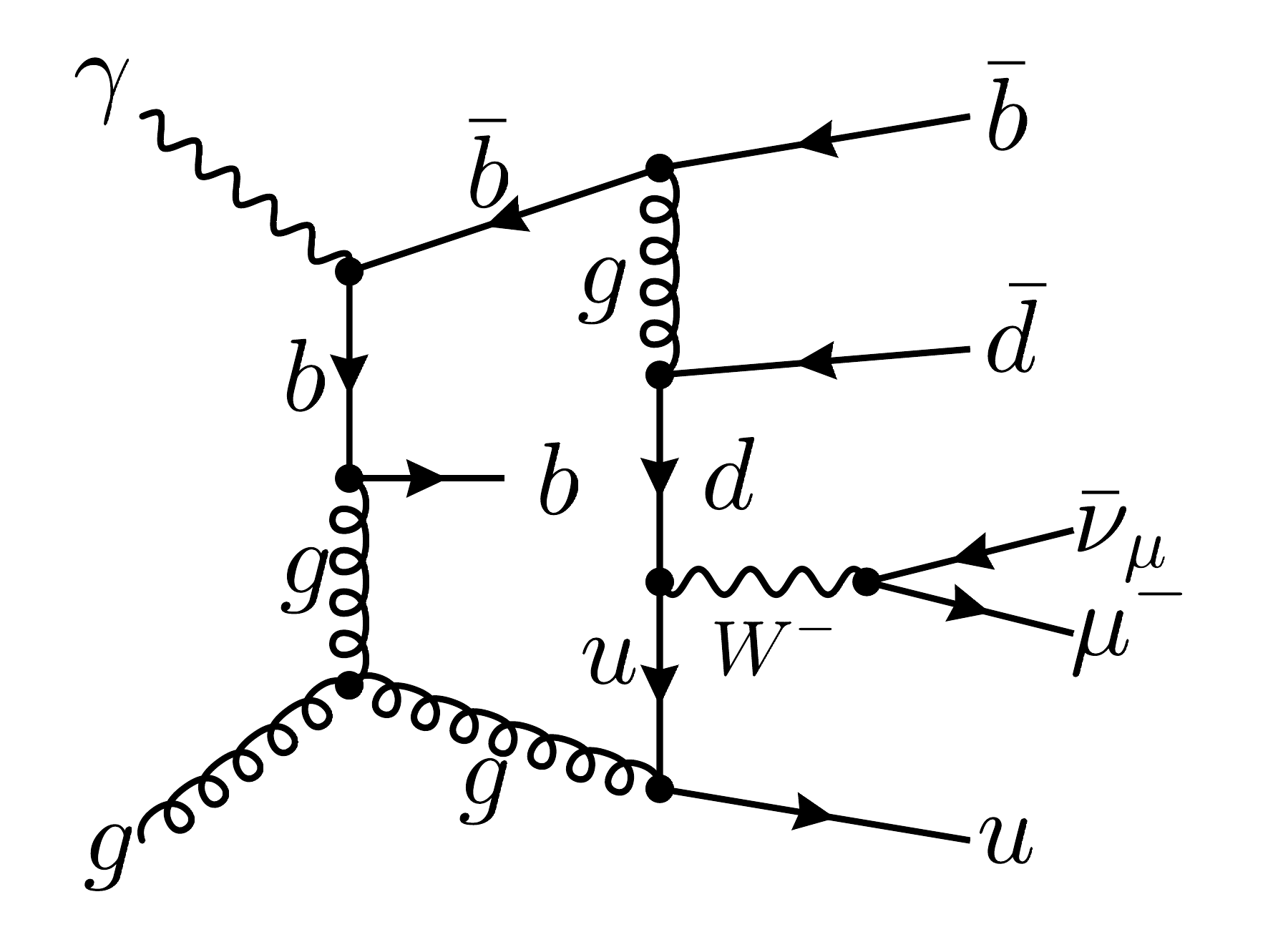}
      \vspace{0.1cm}
      \caption{}
      \label{fig:sub_b}
    \end{subfigure}
  }

  \vspace{6pt}

  \begin{subfigure}[b]{0.3\linewidth}
    \includegraphics[width=\linewidth]{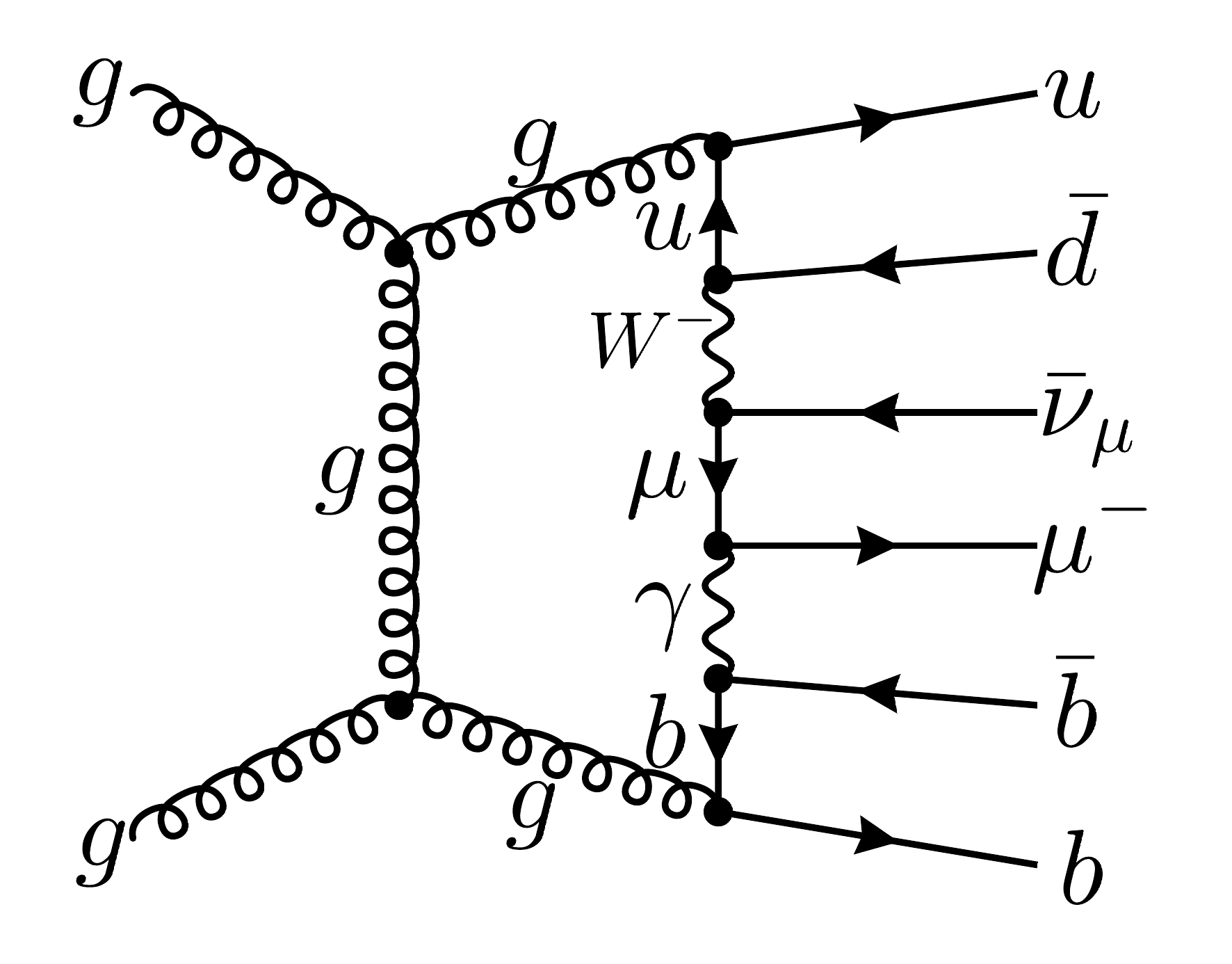}
    \vspace{0.1cm}
    \caption{}
    \label{fig:sub_c}
  \end{subfigure}\hfill
  \begin{subfigure}[b]{0.3\linewidth}
    \includegraphics[width=\linewidth]{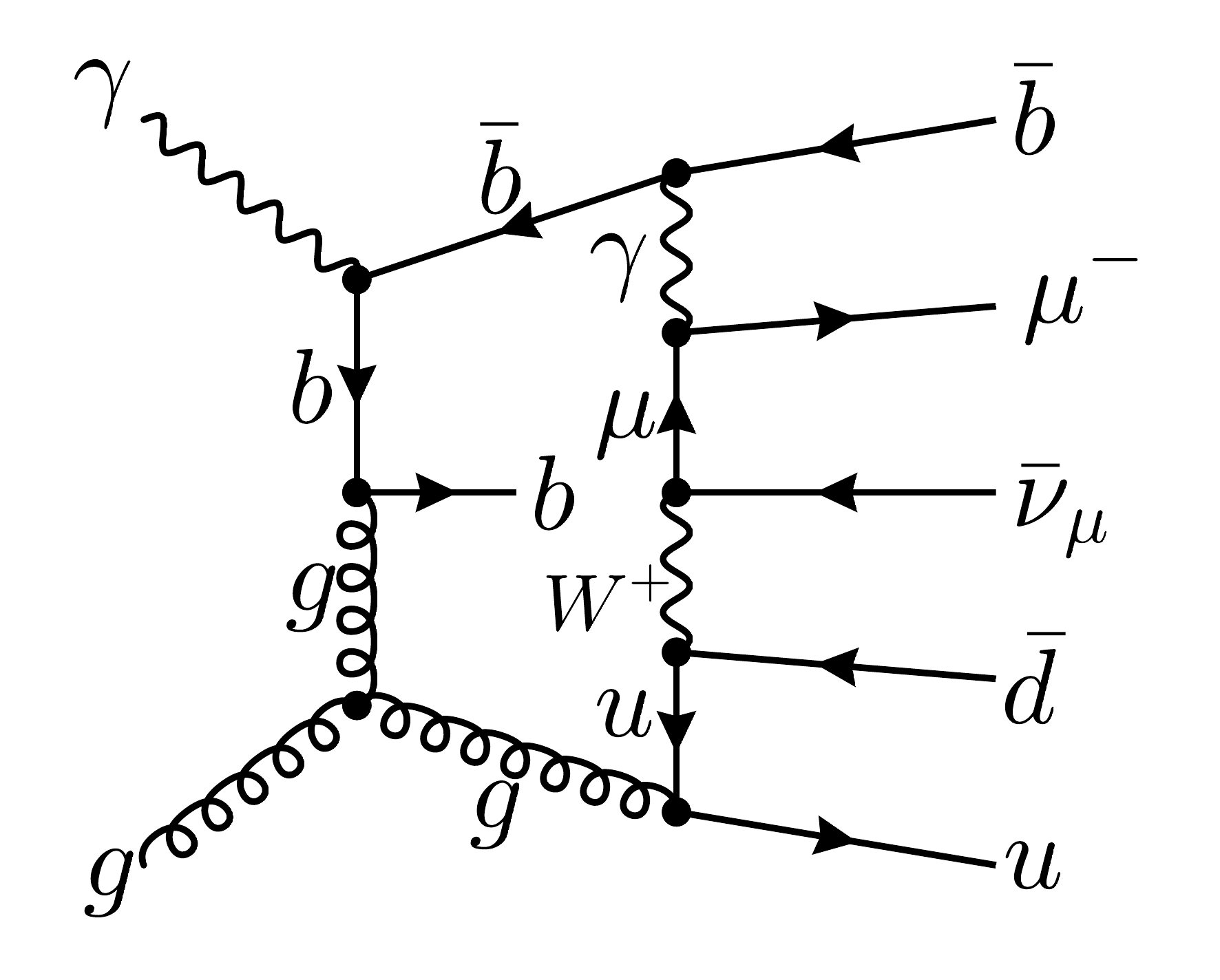}
    \vspace{0.1cm}
    \caption{}
    \label{fig:sub_d}
  \end{subfigure}\hfill
  \begin{subfigure}[b]{0.3\linewidth}
    \includegraphics[width=\linewidth]{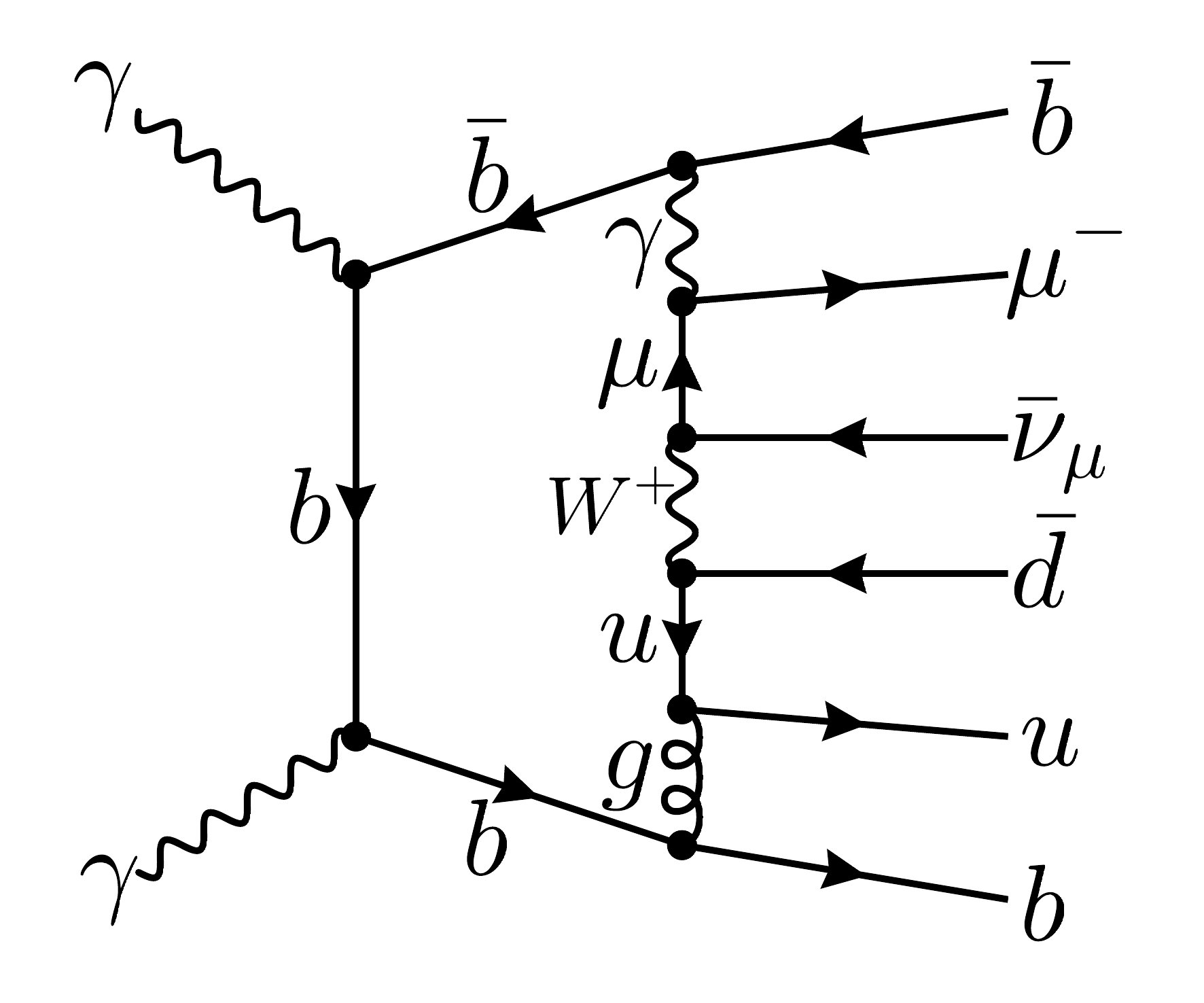}
    \vspace{0.1cm}
    \caption{}
    \label{fig:sub_e}
  \end{subfigure}

  \vspace{6pt}

  \makebox[\linewidth]{
    \begin{subfigure}[b]{0.3\linewidth}
      \includegraphics[width=\linewidth]{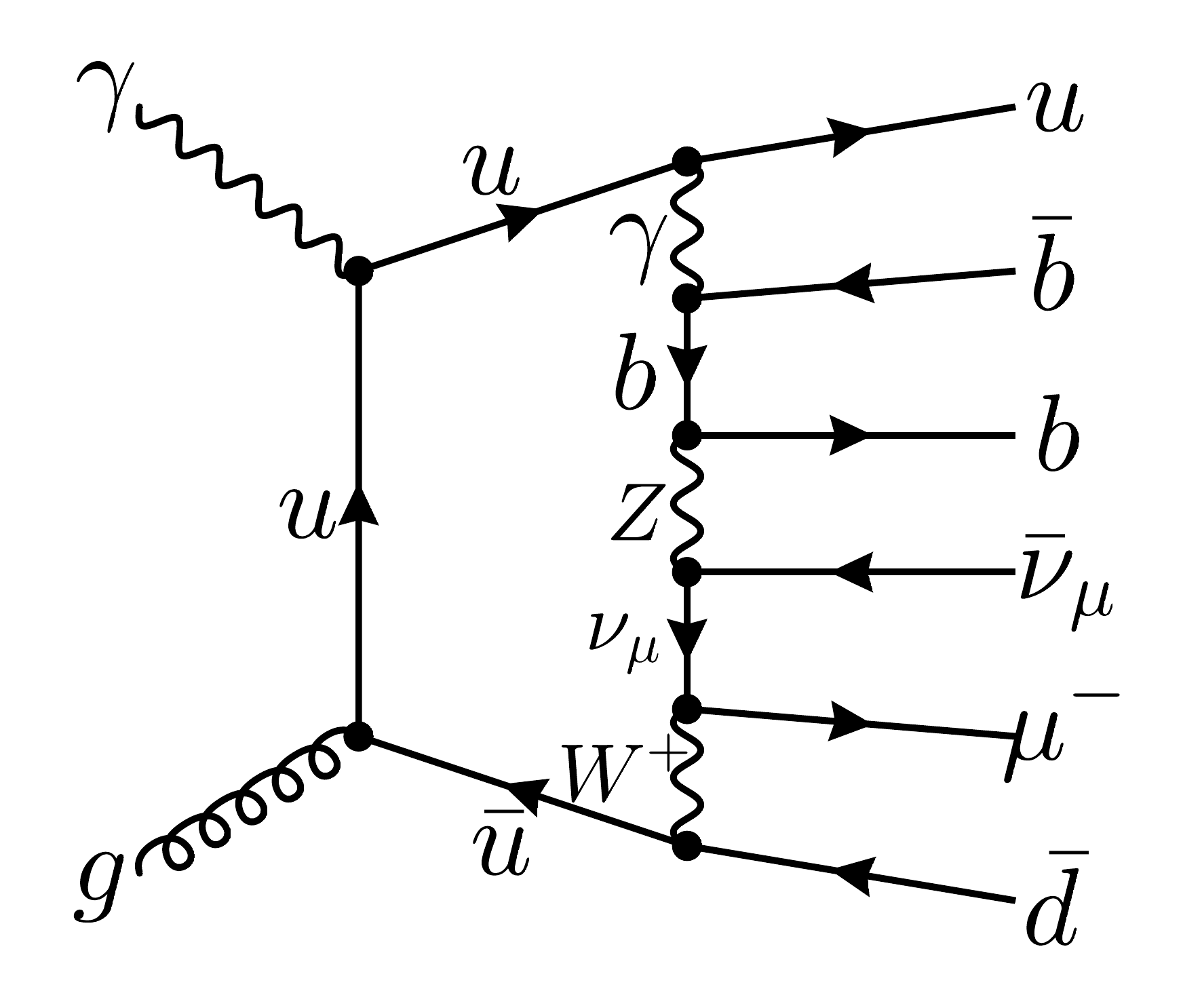}
      \vspace{0.1cm}
      \caption{}
      \label{fig:sub_f}
    \end{subfigure}\hspace{0.04\linewidth}
    \begin{subfigure}[b]{0.3\linewidth}
      \includegraphics[width=\linewidth]{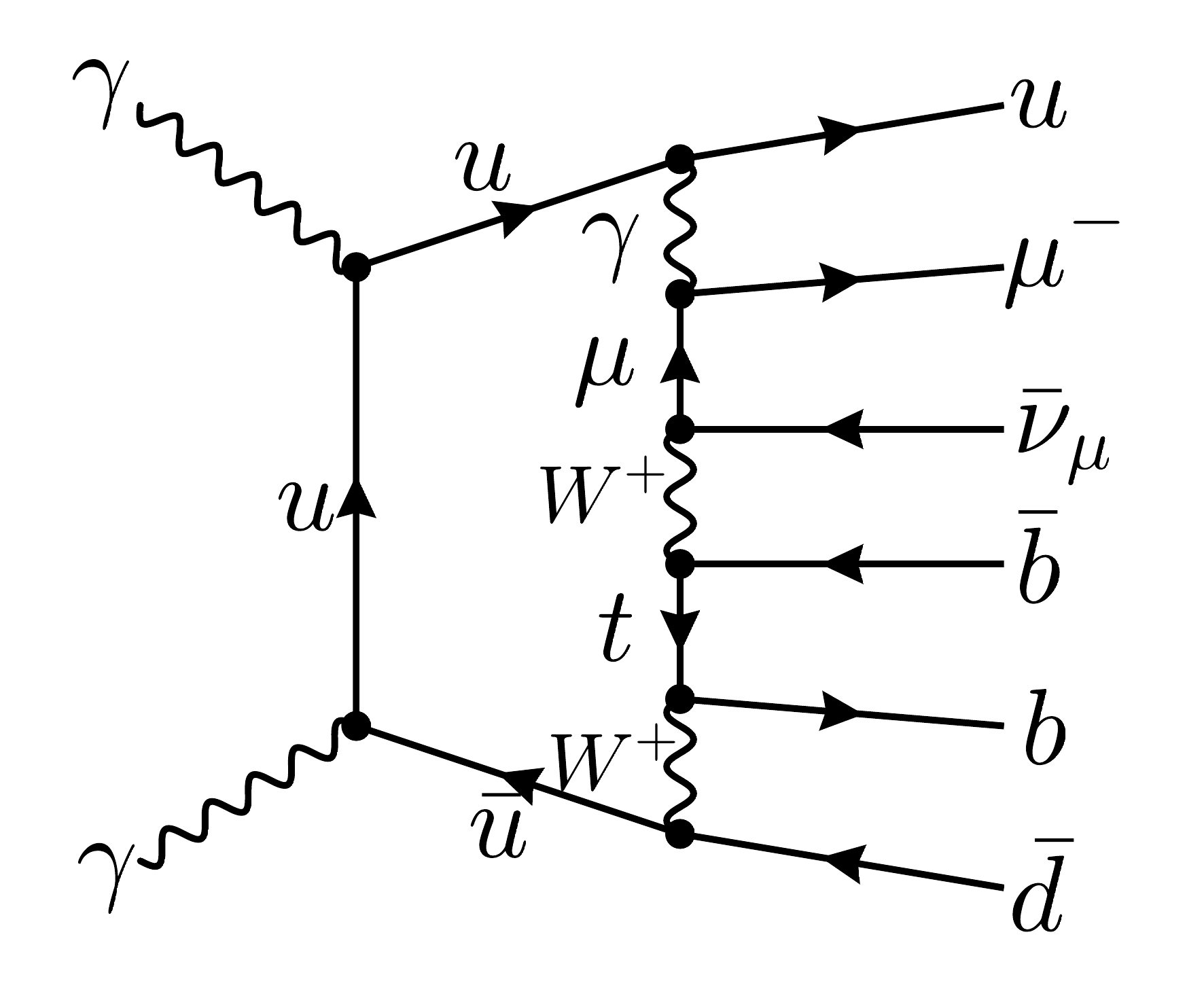}
      \vspace{0.1cm}
      \caption{}
      \label{fig:sub_g}
    \end{subfigure}
  }
  \caption{\it 
  Examples of one-loop Feynman diagrams for the $pp \to \ell^-\bar{\nu}_\ell\,j_b j_b jj+X$  process involving one-loop tensor integrals of various complexities: \eqref{fig:sub_a} 7-point, rank-5 integrals entering at $\mathcal{O}(g_s^6 g^2)$, \eqref{fig:sub_b} 7-point, rank-5 integrals entering at $\mathcal{O}(g_s^5 g^3)$, \eqref{fig:sub_c} 8-point, rank-5 integrals entering at $\mathcal{O}(g_s^4 g^4)$, \eqref{fig:sub_d} 8-point, rank-5 integrals entering at $\mathcal{O}(g_s^3 g^5)$, \eqref{fig:sub_e} 8-point, rank-5 integrals entering at $\mathcal{O}(g_s^2 g^6)$,  \eqref{fig:sub_f} 8-point, rank-5 integrals entering at $\mathcal{O}(g_s^1 g^7)$, \eqref{fig:sub_g} 8-point, rank-5 integrals entering at $\mathcal{O}(g_s^0 g^8)$. }
  \label{fig:oneloop_complex}
\end{figure}

In order to obtain complete NLO corrections, we compute all possible QCD and EW corrections to all the Born-level orders, as illustrated in Figure~\ref{fig:orders}. These include interferences between Born-level and one-loop Feynman diagrams involving QCD and/or EW effects. In Figure~\ref{fig:loop-orders} examples of Feynman diagrams are shown for three specific cases. The diagram on the left contributing at $\mathcal{O}(g_s^6 g^2)$ can be classified as the QCD corrections to $\mathcal{O}(g_s^4 g^2)$. The diagram in the middle contributing at $\mathcal{O}(g_s^2 g^6)$ can be classified as the EW corrections to $\mathcal{O}(g_s^2 g^4)$. Finally, the last diagram on the right at $\mathcal{O}(g_s^4 g^4)$ can be seen as a mixed case, containing both one-loop EW corrections to $\mathcal{O}(g_s^4 g^2)$ or one-loop QCD corrections to $\mathcal{O}(g_s^2 g^4)$. A direct consequence of this is that QCD and EW corrections cannot be considered separately, but must be combined together to ensure proper cancellation of all IR singularities (soft and/or collinear). In Figure \ref{fig:cancellationdiagramsnlo3} we show examples of  Feynman diagrams contributing to  the real corrections to \NLOthree that have to be included for the mixed case displayed in Figure \ref{fig:loop-orders}. To further illustrate the complexity of the one-loop calculations, in Figure \ref{fig:oneloop_complex} we display examples of  Feynman diagrams that show  various degrees of complexity. As can be noted, the most difficult contributions that we must consider comprise $8$-point tensor integrals of rank $5$.

The full list of NLO contributions is as follows:
\begin{itemize}
    \item \textbf{NLO$_1$} ($\mathcal{O}(\alpha_s^5 \alpha^2)$): The pure QCD corrections to the LO process at $\mathcal{O}(\alpha_s^4 \alpha^2)$. It includes real-emission Feynman diagrams at $\mathcal{O}(g_s^5 g^2)$ and virtual corrections arising from the interference between the amplitudes at order $\mathcal{O}(g_s^6 g^2)$ and $\mathcal{O}(g_s^4 g^2)$.

    \item \textbf{NLO$_2$} ($\mathcal{O}(\alpha_s^4 \alpha^3)$): The order that comprises QCD virtual corrections obtained from the interference of the amplitudes at $\mathcal{O}(g_s^6 g^2)$ and $\mathcal{O}(g_s^5 g^3)$ with tree-level amplitudes at $\mathcal{O}(g_s^2 g^4)$ and $\mathcal{O}(g_s^3 g^3)$, respectively. There is an additional contribution from the interference of Born-level diagrams at $\mathcal{O}(g_s^4 g^2)$ with one-loop diagrams at $\mathcal{O}(g_s^4 g^4)$. The latter can be viewed either as the QCD corrections to $\mathcal{O}(g_s^2 g^4)$ or EW corrections to $\mathcal{O}(g_s^4 g^2)$. The real-emission part includes the square of $\mathcal{O}(g_s^4 g^3)$ and the terms arising from the interference between amplitudes at order $\mathcal{O}(g_s^5 g^2)$ and $\mathcal{O}(g_s^3 g^4)$.

    \item \textbf{NLO$_3$} ($\mathcal{O}(\alpha_s^3 \alpha^4)$): The contribution that can be viewed in part as the NLO QCD corrections to the dominant Born-level contribution \LOthree describing QCD production of a $t\bar{t}$ pair. It is therefore expected to have a dominant contribution also at NLO. It contains QCD one-loop amplitudes at $\mathcal{O}(g_s^6 g^2)$ and contributions at $\mathcal{O}(g_s^5 g^3)$ that interfere with Born-level amplitudes at $\mathcal{O}(g_s^0 g^6)$ and $\mathcal{O}(g_s^1 g^5)$, respectively. In addition, there are interferences between tree-level and one-loop amplitudes at $\mathcal{O}(g_s^4 g^4) \times \mathcal{O}(g_s^2 g^4)$,  $\mathcal{O}(g_s^3 g^5) \times \mathcal{O}(g_s^3 g^3)$, and $\mathcal{O}(g_s^2 g^6) \times \mathcal{O}(g_s^4 g^2)$. In these cases the one-loop part can be interpreted as either QCD or EW corrections to different Born-level contributions. The real-emission part comprises the square of $\mathcal{O}(g_s^3 g^4)$, together with interferences at $\mathcal{O}(g_s^5 g^2)\times \mathcal{O}(g_s^1 g^6)$ as well as  $\mathcal{O}(g_s^4 g^3)\times \mathcal{O}(g_s^2 g^5)$.

    \item \textbf{NLO$_4$} ($\mathcal{O}(\alpha_s^2 \alpha^5)$): Similarly to  \NLOthree we have the interference of EW one-loop contributions at $\mathcal{O}(g_s g^7)$ and $\mathcal{O}(g_s^0g^8)$ with the Born-level amplitudes at $\mathcal{O}(g_s^3 g^3)$ and $\mathcal{O}(g_s^4 g^2)$, respectively. In addition, we encounter mixed contributions at order  $\mathcal{O}(g_s^4 g^4) \times \mathcal{O}(g_s^0 g^6)$, $\mathcal{O}(g_s^3 g^5) \times \mathcal{O}(g_s g^5)$, and $\mathcal{O}(g_s^2 g^6) \times \mathcal{O}(g_s^2 g^4)$, with loops containing gluons as well as EW bosons. The real-emission part contains the square of $\mathcal{O}(g_s^2 g^5)$ and interference terms at order $\mathcal{O}(g_s^3 g^4) \times \mathcal{O}(g_s g^6)$ and $\mathcal{O}(g_s^4 g^3) \times \mathcal{O}(g_s^0g^7)$.

    \item \textbf{NLO$_5$} ($\mathcal{O}(\alpha_s \alpha^6)$): The contributions that comprise one-loop amplitudes with EW corrections at $\mathcal{O}(g_s g^7)$ and $\mathcal{O}(g_s^0g^8)$, which interfere with Born-level diagrams at $\mathcal{O}(g_s g^5)$ and $\mathcal{O}(g_s^2 g^4)$, respectively. Furthermore, we have the interference at $\mathcal{O}(g_s^2 g^6) \times \mathcal{O}(g_s^0g^6)$. The real-emission contributions consist of the square of $\mathcal{O}(g_s g^6)$, as well as interference at $\mathcal{O}(g_s^2 g^5)\times \mathcal{O}(g_s^0g^7)$ involving photons in the initial or final state. These interference terms vanish due to the colour structures when the quarks involved belong to different generations.

    \item \textbf{NLO$_6$} ($\mathcal{O}(\alpha_s^0\alpha^7)$): Similarly to \NLOone, this contribution can be uniquely identified as the EW corrections to \LOfive. It includes the interference of one-loop amplitudes at $\mathcal{O}(g_s^0g^8)$ with Born-level amplitudes at $\mathcal{O}(g_s^0g^6)$. For the real corrections, it contains the squared matrix elements of Feynman diagrams at $\mathcal{O}(g_s^0g^7)$.
\end{itemize}
%

\section{Computational framework}
\label{sec:framework}

The calculation of the leading and subleading LO and NLO contributions closely follows the steps outlined in Ref. \cite{Stremmer:2024ecl}. However, we encounter additional complications that can be attributed to the presence of light final-state jets already at the Born level in subprocesses that contain at most one resonant top quark. Below, we will briefly summarise the parts of our computational framework that we have used in the past and instead focus in more detail on the new aspects.

For the calculation of the Born-level and one-loop matrix elements for the $pp \to \ell^-\bar{\nu}_\ell\,j_b j_b jj +X$ process we use  the \textsc{Recola} program \cite{Actis:2012qn,Actis:2016mpe}, which  provides numerical results in the 't Hooft-Feynman gauge and uses the complex-mass scheme for all massive and unstable particles to preserve gauge invariance \cite{Denner:1999gp,Denner:2005fg,Denner:2006ic,Denner:2019vbn}. Additionally, we have implemented in \textsc{Recola} a random polarisation method, see e.g. Refs.~\cite{Draggiotis:1998gr,Draggiotis:2002hm,Bevilacqua:2013iha}, to avoid the explicit helicity summation. In this method, the polarisation state is replaced by a linear combination of helicity eigenstates while the spin summation is replaced by an integration over a phase parameter. This approach leads to a drastic speed improvement, especially for processes with many final-state particles. The one-loop amplitudes are  decomposed into tensor coefficients $c^{(t)}_{\mu_1\cdots\mu_r}$, which are calculated internally in \textsc{Recola}, and tensor integrals $T^{\mu_1\cdots\mu_r}_{(t)}$, which are evaluated with \textsc{Collier} \cite{Denner:2016kdg}, the program  for the numerical evaluation of one-loop scalar and tensor integrals. Alternatively, tensor coefficients can be used as input for an integrand-level decomposition via the OPP method \cite{Ossola:2006us} as implemented  in \textsc{CutTools} \cite{Ossola:2007ax}. In this case, the scalar integrals are then evaluated with \textsc{OneLOop} \cite{vanHameren:2010cp}. We employ this approach to cross-check phase-space points that are identified by \textsc{Collier} as unstable. The phase-space integration is performed using the adaptive multi-channel generators \textsc{Parni} \cite{vanHameren:2007pt} and \textsc{Kaleu} \cite{vanHameren:2010gg}. To cross-check our one-loop amplitudes, we compared the matrix element for 5000 phase-space points for each partonic channel listed in Table \ref{tab:LO orders} with the results obtained in the Background-Field method as implemented in  \textsc{Recola2} \cite{Denner:2017wsf}.  

The real corrections are induced by the radiation of an additional photon or QCD parton. To isolate the soft and collinear divergences, the following two subtraction schemes are applied: Nagy-Soper \cite{Bevilacqua:2013iha} and Catani-Seymour \cite{Catani:1996vz,Catani:2002hc}. We use the first one as our default, while the second scheme is mainly employed for cross-checking purposes. Both schemes, which are implemented in \textsc{Helac-Dipoles} \cite{Czakon:2009ss}, are extended to handle soft or collinear photon emissions \cite{Dittmaier:1999mb,Dittmaier:2008md}. To further cross check our predictions, we have explored the independence of the results from the unphysical phase-space restriction on the unresolved one-particle phase space in the subtraction terms, denoted $\alpha_{max}$, as  implemented in  \textsc{Helac-Dipoles}, see Refs. \cite{Nagy:1998bb,Nagy:2003tz,Bevilacqua:2009zn,Czakon:2015cla} and references therein for more details. In addition, we have verified the cancellation of divergences between the real and virtual corrections. Specifically, for each subprocess and for several phase-space points we checked that the poles in $\varepsilon$ that appear in the ${\cal I}$-operator and those appearing in the virtual part of our calculations cancel. 

Theoretical predictions obtained with the help of the \textsc{Recola} and \textsc{Helac-Dipoles} programs are stored in the form of modified Les Houches Event Files  (LHEF) \cite{Alwall:2006yp}. We store each “event” with supplementary matrix element and PDF information \cite{Bern:2013zja}, which allows us to obtain results for different scale settings and PDF choices by reweighting. Moreover, storing “events” has clear advantages when it is necessary to change observables and binning, or to add more exclusive cuts.

To ensure the IR-safety of our higher-order predictions, we extended the implementation of the photon fragmentation function in \textsc{Helac-Dipoles} \cite{Stremmer:2024zhd} and implemented a photon-to-jet conversion function in both subtraction formalisms. Details of these modifications are presented below.

\subsection{Democratic clustering and parton-to-photon fragmentation function}
\label{sec:fragmentation}

To define the $pp\to \ell^-\bar{\nu}_\ell\,j_b j_b jj+X$ process at NLO, care must be taken if an additional photon is present when higher-order corrections of the order ${\cal O}(\alpha)$ are calculated. In this case, simultaneously occurring photons, charged leptons, and QCD partons must be combined together with the help of a jet algorithm. In particular, we are dealing with photons that are emitted collinearly to a light jet. To ensure IR safety, we employ the democratic clustering approach \cite{Glover:1993xc,Gehrmann-DeRidder:1997fom} and a quark-to-photon fragmentation function. In the democratic clustering approach, photons and partons are treated on the same footing and are combined using the anti-$k_T$ jet algorithm. To separate hard photons from jets, the resulting object is denoted as a “photon jet” (i.e. an isolated photon) if $z_\gamma > z_{\gamma,\text{cut}}$, and as a hadronic jet otherwise, where $z_\gamma$ is defined according to 
\begin{equation}
z_\gamma = \frac{p_{T,\, \gamma}}{p_{T, \,\gamma}+ p_{T, \,i}} \,.\\[0.2cm]
\label{eq:zgamma}
\end{equation}
Here, $p_{T,\, \gamma}$ is the transverse momentum of the photon, $p_{T, \,i}$ is the transverse momentum of the parton $i=q,\bar{q},g$ inside this jet, while $z_{\gamma, \,\text{cut}}$ is some cut value, where typical choices include  $z_{\gamma, \,\text{cut}} \in [0.5-0.9]$. The separation into jets and photon jets is necessary because otherwise soft gluons can be recombined with hard photons in the jet algorithm resulting in a hard jet passing the event selection. The soft singularities associated with such configurations would normally cancel with the virtual corrections of the Born-level configuration where the soft gluon in the final state is removed. However, the Born-level contributions and virtual corrections of such subprocesses are not part of the definition of the Born-level process because the requirement on the minimal number of jets is no longer fulfilled, thus spoiling IR safety. If we apply this recombination procedure, for example, in the $q\,\bar{q}^{\,\prime} \to \ell^- \bar{\nu}_{\ell}\, b\,\bar{b}\, g\,g ({\gamma})$ partonic channel, the recombination of any soft gluon and a hard (collinear) photon will lead to the formation of a photon jet and such events will be discarded as they do not contain the required number of jets. However, since distinguishing quarks and gluons is not possible experimentally, imposing such a strict photon veto makes higher-order calculations, for example for the  $q\,\bar q^{\,\prime} \to \ell^- \bar{\nu}_{\ell}\, b\,\bar{b}\, q^{\,\prime}\,\bar q^{\,\prime}(\gamma)$ partonic channel, IR unsafe. In this case, the collinear $q \to q\gamma$ region will no longer be treated inclusively, and the cancellations between the real emission part and virtual corrections would fail, leaving $1/\epsilon$ poles that are not cancelled. To absorb these final-state collinear singularities, following  the approach proposed e.g. in Refs. \cite{Denner:2009gj,Denner:2010ia,Denner:2014ina,Denner:2014bna}, we incorporate the quark-to-photon fragmentation function \cite{Glover:1993xc,Gehrmann-DeRidder:1997fom} into our computational framework.

In order to ensure proper cancellation of IR singularities, the subtraction terms from standard NLO subtraction schemes have to be restricted by the same requirement on the photon energy fraction in the jet given by $z_\gamma < z_{\gamma,\text{cut}}$. This can be achieved by constructing a proxy variable $\tilde{z}_\gamma$ for the subtraction terms. This variable must be equivalent to $z_\gamma$ in the collinear limit, tends to zero in the soft limit while being defined on the interval $\tilde{z}_\gamma \in [0,1]$. It has to be constructed from the $n+1$ phase space before projecting the momenta to the underlying Born-level $n$-particle phase space by recombining the splitting pair. To this end, we modify the usual jet function featured in the dipole subtraction terms of the real-emission contribution to include the cut on $\tilde{z}_\gamma$:
\begin{equation}
J_n\bigl(\{\tilde{p}\}_n^{(ij)k},\,\tilde{z}_\gamma\bigr)
  = J_n\bigl(\{\tilde{p}\}_n^{(ij)k},\,\tilde{p}_{\rm jet}
  =\tilde{p}_i\bigr)
    \;\Theta(z_{\gamma,\,\text{cut}}-\tilde{z}_\gamma)\,. \\[0.2cm]
\end{equation}
Here $\{\tilde{p}\}_n^{(ij)k}$ denotes the set of $n$-mapped on-shell momenta obtained from the $(n+1)$-parton configuration by combining partons $i$ and $j$ into a single emitter $\tilde p_{i}$ with spectator $k$. In the democratic clustering approach, the jet that results from the parton-photon cluster inherits the momentum of the recombined splitting pair $\tilde{p}_i$. Therefore, the entire $\tilde{z}_\gamma$ dependence can be isolated to the $\Theta (z_{\gamma,\,\text{cut}}-\tilde{z}_\gamma)$ Heaviside function that multiplies the standard jet function. Of course, by limiting the phase space of the local subtraction terms, we also have to limit the integrated part of these subtraction terms in exactly the same way. Thus, the NLO cross section can schematically be written as
\begin{align}
\dd\hat\sigma_{\text{NLO}}
  &= \int_{n+1}\!
     \bigl[\Theta(z_{\gamma, \, \text{cut}}-z_\gamma)\dd\sigma_{\text{real}}-\Theta(z_{\gamma,\,\text{cut}}-\tilde{z}_\gamma)\dd\sigma_{\text{dipole}}\bigr]
\nonumber\\[0.2cm]
  &\quad + \int_n\!
     \Bigl[d\sigma_{\text{virtual}}
           +\!\int_1\Theta(z_{\gamma,\,\text{cut}}-\tilde{z}_\gamma)\,
             \dd\sigma_{\text{dipole}}-\dd\hat{\sigma}_{\rm frag}(z_{\gamma,\,\text{cut}})\Bigr]\,.
\end{align}
The last term represents the convolution of the Born-level cross section with the bare fragmentation function which is given by
\begin{align}
\dd\hat{\sigma}_{\rm frag}(z_{\gamma,\,\text{cut}}) 
  &= \sum_{j}\dd\sigma_{\rm Born}\int^{1}_{z_{\gamma,\,\text{cut}}}\dd z_\gamma\,
       D^{B}_{j\to\gamma}(z_\gamma)\,,
\label{eq:frag_subtr}
\end{align}
where $D^{B}_{j\to\gamma}(z_\gamma)$ is the bare fragmentation function and the sum runs over all (anti-) quarks in the final state.
Since the region $z_{\gamma,\,\text{cut}}>\tilde{z}_\gamma$ of the $q\to q\gamma$ splitting contains both soft and collinear singularities, it is in practice more convenient to rewrite the integrand belonging to the one-particle phase space of the unresolved particle as
\begin{equation}
\Theta(z_{\gamma,\,\text{cut}}-\tilde{z}_\gamma)\,\dd\sigma_{\text{dipole}}
 - \dd\hat{\sigma}_{\rm frag}(z_{\gamma,\,\text{cut}})
= \dd\sigma_{\text{dipole}}
 - \bigl[\Theta( \tilde{z}_\gamma-z_{\gamma,\,\text{cut}})\,\dd\sigma_{\text{dipole}}
 + \dd\hat{\sigma}_{\rm frag}(z_{\gamma,\,\text{cut}})\bigr]\,. \\[0.2cm]
\end{equation}
In this way, the term $d\sigma_{\text{dipole}}$ denotes the standard contribution from the integrated subtraction terms in a general subtraction scheme, while the term $\Theta(\tilde{z}_\gamma-z_{\gamma, \,\text{cut}})\,d\sigma_{\text{dipole}}$ contains by definition only collinear singularities, which are exactly cancelled by $d\hat{\sigma}_{\rm frag}(z_{\gamma,\text{cut}})$. The only subtraction terms that are affected by the  $\tilde{z}_\gamma$ cut are associated with the final-state $q \to q \gamma$ splitting with final- and initial-state spectators. 

The modified $q\to q\gamma$ dipoles in the Catani-Seymour subtraction scheme can be obtained from the corresponding $q \to q g$ ones \cite{Catani:1996vz}, by simple replacement of the colour factors and by adding an additional restriction on the unresolved one-particle phase space. The photon energy fraction of the $q \to q \gamma$ splitting can be written as
\begin{equation}
\tilde{z}_\gamma
  = \frac{p_\gamma\!\cdot p_k}
         {p_q\!\cdot p_k + p_\gamma\!\cdot p_k}\,, \\[0.2cm]
\end{equation}
where $p_q$, $p_\gamma$ and $p_k$ are the momenta of the quark, the photon and the spectator particle $k$, respectively. It is related to the energy fraction of the parton, $\tilde{z} = 1-\tilde{z}_\gamma$, which is also used as an integration variable. The integrated dipoles for final- and initial-state spectators with an additional restriction on the unresolved one-particle phase space parametrised with $\alpha_{max}$ have been calculated in Ref. \cite{Denner:2014ina} and are presented here for completeness. The $q \to q g$ dipole with a final-state spectator is given by
\begingroup
\setlength{\abovedisplayskip}{5pt}
\setlength{\belowdisplayskip}{5pt}
\begin{align}
V^{\text{coll}}_{q\gamma}(\alpha_{max},\epsilon,z_{\gamma,\text{cut}})
  &= Q_q^2
     \int_0^{1-z_{\gamma,\text{cut}}}\!\!\dd\tilde{z}\,
     (\tilde{z}(1-\tilde{z}))^{-\epsilon}
     \int_0^{\alpha_{max}}\!\dd y\,
     y^{-1-\epsilon}(1-y)^{1-2\epsilon}
\nonumber\\[0.2cm]
  &\quad\times
     \Bigl[\frac{2}{1-\tilde{z}+y\tilde{z}}-(1+\tilde{z})-\epsilon(1-\tilde{z})\Bigr]\\[0.2cm]
  &= Q_q^2 \int_0^{1-z_{\gamma,\text{cut}}} \dd\tilde{z} \left\{ \frac{1+\tilde{z}^2}{1-\tilde{z}} \left( -\frac{1}{\epsilon} + \log[\tilde{z}(1-\tilde{z})] + \log\alpha_{max} \right)  \right. \nonumber \\[0.2cm]
&\quad \left. + \,\alpha_{max}(1+\tilde{z}) + 1 - \tilde{z} - \frac{2}{\tilde{z}(1-\tilde{z})} \log\left(\frac{1-\tilde{z}+\alpha_{max} \tilde{z}}{1-\tilde{z}}\right) \right\}\,,
\label{eq:V_ff_full}
\end{align}
\endgroup
where $Q_q$ is the electric charge of the splitting fermion in units of $e$ and $y$ parametrizes the unresolved region of the dipole phase space in the Catani-Seymour subtraction scheme.
On the other hand, for an initial-state spectator it can be written as 
\begingroup
\setlength{\abovedisplayskip}{5pt}
\setlength{\belowdisplayskip}{5pt}
\begin{align}
\mathcal V^{\gamma,\text{coll}}_{q\gamma}(x;\alpha_{max},\epsilon,z_{\gamma,\text{cut}})
&= Q_q^2\,
   \theta(1-x)\,\theta(x-1+\alpha_{max})\,(1-x)^{-1-\epsilon}
   \int_0^{1-z_{\gamma,\text{cut}}}\!\dd\tilde{z}\,(\tilde{z}(1-\tilde{z}))^{-\epsilon}
\nonumber\\[0.2cm]
&\quad\times
   \Bigl[\frac{2}{2-\tilde{z}-x}-(1+ \tilde{z})-\epsilon(1- \tilde{z})\Bigr]\\[0.2cm]
  &= Q_q^2 \int_0^{1-z_{\gamma,\text{cut}}} \dd\tilde{z} \left\{ \left[ \frac{1}{1-x}\left(\frac{2}{2-\tilde{z}-x}-1-\tilde{z}\right)\right]_{1-\alpha_{max}} \right. \nonumber \\[0.2cm]
&\quad +\, \delta(1-x) \left[ \frac{1+\tilde{z}^2}{1-\tilde{z}}\left(-\frac{1}{\epsilon}+\log[\tilde{z}(1-\tilde{z})]+\log\alpha_{max}\right) \right. \nonumber \\[0.2cm]
&\quad \left.\left.  +\, (1-\tilde{z}) - \frac{2}{1-\tilde{z}}\log\left(\frac{1-\tilde{z}+\alpha_{max}}{1-\tilde{z}}\right) \right] \right\}\,,
\label{eq:V_fi_full}
\end{align}
\endgroup
where $x$ denotes the longitudinal momentum fraction of the incoming emitter after the unresolved emission, defined by $\tilde p_a = x\,p_a$, which satisfies  $x \to 1$ in the collinear limit. The $(1-\alpha_{max})$ distribution is defined as
\begin{equation}
\int^1_0 \dd x\, \left(f(x)\right)_{1-\alpha_{max}}\,g(x)=\int^1_{1-\alpha_{max}}\dd x f(x)\left(g(x)-g(1)\right)\,.\\[0.2cm]
\end{equation}
In this way, $\alpha_{max}=1$ corresponds to the usual plus distribution. 

In the Nagy-Soper subtraction scheme we define the photon energy fraction as
\begin{equation}
\tilde{z}_\gamma(e,c)
= \frac{p_\gamma \cdot Q}{p_q \cdot Q + p_\gamma \cdot Q} \,,
\end{equation}
where $Q$ is the total momentum of the process and the variables $e$ and $c$ parametrize the soft ($e \to 0$) and collinear ($c \to 1$) phase space of the splitting, respectively. This definition coincides with $\tilde{z}_\gamma$ from the Catani-Seymour subtraction scheme when the following substitution is performed $p_k \to Q$. The Nagy-Soper subtraction terms are split into the diagonal $W^{(ii,j)}$ and interference term $W^{(ik,j)}$ according to 
\begin{equation} \label{eq:nsstruct}
\mathcal{D}^{(ijk)}_{\tilde{s}_1\tilde{s}_2}=W^{(ii,j)}_{\tilde{s}_1\tilde{s}_2}\delta_{ik}+W^{(ik,j)}_{\tilde{s}_1\tilde{s}_2}(1-\delta_{ik})\, \delta_{\tilde{s}_1\tilde{s}_2}\,,\\[0.2cm]
\end{equation}
where $i,j$ correspond to the emitter pair from the splitting $\widetilde{ij}\to i+j$, $k$ is the spectator particle and $\tilde{s}_1,\tilde{s}_2$ are spin indices of the splitting particle $\widetilde{ij}$. By construction, the diagonal part $W^{(ii,j)}$ contains soft and collinear singularities, while only soft divergences are present in the interference term $W^{(ik,j)}$. 

Since the democratic clustering approach affects only the collinear limit, we restrict the phase space of the diagonal terms only. In practise, we impose the condition $\tilde{z}_{\gamma,\rm{cut}}>\tilde{z}_{\gamma}$, which is implemented through the following replacement
\begin{equation}
W^{(ii,j)}_{\tilde{s}_1\tilde{s}_2} \to W^{(ii,j)}_{\tilde{s}_1\tilde{s}_2}\Theta(z_{\gamma,\text{cut}}-\tilde{z}_\gamma)\,,\\[0.2cm]
\end{equation}
while the interference term remains unchanged. We note that this term is completely independent of the spectator particle $k$, since also the momentum mapping does not depend on it. Therefore, we have only a single new integrated dipole with respect to the standard Nagy-Soper subtraction scheme, which has already been calculated in Ref. \cite{Stremmer:2024zhd} based on a semi-numerical approach.

Following the notation given in Ref. \cite{Gehrmann:2022cih}, see also Ref. \cite{Stremmer:2024zhd}, the bare parton-to-photon fragmentation function given in Eq. \eqref{eq:frag_subtr} can be obtained by inverting the following relation
\begin{align}
D_{i\to\gamma}(z_\gamma,\mu^2_{Fr})
  &= \sum_{j}\;
     \Gamma_{i\to j}(z_\gamma,\mu^2_{Fr})
     \,\otimes\,D^{B}_{j\to\gamma}(z_\gamma)\,,
\label{eq:mass_fact}
\end{align}
where $D_{i\to\gamma}$ are the renormalised fragmentation functions, $\mu_{Fr}$ is the factorisation scale for the fragmentation process while the sum effectively runs over all final state quarks. The mass-factorization kernels $\Gamma_{i\to j}$ can be expanded in a series up to NLO and take the following form
\begin{align}
\Gamma_{i\to j}(z_\gamma,\mu_{Fr}^2)
  &= \delta_{ij}\,\delta(1-z_\gamma)
     + \frac{\alpha}{2\pi} \Gamma^{(0)}_{i\to j}(z_\gamma,\mu_{Fr}^2)\,.
\end{align}
In particular, we only need the leading term $\Gamma^{(0)}_{q\to \gamma}$, which is provided by
\begin{align}
\Gamma^{(0)}_{q\to \gamma}(z_\gamma,\mu_{Fr}^2)
  &= -\frac{1}{\epsilon}\, Q_q^2\,
      \frac{(4\pi)^\epsilon}{\Gamma(1 - \epsilon)}\,
      \left( \frac{\mu_R^2}{\mu_{Fr}^2} \right)^{\!\epsilon}
      P_{q\to \gamma}(z_\gamma)\,,
\end{align}
where $\mu_R$ is the renormalization scale and $P_{q\to \gamma}(z_\gamma)$ is the quark-to-photon splitting function given by
\begin{equation}
P_{q\to\gamma}(z_\gamma) = \frac{1+(1-z_\gamma)^2}{z_\gamma}\,.\\[0.2cm]
\end{equation}
Finally, the bare quark-to-photon fragmentation function can be written as follows 
\begin{align}
D^{B}_{q\to\gamma}(z_\gamma)
  &= D_{q\to\gamma}(z_\gamma, \mu^2_{Fr})
     +\frac{\alpha Q_q^2}{2\pi}\,
     \frac{(4\pi)^\epsilon}{\Gamma(1 - \epsilon)}
      \frac{1}{\epsilon}\,
      \left(\frac{\mu_R^2}{\mu^2_{Fr}}\right)^{\!\epsilon}\,
      P_{q\to\gamma}(z_\gamma)\,.
\label{eq:bare_inverted}
\end{align}
The finite remainder $D_{q\to\gamma}(z,\mu_{Fr}^2)$ encodes the physical fragmentation probability.
To calculate the scale dependence of the fragmentation function, we use the renormalization group approach. Requiring the bare function to be scale independent,
\begin{equation}
  \mu^2_{Fr}\, \frac{\dd D^{B}_{q\to\gamma}(z_\gamma)}{\dd\mu^2_{Fr}}=0\,,\\[0.2cm]
\end{equation}
one obtains the leading–order DGLAP equation
\begin{align}
\mu^2_{Fr}\,
\frac{\dd D_{q\to\gamma}(z_\gamma,\mu^2_{Fr})}{\dd\mu^2_{Fr}}
  &= \frac{\alpha Q_q^2}{2\pi}\,
  \frac{(4\pi)^\epsilon}{\Gamma(1 - \epsilon)}
     P_{q\to\gamma}(z_\gamma)\,.
\end{align}
Its leading logarithmic solution reads
\begin{align}
D_{q\to\gamma}(z_\gamma,\mu^2_{Fr})
  &= D^{\text{np}}_{q\to\gamma}(z_\gamma,\mu_0^2)
     +\frac{\alpha Q_q^2}{2\pi}\,
     \frac{(4\pi)^\epsilon}{\Gamma(1 - \epsilon)}
      P_{q\to\gamma}(z_\gamma)\,
      \log\left(\frac{\mu_{Fr}^2}{\mu_0^2}\right)\,,
\label{eq:dglap_sol}
\end{align}
where $ D^{\text{np}}_{q\to\gamma}(z_\gamma,\mu_0^2)$ is the non–perturbative input that needs to be taken from experimental measurements. This non-perturbative input is obtained from the ALEPH data for $e^+e^- \to \gamma + jets$ \cite{ALEPH:1995zdi}. In particular, the analysis consisted of measuring hadronic $Z$ decay events  with at least one jet containing a large photon energy fraction of $z_\gamma>0.7$. Its explicit form obtained from the lowest-order fit is given by 
\begin{align}
D^{\text{np}}_{q\to\gamma}(z_\gamma,\mu_0^2)
  &= \frac{\alpha Q_q^2}{2\pi}\,
  \frac{(4\pi)^\epsilon}{\Gamma(1 - \epsilon)}
     \bigl[
       -P_{q\to\gamma}(z_\gamma)\,\log\left[(1-z_\gamma)^2\right]
       +C
     \bigr]\,,
\end{align}
where $\mu_0=0.14~\text{GeV}$ and  $C=-13.26$.

\subsection{Photon--to--jet conversion function}
\label{sec:photon-jet-conversion}

\begin{figure}[t!]
    \centering
    \hspace*{\fill}
    \begin{minipage}{0.36\linewidth}
        \centering
        \includegraphics[width=\linewidth]{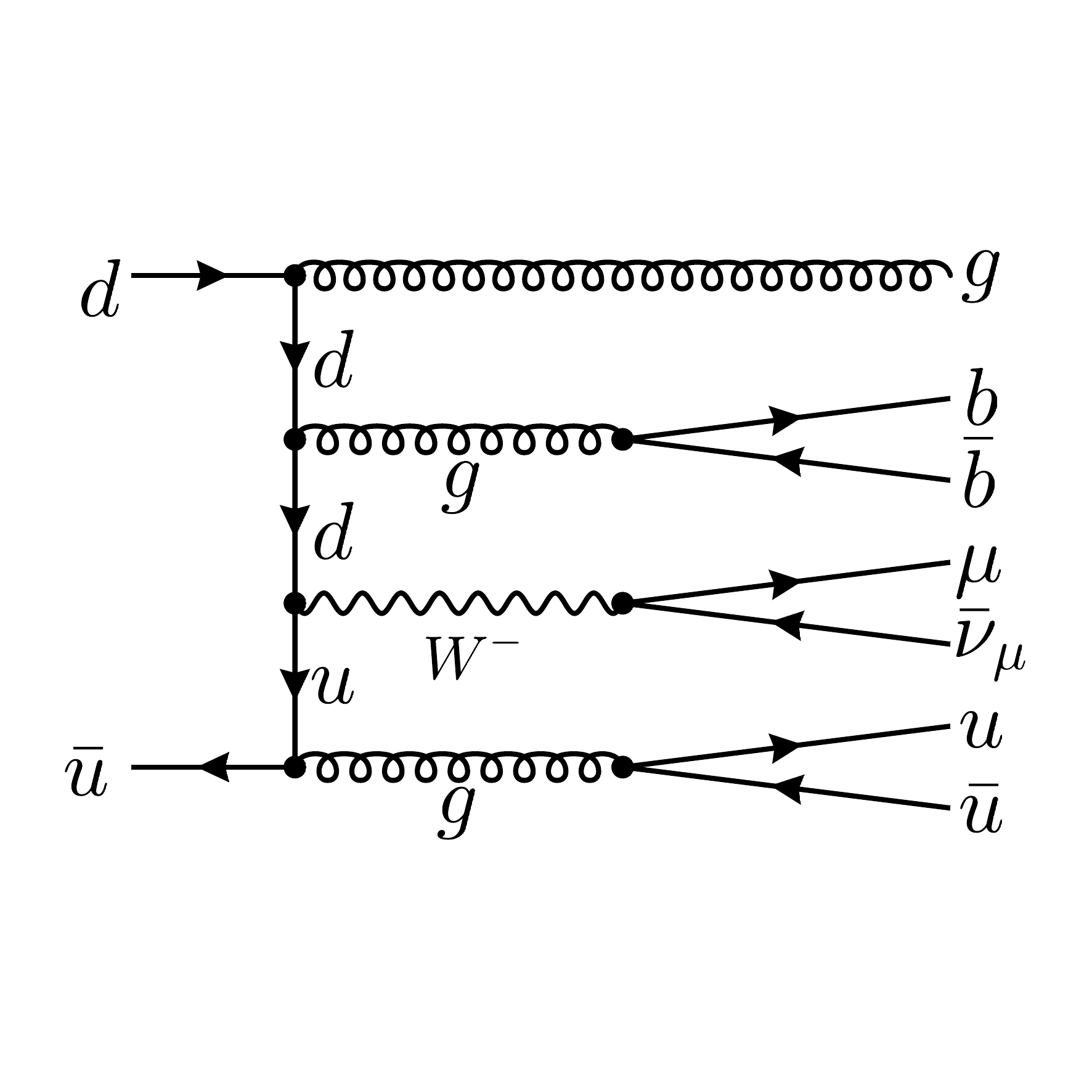}
    \end{minipage}
    \hspace*{\fill}
    \begin{minipage}{0.36\linewidth}
        \centering
        \includegraphics[width=\linewidth]{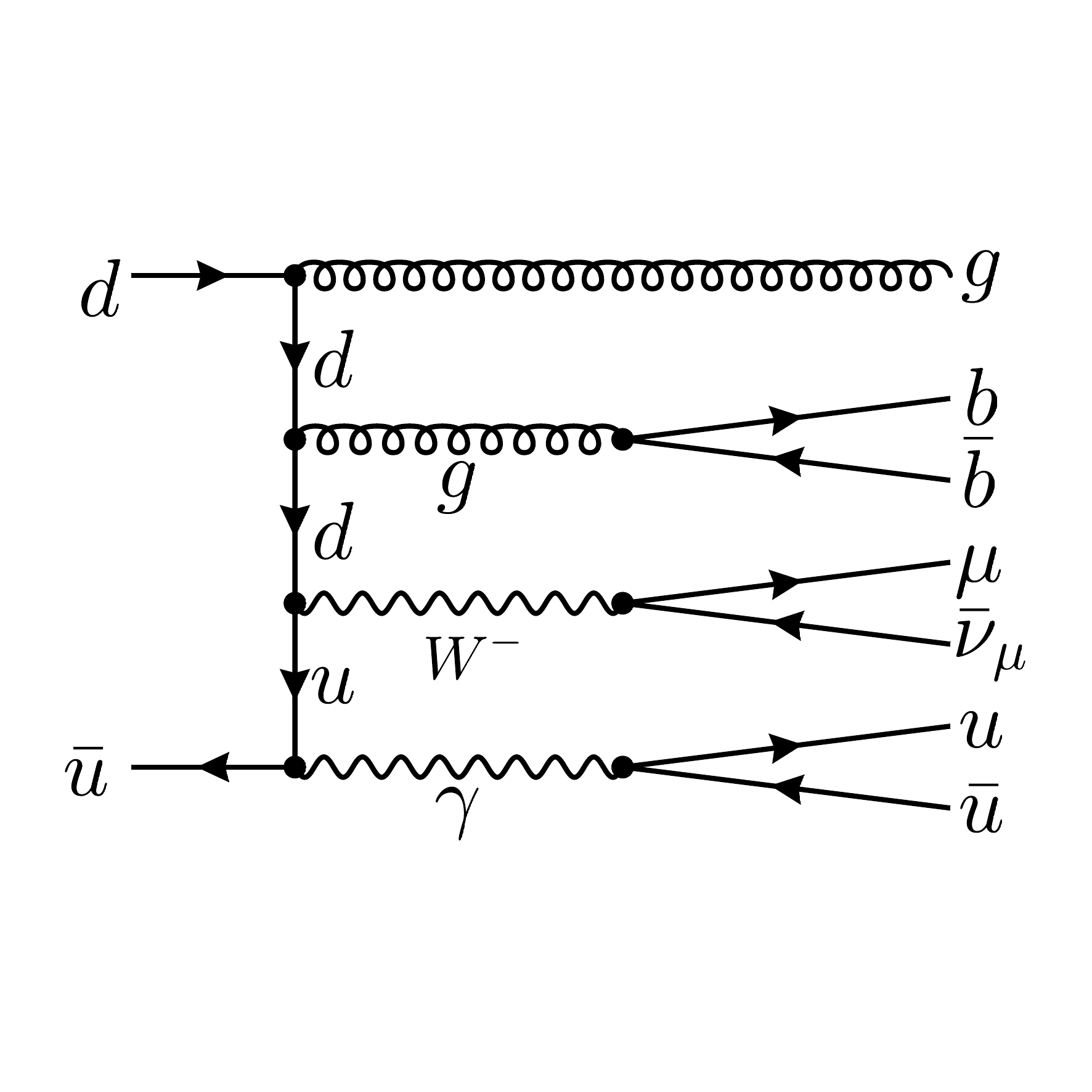}
    \end{minipage}
    \hspace*{\fill}
    \vspace{-1cm}
    \caption{\it Examples of real-emission Feynman diagrams contributing to the $pp \to \ell^-\bar{\nu}_\ell\,j_b j_b jj +X$ process at $\text{NLO$_1$}$ via final state gluon splitting (left) and at $\text{NLO$_3$}$ via final state photon splitting (right). }
    \label{fig:real-emission-splitting}
\end{figure}

In addition to the problems mentioned earlier, the mixing of EW and QCD corrections introduces further complications for the $pp \to \ell^-\bar{\nu}_\ell\,j_b j_b jj +X$ process. Specifically, some mechanisms for producing gluon jets have a direct counterpart in photon production. As an example in Figure \ref{fig:real-emission-splitting} we show two real-emission Feynman diagrams. The diagram on the left contributing to $\text{NLO$_1$}$ at $\mathcal{O}(\alpha_s^5\alpha^2)$ through gluon splitting ($g\to u\bar{u}$) in the final state can be classified as the QCD corrections to $\mathcal{O}(\alpha_s^4\alpha^2)$. In this case, if the quark and antiquark are very close they are merged to one jet. The resulting collinear singularity in the $d\bar{u} \to \ell^-\bar{\nu}_\ell\,b \bar{b} g (g\to u\bar{u}) +X$ partonic channel is cancelled by adding the virtual QCD corrections to the  $d\bar{u} \to \ell^-\bar{\nu}_\ell\,b \bar{b} gg +X$ subprocess. On the other hand, the diagram on the right contributes to $\text{NLO$_3$}$ at  $\mathcal{O}(\alpha_s^3\alpha^4)$ via photon splitting. The resulting collinear singularity in the $d\bar{u} \to \ell^-\bar{\nu}_\ell\,b \bar{b} g (\gamma\to u\bar{u}) +X$ partonic channel would normally be cancelled by adding the virtual EW corrections to the  $d\bar{u} \to \ell^-\bar{\nu}_\ell\,b \bar{b} g\gamma +X$ partonic channel. However, this underlying Born-level subprocess, in which a $u\bar{u}$ pair is replaced with a photon, is excluded by our event selection simply  because it is not part of the  $pp \to \ell^-\bar{\nu}_\ell\,j_b j_b jj +X$ process. Consequently, the collinear singularity from the $\gamma\to q\bar{q}$  splitting  and its non-perturbative contribution, arising from the integration over the virtuality of the intermediate photon down to the mass scale of the order of $\Lambda_{\rm{QCD}}$, do not cancel in the cross-section predictions and require an additional treatment. 

The corresponding subtraction terms for the $\gamma\!\to\!q\bar q$ splitting for the Catani-Seymour and Nagy-Soper subtraction schemes can be straightforwardly obtained from the $g\to q\bar{q}$ case with simple replacements of colour factors. We absorb the remaining  divergence into a non‑perturbative photon-to-jet fragmentation function, $D_{\gamma\to\text{jet}}$, also called a photon-to-jet conversion function, first introduced in Ref. \cite{Denner:2019zfp}. Following the notation of Ref. \cite{Denner:2019zfp}, any partonic channel with a $q\bar q$ pair in the final state, denoted as $pp\to X+q\bar q$ for brevity, can be written as
\begin{equation}
\mathrm d\sigma_{pp\to X+q\bar q}
  = \mathrm d\sigma^{\text{pert}}_{pp\to X+q\bar q}
  + \mathrm d\sigma^{\text{conv}}_{pp\to X+\text{jet}}\,,\\[0.2cm]
  \label{eq:sum}
\end{equation}
where $d\sigma^{\text{pert}}_{pp\to X+q\bar q}$ describes the perturbative contribution to the  $pp \to \ell^-\bar{\nu}_\ell\,j_b j_b jj +X$ process, which includes, among others, the appropriate subtraction terms to the underlying Born-level subprocess containing a hard photon in the final state, while $d\sigma^{\text{conv}}_{pp\to X+\text{jet}}$ comprises the photon-to-jet conversion function. The second term can be written as 
\begin{equation}
\mathrm d\sigma^{\text{conv}}_{X+\text{jet}}
  = \mathrm d\sigma^{\text{LO}}_{X+\gamma}
    \int_0^1\!\dd z\;D^{B}_{\gamma\to\text{jet}}(z)\,,\\[0.2cm]
\end{equation}
where  $\mathrm d\sigma^{\text{LO}}_{X+\gamma}$ is the Born-level cross section with an external photon replacing the $q\bar q$ pair, $D^{B}_{\gamma\to\text{jet}}(z)$ is the bare conversion function, which contains collinear singularities so the result of Eq. \eqref{eq:sum} is finite and $z$ is the splitting parameter that controls how the photon momentum is shared by the quarks. The bare photon-to-jet conversion function is given by
\begin{equation}
D^{B}_{\gamma\to\text{jet}}(z)
  = D_{\gamma\to\text{jet}}(z,\mu_{Fr}^2)
     -\frac{\alpha}{2\pi}\,
      \Gamma^{(0)}_{\gamma\to\text{jet}}(z,\mu_{Fr}^2)
     +\mathcal O(\alpha^2)\,,\\[0.2cm]
\end{equation}
with
\begin{equation}
\Gamma^{(0)}_{\gamma\to\text{jet}}(z,\mu_{Fr}^2)
  = -\frac{1}{\epsilon}\,
      \frac{(4\pi)^\epsilon}{\Gamma(1 - \epsilon)}\,
     \sum_{q} N_c Q_q^2\,
     \left(\frac{\mu_R^2}{\mu_{Fr}^2}\right)^{\!\epsilon}
     P_{\gamma\to\text{jet}}(z)\,,\\[0.2cm]
\end{equation}
and the $\gamma \to q\bar{q}$ splitting function provided by 
 \begin{equation}    
P_{\gamma\to\text{jet}}(z)=(1-z)^2+z^2\,,\\[0.2cm]
\end{equation}
where $N_c=3$ counts colours, the quark sum runs over all massless quark flavours, $\mu_{Fr}$ is a factorisation scale for the fragmentation process and the $P_{\gamma\to\text{jet}}(z)$ splitting function integrates to $\int_0^1dz\,P_{\gamma\to\text{jet}}(z)=2/3$. Carrying out the $z$–integration we end up with 
\begin{equation}
\int_0^1\!\dd z\;D^{B}_{\gamma\to\text{jet}}(z)
  = \int_0^1\!\dd z\;D_{\gamma\to\text{jet}}(z,\mu_{Fr}^2)
    +\frac{2}{3\epsilon}\,
     \frac{\alpha}{2\pi}\frac{(4\pi)^\epsilon}{\Gamma(1 - \epsilon)}\,\sum_{q} N_c  Q_q^2
     \left(\frac{\mu_R^2}{\mu_{Fr}^2}\right)^{\!\epsilon}\,.
\label{eq:intD_bare}
\end{equation}
The second term cancels the remaining $1/\epsilon$ pole from the $\gamma\to q\bar q$ splitting that appears in the $\mathrm d\sigma^{\text{pert}}_{X+q\bar q}$ part of Eq. \eqref{eq:sum}. Furthermore, the renormalised photon-to-jet conversion function obeys
\begin{equation}
\mu_{Fr}^2\,\frac{\dd D_{\gamma\to\text{jet}} (z,\mu_{Fr}^2)}{\dd\mu_{Fr}^2}
  = \frac{\alpha }{2\pi}\frac{(4\pi)^\epsilon}{\Gamma(1 - \epsilon)}\,\sum_q N_cQ_q^2 P_{\gamma\to\text{jet}}(z)\,.
\end{equation}
The leading-logarithmic solution can be written as 
\begin{equation}
D_{\gamma\to\text{jet}}(z,\mu_{Fr}^2)
  = D^{\text{np}}_{\gamma\to\text{jet}}(z,\mu_0^2)
     + \frac{\alpha}{2\pi}\frac{(4\pi)^\epsilon}{\Gamma(1 - \epsilon)}\sum_q
        N_c Q_q^2P_{\gamma\to\text{jet}}(z)
       \log \left(\frac{\mu_{Fr}^2}{\mu_0^2}\right)\,,
\end{equation}
where $D^{\text{np}}_{\gamma\to\text{jet}}(z,\mu_0^2)$ is  the non-perturbative contribution to $D_{\gamma\to\text{jet}}(z,\mu_{Fr}^2)$ and $\mu_0$ is some reference scale. For $D^{\text{np}}_{\gamma\to\text{jet}}(z,\mu_0^2)$  we need to employ the input from experimental measurements once again. Choosing, for example,  $\mu_0=M_Z$ and relating the $z$-integral of $D^{\text{np}}_{\gamma\to\text{jet}}(z,\mu_0^2)$ to the hadronic contributions to the effective QED coupling at $M_Z$ in the $5$-flavour scheme, denoted as $\Delta\alpha^{(5)}_{\text{had}}(M_Z^2)$, which can be obtained from the combination of $e^+e^- \to hadrons$ cross-section data, see e.g. Refs. \cite{Eidelman:1995ny,Keshavarzi:2018mgv}, we would attain the following result 
\begin{equation}
\int_0^1\!\dd z\;D^{\text{np}}_{\gamma\to\text{jet}}(z,\mu_0^2=M_Z^2)
  = \Delta\alpha^{(5)}_{\text{had}}(M_Z^2)
    +\frac{\alpha}{3\pi}\sum_q N_cQ_q^2\,\frac{5}{3}\,.\\[0.2cm]
\end{equation}
Inserting this back into Eq. \eqref{eq:intD_bare} yields the factorized photon–to–jet conversion integral in the $\overline{\text{MS}}$ scheme
\begin{equation}
\int_0^1 \dd z\, D^{B}_{\gamma \to \text{jet}}(z)
= \Delta\alpha^{(5)}_{\text{had}}(M_Z^2)
+ \frac{\alpha}{3\pi} 
 \sum_{q} N_c Q_q^2 
  \left[
    \frac{1}{\epsilon}
    + \log \left(\frac{\mu_{R}^2}{M_Z^2}\right)
    + \frac{5}{3}
  \right]
+ \mathcal{O}(\epsilon)\,.\\[0.2cm]
\end{equation}
In particular, the value of the hadronic contribution to the effective QED coupling is determined based on the low-energy $e^+e^- \to hadrons$ data, where $m_\pi \le \sqrt{s} \le 11.1985$ GeV, for the hadronic $R$-ratio 
\begin{equation}
R(s) = \frac{\sigma(e^+e^-\!\to hadrons)}{\sigma(e^+e^-\!\to \mu^+\mu^-)}\,,\\[0.2cm]
\end{equation}
which is related to the momentum-dependent QED coupling, $\Delta \alpha^{(5)}_{\text{had}}(q^2)$, through the following dispersion relation
\begin{equation}
\Delta \alpha^{(5)}_{\text{had}}(q^2) 
= -\frac{\alpha q^2}{3\pi} \, PV \int_{s_{\text{th}}}^{\infty} 
\dd s \, \frac{R(s)}{s(s - q^2)}\,,
\end{equation}
where $PV$ stands for the Cauchy principal value of the given integral and $s_{\text{th}} = m_{\pi}^2$. This relationship is then used to determine the effective QED coupling at the mass of the boson $Z$, which ultimately gives the following value \cite{Keshavarzi:2018mgv}
\begin{equation}
\Delta \alpha_{\mathrm{had}}^{(5)}(M_Z^2) 
= (276.11 \pm 1.11) \times 10^{-4} \,,\\[0.2cm]
\end{equation}
which we use in our calculations. We note here that the explicit dependence of $D^{\text{np}}_{\gamma \to \text{jet}}$ on $z$ is neither known in this approach nor really needed for our purposes.  Indeed, the photon-to-jet conversion function that has been introduced here is inclusive in $z$. Such an explicit dependence on $z$ would require the introduction of some kind of hadronization model for the formation of jets from the low-virtuality photon. Instead, this non-perturbative $z$-dependence is approximated by a constant that reproduces the correct $z$-integral. However, since neither the dependence of the photon-to-jet conversion function on $z$ is measurable at the LHC, nor its differential form in $z$ is required in this analysis, the current approach is more than sufficient for the problem at hand.  Ultimately, the photon-to-jet conversion function is combined with the corresponding integrated subtraction terms for the $\gamma \to q\bar{q}$ splitting, cancelling all $1/\epsilon$ poles and rendering the calculations IR (soft and collinear) finite.

\section{Input parameters}
\label{sec:parameters}

We consider the $pp \to \ell^-\bar{\nu}_\ell\,j_b j_b jj+X$ process, where $\ell^-=e^-,\,\mu^-$, and provide LO and NLO theoretical predictions for LHC Run III center-of-mass energy of $\sqrt{s}=13.6$ TeV. In practise we generate the results for $pp \to \mu^-\bar{\nu}_\mu\,j_b j_b jj+X$ and multiply them with a lepton-flavour factor of 2.  When calculating the complete NLO corrections we need to include both the  QCD and EW higher-order corrections, as well as the photon-initiated subprocesses. Therefore, it is important to  take into account the distribution of photons in the proton using the photon parton distribution function (PDF). In our study, we utilize the NLO NNPDF3.1luxQED PDF set \cite{Manohar:2016nzj,NNPDF:2017mvq,Manohar:2017eqh,Bertone:2017bme} with $\alpha_s(m_Z)=0.118$, which accounts for QED effects in the DGLAP evolution. The two-loop running of $\alpha_s$ is performed with the help of the LHAPDF interface \cite{Buckley:2014ana}. The number of active flavours is set to  $N_F = 5$.  The $\overline{\text{MS}}$ factorisation scheme for initial-state collinear singularities is understood for both QCD and EW corrections. Furthermore, in the complete calculation, the same PDF set is used in both the LO and NLO predictions. The electromagnetic coupling $\alpha$ is calculated in the $G_\mu$-scheme according to 
\begin{equation}
	\alpha_{G_\mu} =\frac{\sqrt{2}}{\pi} \,G_\mu \, m_W^2\,\left(1-\frac{m_W^2}{m_Z^2}\right)\,,
	\quad \quad \quad \quad \quad \quad 
	G_{ \mu}=1.1663787 \cdot 10^{-5} \textrm{ GeV}^{-2}\,.
\end{equation}
The on-shell masses and widths of the $W^\pm$ and $Z$ gauge bosons are taken from Ref. \cite{ParticleDataGroup:2022pth}
\begin{equation}
\begin{array}{lll}
 m^{\rm OS}_{W}= 80.377 ~{\rm GeV}\,,&\quad\quad\quad\quad & \Gamma^{\rm OS}_{W} = 2.085 ~{\rm GeV}\,, 
\vspace{0.2cm}\\[0.2cm]
 m^{\rm OS}_{Z}= 91.1876 ~{\rm GeV}\,,&\quad\quad\quad\quad & \Gamma^{\rm OS}_{Z} = 2.4955 ~{\rm GeV}\,,
\end{array}
\end{equation}
and converted into their pole values through the following  relations  \cite{Bardin:1988xt}
\begin{equation}
 m_V=\frac{m^{\textrm{OS}}_V}{\sqrt{1+\left(\Gamma^{\textrm{OS}}_V/m^{\textrm{OS}}_V\right)^2}} \,, \quad\quad\quad\quad \quad\quad
 \Gamma_V=\frac{\Gamma^{\textrm{OS}}_V}{\sqrt{1+\left(\Gamma^{\textrm{OS}}_V/m^{\textrm{OS}}_V\right)^2}}\,,\\[0.2cm]
\end{equation}
where $V=W^\pm,Z$. The top-quark mass is set to $m_t=172.5$ GeV. For the mass and width of the Higgs boson, the following values are adopted: $m_{H}= 125$ GeV and $\Gamma_{H} = 4.07\cdot 10^{-3}$ GeV. All other particles, including bottom quarks, are considered massless. The top-quark width is calculated assuming an unstable $W$ gauge boson while neglecting the bottom-quark mass according to the formulas given in Refs.  \cite{Jezabek:1988iv,Denner:2012yc}, using $\alpha_s(\mu_R=m_t)$. The resulting LO and NLO QCD top-quark widths are given by
\begin{equation}
 \Gamma_{t}^{\rm LO}= 1.4580658 ~{\rm GeV}\,, \quad \quad \quad 
 \Gamma_{t}^{\rm NLO, \,QCD}= 1.3329042~{\rm GeV}\,.\\[0.2cm]
\end{equation}
For the complete NLO calculation, however,  we employ the full NLO QCD+EW result, denoted as $\Gamma_t^{\rm NLO}$. Its numerical value is obtained from  $\Gamma_{t}^{\rm NLO,\,QCD}$ by employing the result from Ref. \cite{Basso:2015gca}. Specifically,   $\Gamma_t^{\rm NLO}=\Gamma_{t}^{\rm NLO,\, QCD}+\delta^\alpha \,\Gamma_t^{\rm LO}$, where $\delta^\alpha=+1.3\%$ is the relative EW correction factor. In this way, we ultimately obtain
\begin{equation}
\Gamma_t^{\rm NLO}=1.3518591~{\rm GeV}\,.\\[0.2cm]
\end{equation}

The central renormalisation, $\mu_R$, and factorisation, $\mu_F$, scales are set to a common value $\mu_R=\mu_F=\mu_0$. The QCD-scale uncertainties are estimated with a $7$-point scale variation of $\mu_0$, which are evaluated according to 
\begin{equation}
\label{scan}
\left(\frac{\mu_R}{\mu_0}\,,\frac{\mu_F}{\mu_0}\right) = \Big\{
\left(2,1\right),\left(0.5,1  
\right),\left(1,2\right), (1,1), (1,0.5), (2,2),(0.5,0.5)
\Big\} \,.\\[0.2cm]
\end{equation}
By searching for the minimum and maximum of the resulting cross sections we estimate an uncertainty band. We consider the following dynamical scale setting $ \mu_0 = E_T/4$, where $E_T$ is the sum of the transverse energies of the top and anti-top quarks, defined as
\begin{equation}
        E_T = \sqrt{m_t^2 + p_{T,\,t}^2} + \sqrt{m_t^2 + p_{T,\,\bar{t}}^2}\,\,.\\[0.2cm]
\end{equation}
This scale setting relies on reconstructing the momenta of the top quarks. However, in a full off-shell calculation, especially in the semi-leptonic decay channel, the decay histories of the top quarks are not uniquely defined. At NLO an additional resolved light-jet ($b$-jet) can be produced. In practice, to closely mimic what is done at the LHC, only the additional light-jet ($b$-jet) that passes all the cuts is taken into account. First, we assign light jets to the decay of the $W$ gauge boson, considering all combinations and choosing the one that minimizes the value of $|M_{jj} -m_W|$, where $M_{jj}$ is the invariant mass of the two light jets. After completing this part we continue with the $b$-jet assignment. To assign the $b$-jets to the top or anti-top quark, we again consider all the combinations and choose the one that minimises the quantity $\mathcal{Q}$, given by 
\begin{equation}
        \mathcal{Q} = |M_t - m_t| + |M_{\bar{t}} - m_t|\,,\\[0.2cm]
\end{equation}
where $M_t$ and $M_{\bar{t}}$ are the invariant masses of the reconstructed candidate top quarks, and $m_t$ is the input top-quark mass. This last part is mainly important in the case of three $b$-jets in the final state or for the partonic subprocesses $bb$ and $\bar{b}\bar{b}$. Note here that in the procedure we assume a full reconstruction of the neutrino momentum \cite{Sonnenschein:2005ed,Raine:2023fko}. As an alternative scale choice we use $\mu_0 = H_T/4$, where $H_T$ is the scalar sum of the transverse momenta of all final-state particles present at the Born level, defined as 
\begin{equation}
     H_T =  p_{T,\,\ell^-} + p_{T,\,b_1} + p_{T,\,b_2} + p_{T,\,j_1} + p_{T,\,j_2} + p_{T}^{miss}\,.\\[0.2cm]
\end{equation}
In the case with three $b$- or light-jets, only the two hardest jets, as ordered in $p_T$, are included in the definition.  The missing transverse momentum $p_{T}^{miss}$ originates solely from the anti-neutrino $\bar{\nu}_\ell$. This scale choice does not rely on identifying intermediate top-quark resonances and is therefore arguably more suited for the predictions with full off-shell effects included. We consider this scale setting to be the default setting in our complete LO and NLO calculations.

We cluster all final state particles in the pseudorapidity range of $|\eta| <5$ into jets with the IR-safe {\it anti}$-k_T$ jet algorithm \cite{Cacciari:2008gp} where the resolution parameter $R$ is set to $R= 0.4$. While quarks and/or gluons are recombined into jets, leptons and photons are recombined into leptons, and quarks/gluons and photons are recombined into jets. From the experimental point of view the $b$-jet tagging algorithms are sensitive mainly to the absolute flavour and  not to the charge of a single $b$-jet. For the jet flavour assignment we use, therefore, the so-called charge-blind scheme \cite{Bevilacqua:2021cit}. In short, for massless bottom quarks the important parton recombination rules are  $bg \to b$, $\bar{b}g \to \bar{b}$ and $b\bar{b}\to g$. They are required to guarantee the IR-safety of the jet algorithm. Since we employ the charge-blind $b$-jet tagging, any combination that contains an even number of $b$ or $\bar{b}$ quarks should also be considered to carry zero flavour. Consequently, we also add the following recombination rules $bb\to g$ and $\bar{b}\bar{b}\to g$. As explained earlier, in the case of an additional photon in the final state, the resulting object containing a parton and a photon is classified as a photon if $z_\gamma > z_{\gamma,\text{cut}} = 0.7$ and as a hadronic jet otherwise. To assess the dependence on the $z_{\gamma,\text{cut}}$ cut, we have also generated results for  $z_{\gamma,\text{cut}} = 0.5$ and $z_{\gamma,\text{cut}} = 0.9$. 
The obtained integrated cross-section predictions differed from the result generated with the default $z_{\gamma,\text{cut}} = 0.7$ setting by less than one per-mille. We observed differences of the same magnitude for the differential cross-section results we examined. 

To define the  $pp \to \ell^-\bar{\nu}_\ell\,j_b j_b jj +X$ process, we require at least two $b$-jets, at least two light-jets, and one negatively charged lepton. All final states have to
fulfil the subsequent selection criteria that mimic as closely as possible the ATLAS and the
CMS detector acceptances. In particular, the charged lepton is required to have
\begin{equation}
p_{T,\,\ell}>25 ~{\rm GeV}\,,  \quad \quad\quad \quad |y_\ell|<2.5 \,.\\[0.2cm]
\end{equation}
The flavoured jets with 
\begin{equation}
p_{T,\,b}>25 ~{\rm GeV}\,,   \quad \quad \quad \quad |y_b|<2.5\,,\\[0.2cm]
\end{equation}
are selected, and only the  $b$-jets that are well separated from the lepton in the rapidity-azimuthal angle plane, $\Delta R_{\ell b}> 0.4$, are taken into account. The light jets are required to have
\begin{equation}
p_{T,\,j}>25 ~{\rm GeV}\,,   \quad \quad \quad \quad |y_j|<2.5\,.\\[0.2cm]
\end{equation}
They have to be well isolated from the charged lepton,  $\Delta R_{\ell j}> 0.4$,
and the two $b$-jets, $\Delta R_{b j}> 0.4$. We do not impose any restrictions on the missing transverse momentum, $p_{T}^{miss}$, or on the kinematics of the additional (light- or $b$-) jet if it is resolved by the jet algorithm.  Following Refs. \cite{Denner:2017kzu,Stremmer:2023kcd},  we require that each event contains at least one pair of light jets for which the following condition is satisfied 
\begin{equation}
|M_{jj}-m_W|< {\cal Q}_{cut} \,.\\[0.2cm]
\end{equation}
The purpose of this additional restriction is twofold. On the one hand, it reduces the QCD background, and on the other hand, it reduces the size of the NLO QCD corrections. Our default setting is ${\cal Q}_{cut} = 20$ GeV. However, for comparison purposes, we generate the complete NLO results with and without this restriction.

\section{Integrated cross sections}
\label{sec:results-int}

%
\begin{table}[t!]
    \centering
    \renewcommand{\arraystretch}{1.2}
    \begin{tabular}{ll@{\hskip 10mm}l@{\hskip 10mm}l@{\hskip 10mm}}
        \hline
        \noalign{\smallskip}
         &&$\sigma_{i}$ [pb] & Ratio to ${\rm Born}_3$  \\
        \noalign{\smallskip}\midrule[0.5mm]\noalign{\smallskip}
        \Bornone&\asLOone& $ 0.778(4) $ & $ 2.40\% $ \\
        \Borntwo&\asLOtwo& $ 0.0021(1) $ & $ 0.01\% $ \\
        \Bornthree&\asLOthree& $ 32.388(4) $ & $ 100.00\% $ \\
        \Bornfour&\asLOfour& $ 0.1365(2) $ & $ 0.42\% $ \\
        \Bornfive&\asLOfive& $ 0.09980(6) $ & $ 0.31\% $ \\
        \noalign{\smallskip}\hline\noalign{\smallskip}
        \NLOone&\asNLOone& $ +\,0.220(6) $ & $ +\,0.68\% $\\
        \NLOtwo&\asNLOtwo& $ -\,0.0172(8) $ & $ -\,0.05\% $\\
        \NLOthree&\asNLOthree& $ +\,16.56(2) $ & $ +\,51.15\% $\\
        \NLOfour&\asNLOfour& $ +\,0.192(2) $ & $ +\,0.59\% $\\
        \NLOfive&\asNLOfive& $ +\,0.276(1) $ & $ +\,0.85\% $\\
        \NLOsix&\asNLOsix& $ +\,0.00235(2) $ & $ +\,0.01\% $\\
        \noalign{\smallskip}\hline\noalign{\smallskip}
        \LOfull&& $ 28.896(2)^{+31.2\%}_{-22.2\%} $ & $ 0.89 $ \\
        \Bornfull&& $ 33.404(4)^{+31.2\%}_{-22.2\%} $ & $ 1.03 $ \\
        \NLOqcd&& $ 49.98(2)^{+15.1\%}_{-13.5\%} $ & $ 1.54 $ \\
        \NLOfull&& $ 50.64(2)^{+14.2\%}_{-13.2\%} $ & $ 1.56 $ \\
        \noalign{\smallskip}\hline\noalign{\smallskip}
    \end{tabular}
    \caption{\it Integrated fiducial cross-section results for the $pp \to \ell^-\bar{\nu}_\ell\,j_bj_b\,jj +X$ process at the LHC with $\sqrt{s}=13.6$ TeV. Results are shown for \LOfull, \Bornfull, \NLOqcd and \NLOfull with the corresponding scale uncertainties. Monte Carlo  integration errors are given in parentheses. In addition, all individual ${\rm Born}_i$ and ${\rm NLO}_i$ contributions are displayed. Finally, the ratios to the leading Born-level contribution \Bornthree are also presented for each case. The results are provided for $\mu_0=H_T/4$ and the NLO NNPDF3.1luxQED PDF set.}
    \label{tab:integratednoqcut}
\end{table}
\begin{table}[t!]
    \centering
    \renewcommand{\arraystretch}{1.2}
    \begin{tabular}{ll@{\hskip 10mm}l@{\hskip 10mm}l@{\hskip 10mm}}
        \hline
        \noalign{\smallskip}
         &&$\sigma_{i}$ [pb] & Ratio to ${\rm Born}_3$  \\
        \noalign{\smallskip}\midrule[0.5mm]\noalign{\smallskip}
        \Bornone&\asLOone& $ 0.138(2) $ & $ 0.44\% $ \\
        \Borntwo&\asLOtwo& $ 0.0002(1) $ & $ 0.00\% $ \\
        \Bornthree&\asLOthree& $ 31.264(4) $ & $ 100.00\% $ \\
        \Bornfour&\asLOfour& $ 0.1336(1) $ & $ 0.43\% $ \\
        \Bornfive&\asLOfive& $ 0.0968(1) $ & $ 0.31\% $ \\
        \noalign{\smallskip}\hline\noalign{\smallskip}
        \NLOone&\asNLOone& $ +\,0.144(4) $ & $ +\,0.46\% $\\
        \NLOtwo&\asNLOtwo& $ -\,0.0022(4) $ & $ -\,0.01\% $\\
        \NLOthree&\asNLOthree& $ -\,1.90(2) $ & $ -\,6.10\% $\\
        \NLOfour&\asNLOfour& $ +\,0.112(2) $ & $ +\,0.36\% $\\
        \NLOfive&\asNLOfive& $ +\,0.1864(6) $ & $ +\,0.60\% $\\
        \NLOsix&\asNLOsix& $ +\,0.00228(2) $ & $ +\,0.01\% $\\
        \noalign{\smallskip}\hline\noalign{\smallskip}
        \LOfull&& $ 27.241(2)^{+30.7\%}_{-22.0\%} $ & $ 0.87 $ \\
        \Bornfull&& $ 31.634(4)^{+30.7\%}_{-22.0\%} $ & $ 1.01 $ \\
        \NLOqcd&& $ 29.72(2)^{+1.4\%}_{-6.0\%} $ & $ 0.95 $ \\
        \NLOfull&& $ 30.16(2)^{+1.5\%}_{-5.6\%} $ & $ 0.96 $ \\
        \noalign{\smallskip}\hline\noalign{\smallskip}
    \end{tabular}
    \caption{\it Same as Table \ref{tab:integratednoqcut}, but with the additional restriction given by $|M_{jj}-m_W|< {\cal Q}_{cut}$ where ${\cal Q}_{cut}=20~{\rm GeV}$.}
    \label{tab:integratedqcut}
\end{table}

We begin the presentation of our findings with the integrated fiducial cross-section results, which are provided for the input parameters and cuts specified in the previous section, but first without the additional restriction given by $|M_{jj}-m_W|< {\cal Q}_{cut}$. In Table \ref{tab:integratednoqcut} we show the \LOfull, \Bornfull, \NLOqcd and \NLOfull results, together with the theoretical error as obtained from the 7-point scale variation. All theoretical predictions are evaluated  for $\mu_0=H_T/4$.  The  \LOfull and \Bornfull contributions are defined according to
\begin{equation}
\text{LO} = \sum_{i=1}^{5} \text{LO}_i \,,
\quad \quad \quad \quad \quad \quad 
\text{Born} = \sum_{i=1}^{5} \text{Born}_i \,.\\[0.2cm]
\end{equation}
The individual terms ${\rm LO}_i$ and ${\rm Born}_i$ differ only in the numerical value used for the top-quark width, where in the latter case we use the NLO value, while in the former the LO value is utilised instead. We present the individual ${\rm Born}_i$ components to show how the full  \NLOfull result can be calculated by summing them with all the remaining ${\rm NLO}_i$ contributions.  Furthermore, we define the NLO QCD cross section, denoted \NLOqcd, as follows
\begin{equation}
\text{NLO}_{\text{QCD}} = \text{Born} + \text{NLO}_3\,,\\[0.2cm]
\end{equation}
which can be seen as the NLO QCD corrections to the dominating production mechanism for this process. However, we remind the reader that  \NLOqcd  is not simply  given by the ${\cal O}(\alpha_s)$ correction  to $\text{LO}_3$, but receives QCD and EW corrections from various  orders. Finally, the complete  \NLOfull is defined according to 
\begin{equation}
\text{NLO} = \text{Born} + \sum_{i=1}^{6} \text{NLO}_i\,,\\[0.2cm]
\end{equation}
and includes all partonic channels, Feynman diagrams, interference terms and resonance structures. All individual ${\rm Born}_i$ and ${\rm NLO}_i$ contributions are also presented in Table \ref{tab:integratednoqcut}. In addition, the last column provides the ratios of these contributions to \Bornthree. 

First, we see that the hierarchy in the size of the different $\text{Born}_{i}$ contributions does not follow the hierarchy in powers in $\alpha_s$.  Indeed, \Bornthree at ${\cal O} (\alpha_s^2\alpha^4)$ dominates the complete Born-level result by about $97\%$. As expected, the $gg$ partonic channel dominates here, accounting for approximately $84\%$ of the \Bornthree contribution. On the other hand, the largest subleading contribution denoted as \Bornone at ${\cal O} (\alpha_s^4\alpha^2)$, which can be classified as the irreducible QCD background for the production of the $t\bar{t}$ pair, is of the order of $2.4\%$ only. The  \Borntwo contribution, consisting of photon-initiated subprocesses and comprising the interference terms between various orders, is completely negligible. Finally, the contributions of both  \Bornfour, which is dominated by the $g\gamma$ and $b\bar{b}$ partonic channels, and \Bornfive, which is generated by purely electroweak-induced contributions, are less than $0.5\%$. The reason why both \Bornfour and  \Bornfive are similar in size is as follows. In the case of \Bornfour  the $b$-quark initiated subprocesses contribute negatively via interference effects reducing the overall contribution by approximately $44\%$. For \Bornfive we have substantial contributions coming from the $b$-initiated partonic channels accounting for up to $79\%$ of \Bornfive. The latter contributions are enhanced due to the presence of two top-quark resonances that are produced in the $t$-channel Feynman diagrams involving intermediate $W$ bosons. At last, we can see that the complete LO result provides only a rough estimate of the cross section for the  $pp \to \ell^-\bar{\nu}_\ell\,j_bj_b\,jj$ process due to the large scale uncertainties that are of the order of $31\%$. 

Moving on to the size of the higher-order effects, we can see that  \NLOthree, which consists mainly of  ${\cal O}(\alpha_s)$ corrections to \Bornthree, is by far the dominant correction. It increases the complete Born-level cross section by about $50\%$. All the other $\text{NLO}_i$ contributions are below $1\%$ accounting for the overall higher-order effect of around $2\%$. The ${\cal K}$-factors defined as  ${\cal K}_{\text{QCD}}= \text{NLO}_\text{QCD}/\text{LO}=1.73$ and ${\cal K}= \text{NLO}/\text{LO}=1.75$ are rather similar. Indeed, the overall size of the NLO corrections differs by $2.3\%$ only. The theoretical uncertainties due to scale dependence for these two cases are also similar, and of the order of $15\%$. Although at the integrated fiducial cross-section level, the inclusion of subleading contributions is not phenomenologically significant, it is nevertheless vital to study the effects of those contributions, especially for the unusual case where \Bornthree   instead of \Bornone   is the dominating Born-level contribution. 

The size of NLO corrections can be substantially reduced when an additional restriction on the invariant mass of the light jets is employed, see e.g. Refs. \cite{Denner:2017kzu,Stremmer:2023kcd}. Requesting $|M_ {jj}-m_W|< {\cal Q}_{cut}$ in the NLO calculation, where ${\cal Q}_{cut}=20$ GeV, increases the probability of correctly assigning two light-jets to the $W$ gauge boson and thus to the top quark. In Table \ref{tab:integratedqcut} we present the LO and NLO results taking into account the ${\cal Q}_{cut}$ cut. Already at the Born level we observe significant differences. Indeed,  \Bornone is reduced by almost a factor of $6$. The contribution of \Borntwo decreases by an order of magnitude and remains negligible, while \Bornthree decreases by only $3.5\%$ and continues to comprise the largest contribution to the complete Born-level result. The other two contributions, \Bornfour and \Bornfive, decrease only slightly, and their relative sizes remain at the same level as before. In the case of NLO, all subleading contributions are again small,  below $1\%$. The largest but expected change occurs for \NLOthree. Its contribution is reduced in absolute terms from $50\%$ to $6\%$ and its sign is flipped.  This is due to the fact that the $|M_ {jj}-m_W|< {\cal Q}_{cut}$ condition substantially restricts the non-$t\bar{t}$ background. In practice, it reduces  the probability that the additional light jet coming from the real-emission part of the NLO calculation is incorrectly attributed to one of the decay products of the $W$ gauge boson. The \NLOtwo and \NLOsix contributions are again completely negligible, while \NLOone,  \NLOfour and \NLOfive are reduced by about $32\%$, $42\%$ and $33\%$,  respectively. Furthermore, the NLO corrections are significantly reduced in this case. Indeed, we obtain the following ${\cal K}$-factors  ${\cal K}_{\text{QCD}}= \text{NLO}_\text{QCD}/\text{LO}=1.09$ and ${\cal K}= \text{NLO}/\text{LO}=1.11$. After incorporating the extra cut, the LO uncertainties remain unchanged, but the NLO  uncertainties are substantially reduced from  $14\%-15\%$ to $6\%$. 

Overall, the $|M_ {jj}-m_W|< {\cal Q}_{cut}$ requirement substantially improves the perturbative stability of the higher-order calculations. Nevertheless, the phenomenological picture related to the subleading contributions does not change qualitatively. At the integrated cross-section level the subleading NLO corrections are less than $1\%$ and are  therefore negligible compared to the size of theoretical uncertainties.

\section{Differential distributions}
\label{sec:results-diff}

%
\begin{figure}[t!]
    \begin{center}
	\includegraphics[width=0.49\textwidth]{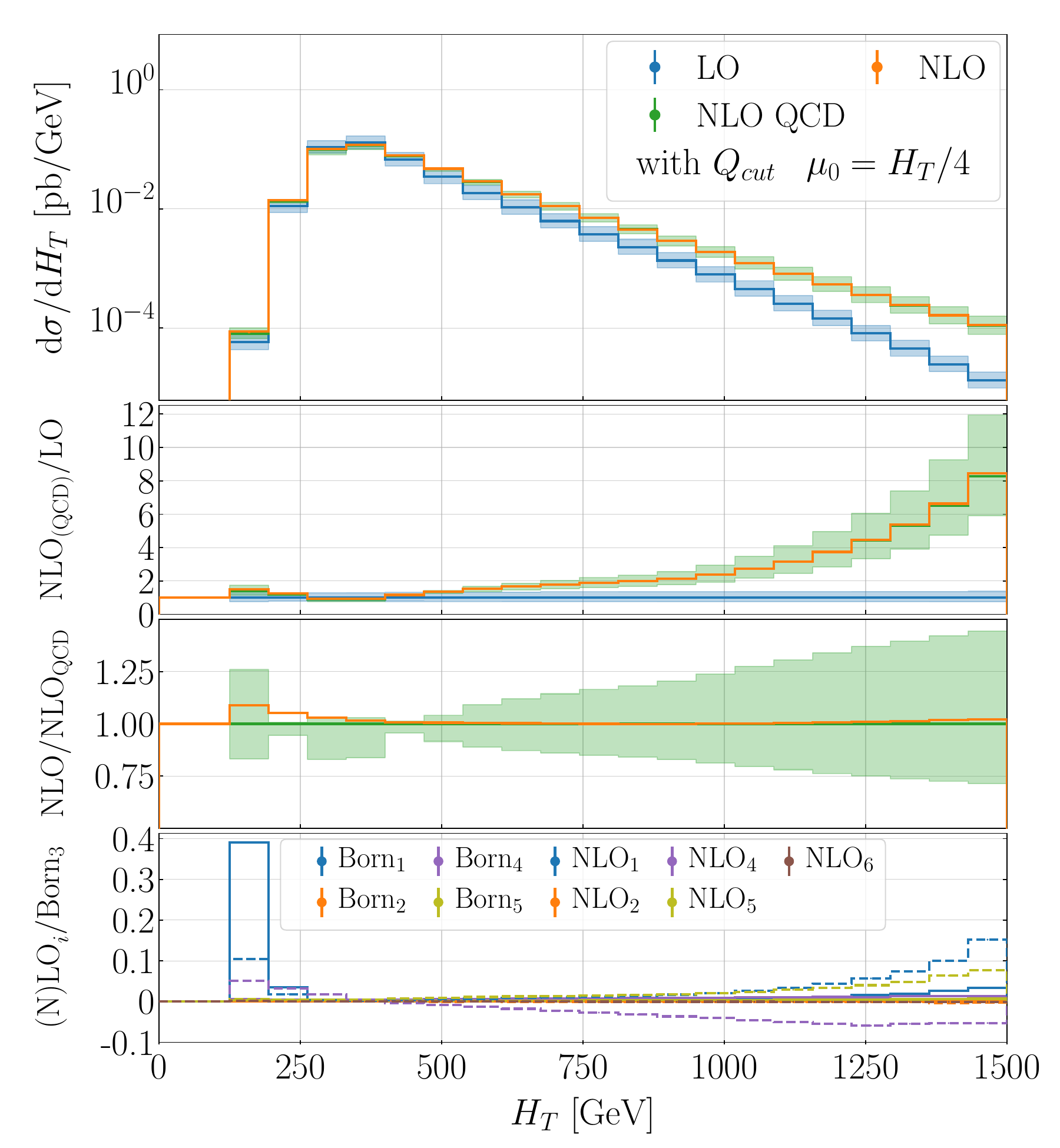}
    \includegraphics[width=0.49\textwidth]{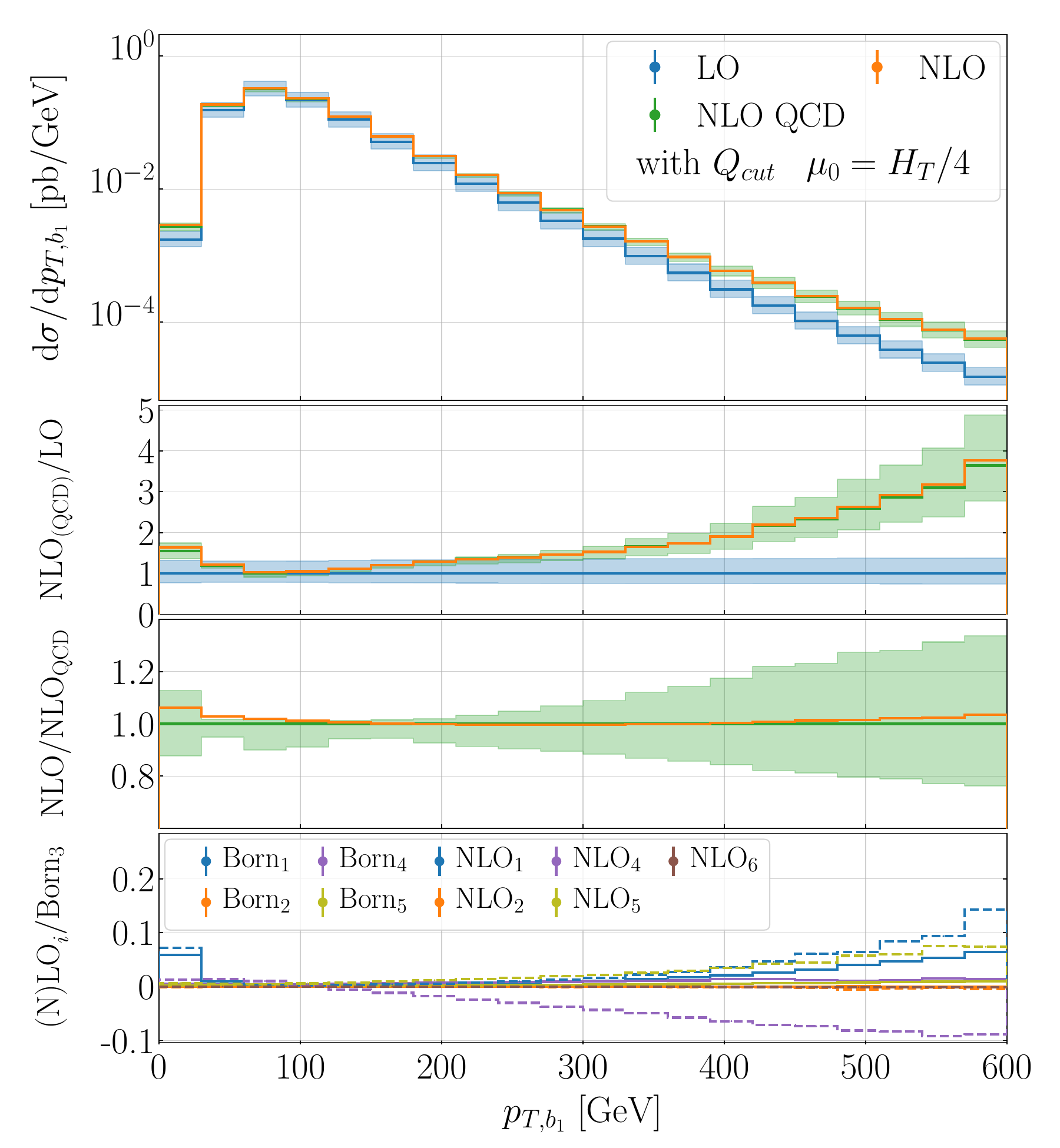}
	\includegraphics[width=0.49\textwidth]{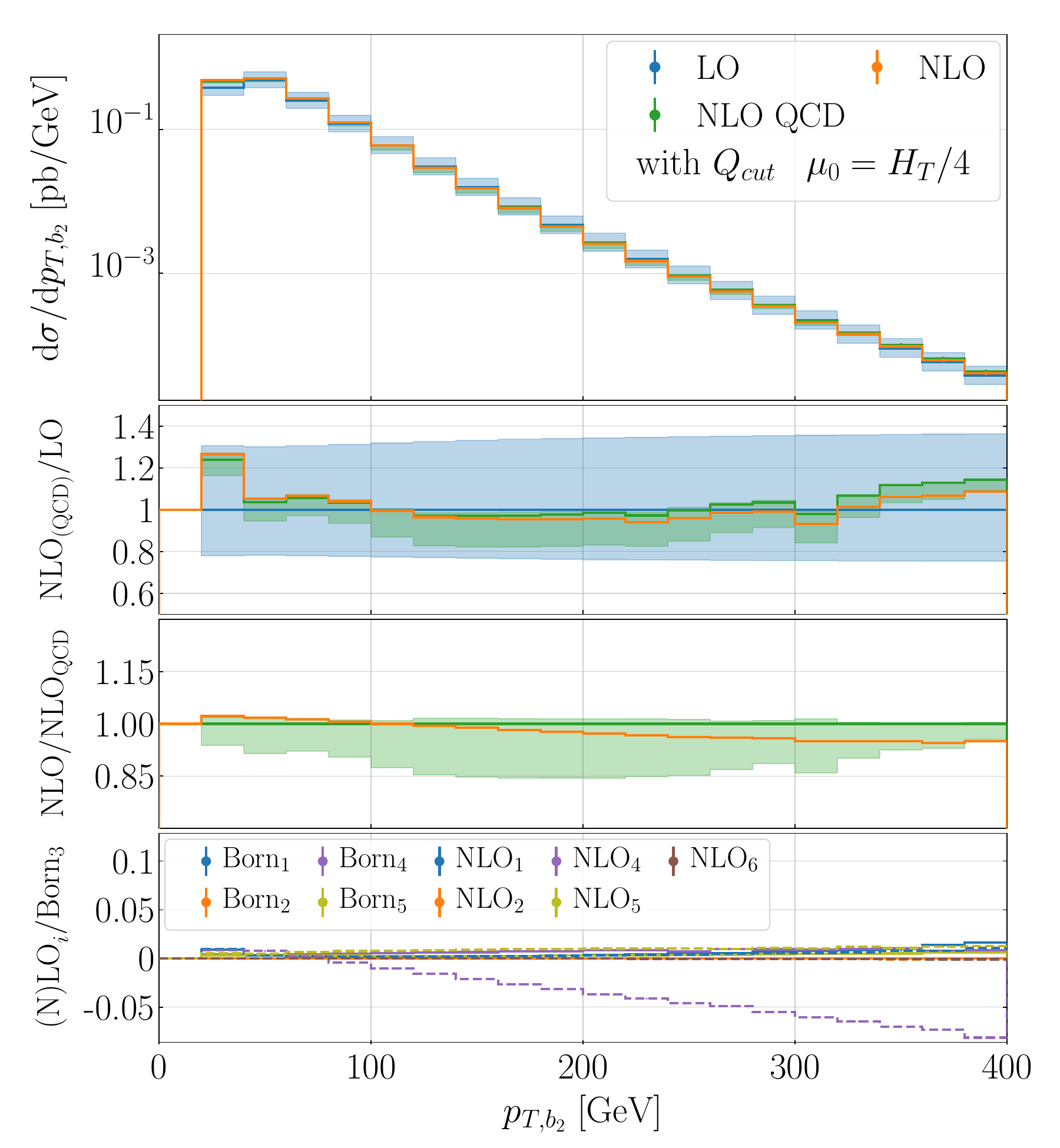}
    \includegraphics[width=0.49\textwidth]{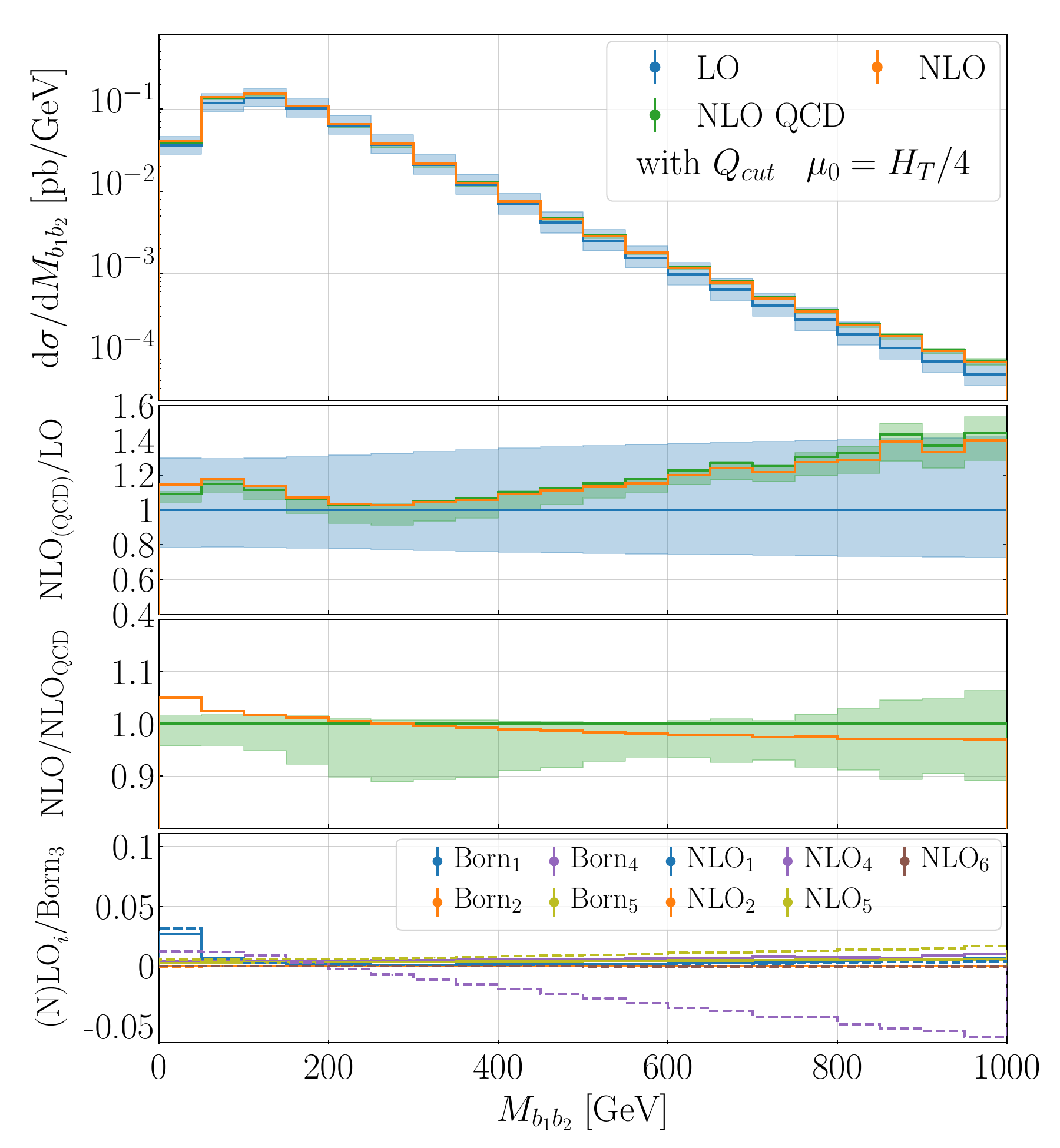}
    \end{center}
    \caption{\label{fig-ttsemi:nlo1} \it Differential cross-section distributions for the $pp \to \ell^-\bar{\nu}_\ell\,j_b j_b jj +X$ process at the LHC with $\sqrt{s}=13.6$ TeV as functions of $H_T$, $p_{T,\,b_1}$, $p_{T,\,b_2}$ and $M_{b_1b_2}$. Results are given for $\mu_0 = H_T/4$ and the NLO NNPDF3.1luxQED PDF set with $|M_{jj}-m_W|< {\cal Q}_{cut}$, where  ${\cal Q}_{cut}=20$ GeV. The upper panels show the absolute \LOfull, \NLOqcd and \NLOfull predictions together with the \LOfull and \NLOqcd uncertainty bands. The second panels present the ratios of \NLOqcd and \NLOfull to \LOfull,  while the ratios of \NLOfull to \NLOqcd are displayed in the third panels. The last panels depict ${\rm Born}_i$ (solid lines) and ${\rm NLO}_i$ (dashed lines) normalised to the dominant \Bornthree contribution. }
\end{figure}

Although at the integrated cross-section level the impact of subleading effects is negligible, this is not necessarily the case at the differential level. In the following, we analyse several differential cross-section distributions to see whether the subleading effects are significant in certain fiducial phase-space regions or whether there are accidental cancellations between different contributions. In Figure \ref{fig-ttsemi:nlo1}  we display the $H_T$ observable defined as the scalar sum of the transverse momenta of the observed final state particles, i.e. $H_T = p_{T,\,\ell} + p_{T,\,b_1} + p_{T,\,b_2} + p_{T,\,j_1} + p_{T,\,j_2} + p_{T}^{miss}$, where in the case of the additional light-jet ($b$-jet)  only the two jets with the highest $p_T$ will contribute. Also presented in Figure \ref{fig-ttsemi:nlo1} are the transverse momentum of the first and the second hardest $b$-jet, $p_{T,\,b_1}$ and $p_{T,\,b_2}$, respectively, and the invariant mass of the two-$b$-jet system, $M_{b_1b_2}$. The upper panels show the absolute predictions for \LOfull, \NLOqcd and \NLOfull with the \LOfull and \NLOqcd uncertainty bands. The second panels depict the following differential ratios \NLOfull/\LOfull and \NLOqcd/\LOfull with the \NLOqcd uncertainty bands. The third panels provide  the  \NLOfull/\NLOqcd differential ratios. Finally, the lower panels display the  $\text{Born}_i$ and $\text{NLO}_i$ contributions normalised to the dominant Born-level contribution \Bornthree. All results are presented for $\mu_R=\mu_F=\mu_0=H_T/4$, the NLO NNPDF3.1luxQED PDF set and with the additional $|M_{jj}-m_W|< {\cal Q}_{cut}$ cut, where ${\cal Q}_{cut}=20$ GeV. 

Even after applying the ${\cal Q}_{cut}=20$ GeV cut, we can observe that some fiducial phase-space regions are strongly affected by higher-order effects. This is, for example, the case for the $H_T$ and $p_{T,\, b_1}$ observables, where in the high-$p_T$ tails gigantic NLO corrections of up to $+750\%$ for $H_T$ and $+280\%$ for $p_{T,\, b_1}$ are observed. These huge NLO QCD corrections result from the hard jet recoiling against the LO system. This opens up a new phase-space region because the $t\bar{t}$ system no longer needs to be back-to-back in the transverse plane. For $H_T$ the \NLOqcd scale uncertainties reach up to $44\%$ in this phase space region, while the ratio of \NLOfull to \NLOqcd amounts to  $1.02$. This relatively small difference between the complete NLO result and the NLO QCD part can be explained, on the one hand, by the large ratio between \NLOqcd and \LOfull, driven by \NLOthree. On the other hand, the QCD background from \NLOone contributes about $+15\%$ (compared to \Bornthree), while electroweak Sudakov logarithms from \NLOfour amount to $-5\%$. The  \NLOfive subleading part gives a contribution of the order of $+8\%$. Due to a cancellation among various $\text{NLO}_i$ the overall impact of the subleading corrections is, therefore, quite limited. For $p_{T,\,b_1}$ the \NLOqcd scale uncertainties reach up to  $34\%$ and become of the order of the \LOfull ones. Also in this case we could find small higher-order corrections to subleading LO contributions. In particular, for \NLOone we have effects of the order of $+14\%$ with respect to \Bornthree, while \Bornone is about $+8\%$. The electroweak Sudakov logarithms present in \NLOfour are of similar size as \Bornone but with an opposite sign. In this case, too, \NLOthree has the dominant impact, resulting in a difference of $3\%$ between \NLOqcd and \NLOfull, which is an order of magnitude smaller than the uncertainties associated with the scale variation. 

In the case of $p_{T,\,b_2}$, the complete NLO corrections are significantly smaller, only of the order of $+10\%$. In the tail of the distribution, the \NLOqcd (complete \NLOfull) uncertainties also decrease significantly to approximately $4\%$ $(7\%)$. In addition, subleading higher-order contributions cancel each other out again. Indeed,  \NLOfour contributes with $-8\%$, while the \NLOqcd corrections are of the order of $+15\%$. Overall, the difference between \NLOqcd and \NLOfull is $-5\%$, which is of the order of the theoretical uncertainties.  Finally, $M_{b_1b_2}$ shows rather moderate \NLOfull corrections up to $+40\%$ in the tails of the distribution. In these phase-space regions, the difference between \NLOqcd and \NLOfull is around $-3\%$, while the \NLOqcd uncertainties are of the order of $11\%$. The dominant subleading contribution is given by \NLOfour, which contributes $-6\%$ compared to $\text{Born}_3$. All other subleading Born-level and NLO contributions are very small, below $2\%$. 
\begin{figure}[t!]
    \begin{center}
	\includegraphics[width=0.49\textwidth]{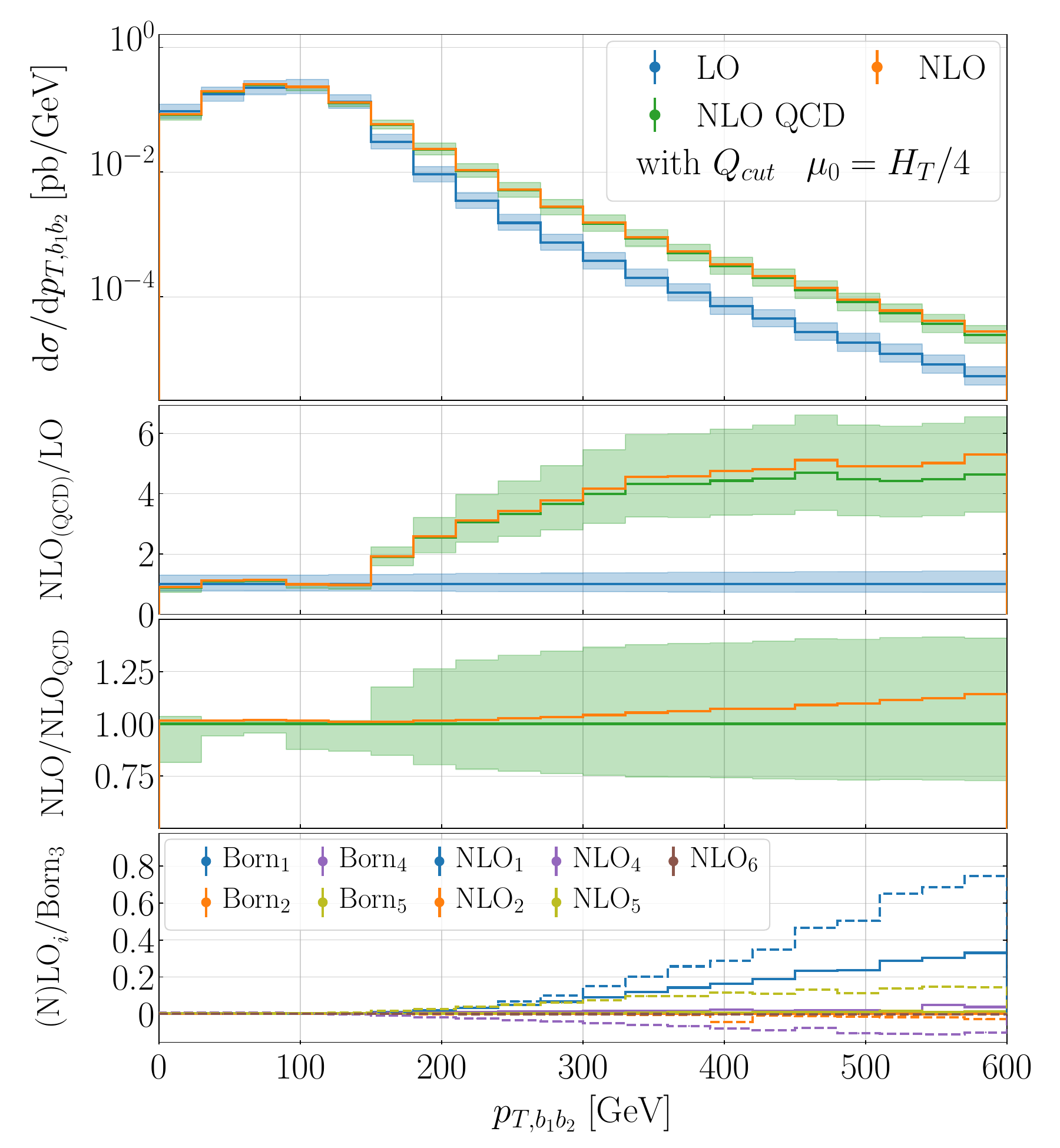}
	\includegraphics[width=0.49\textwidth]{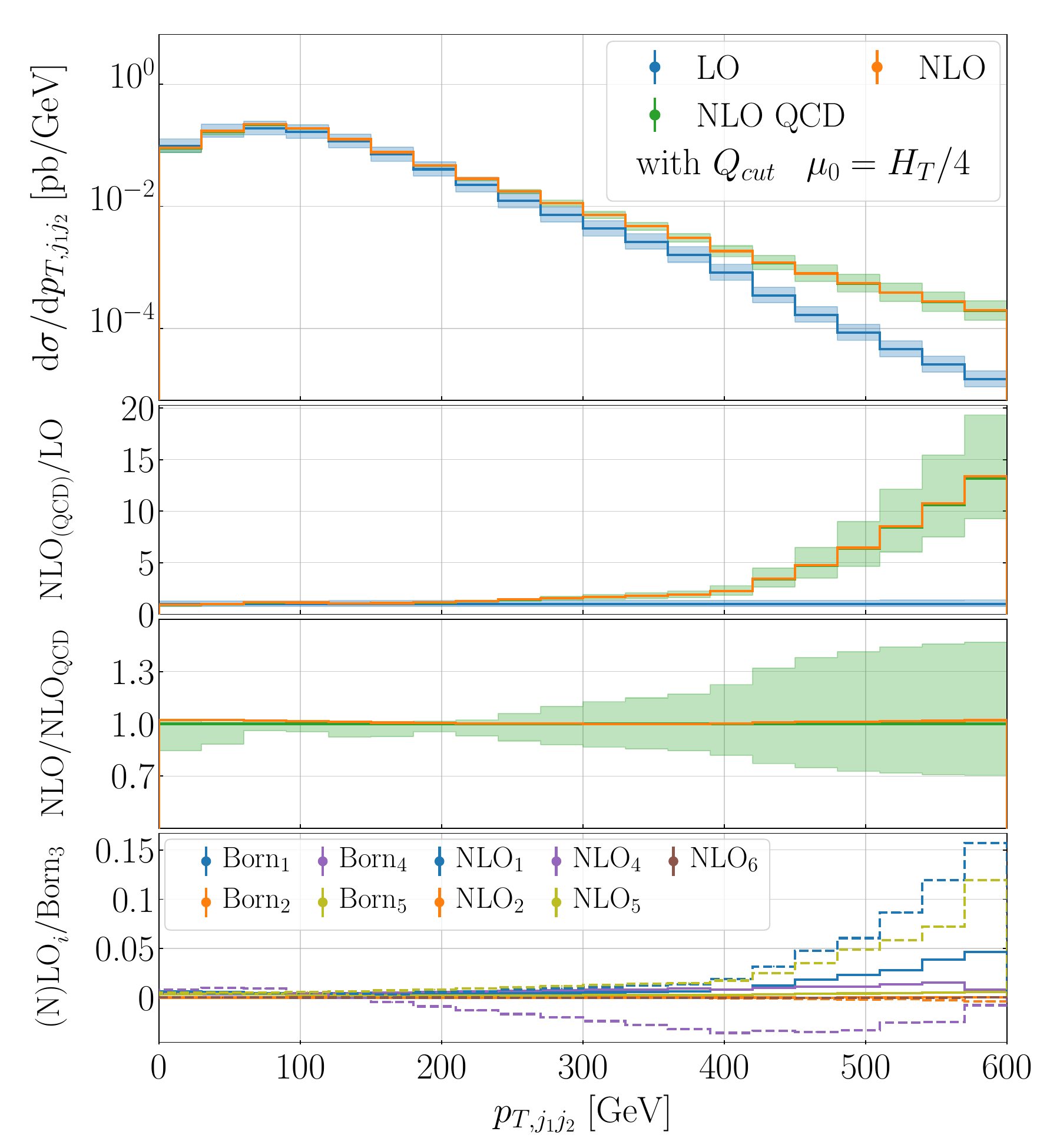}
	\includegraphics[width=0.49\textwidth]{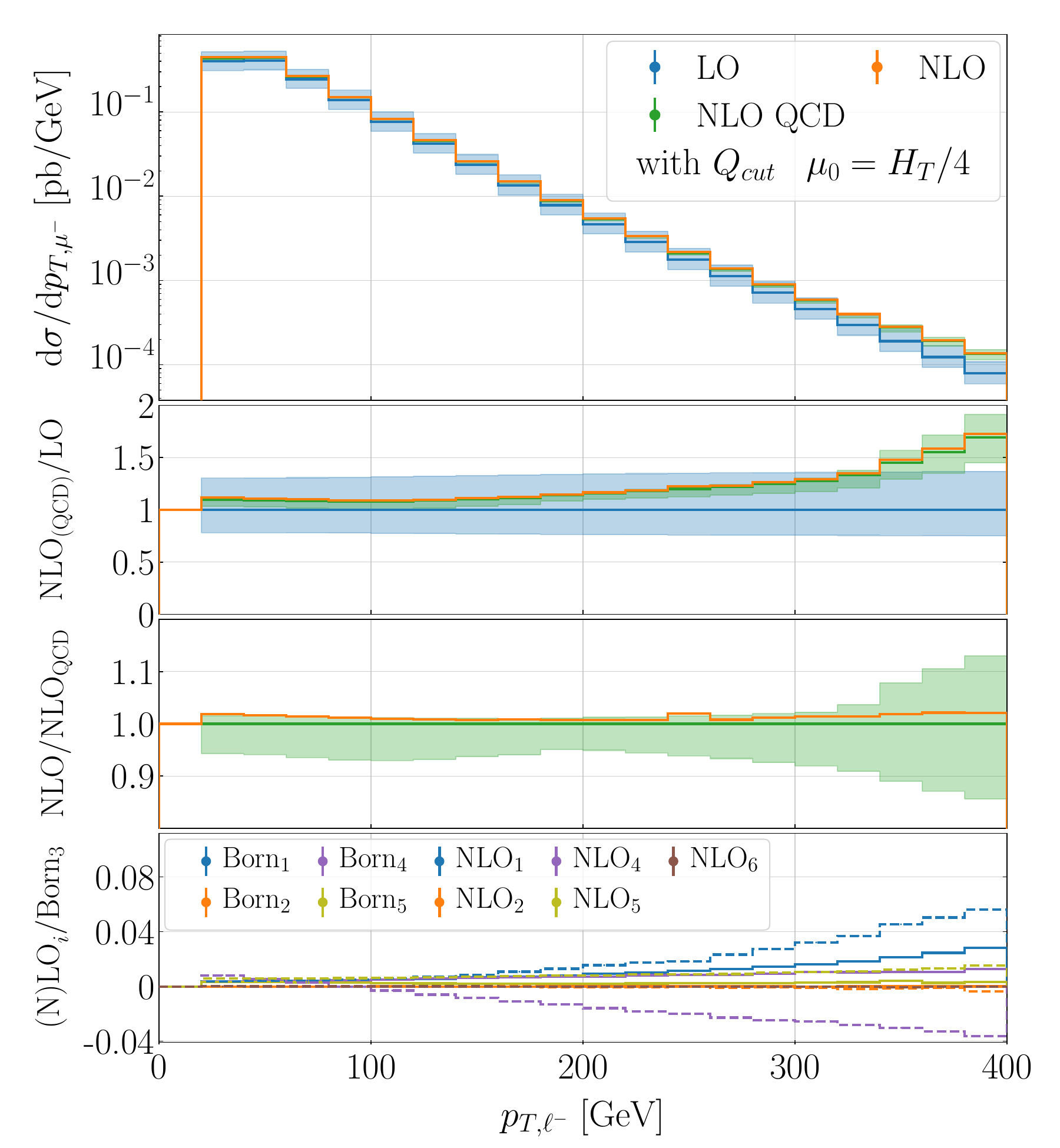}
	\includegraphics[width=0.49\textwidth]{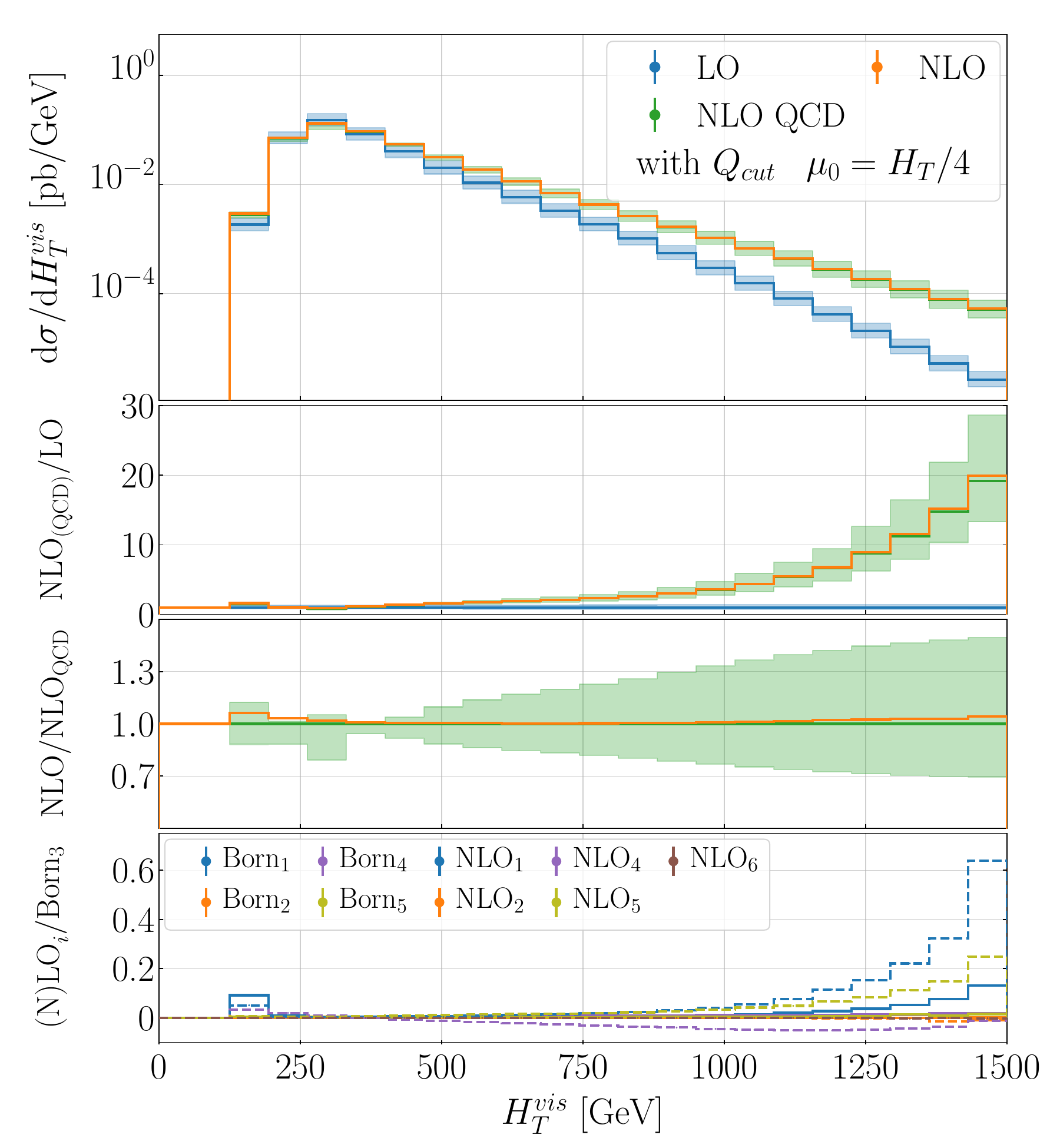}
    \end{center}
    \caption{\label{fig-ttsemi:nlo2} \it Same as Figure \ref{fig-ttsemi:nlo1}, but for $p_{T,\,b_1b_2}$, $p_{T,\,j_1j_2}$, $p_{T,\, \ell}$ and $H_T^{vis}$.}
\end{figure}

In Figure \ref{fig-ttsemi:nlo2} we show the transverse momentum of the $b_1b_2$ and $j_1j_2$ systems, denoted as $p_{T,\,b_1b_2}$ and $p_{T,\,j_1j_2}$, respectively. Also given are the transverse momentum of the charged lepton, $p_{T, \,\ell^-}$, and the scalar sum of all transverse momenta of the visible particles in the final state, denoted as $H_T^{vis}$, and defined according to $H_T^{vis}=p_{T,\,\ell} + p_{T,\,b_1} + p_{T,\,b_2} + p_{T,\,j_1} + p_{T,\,j_2}$. Both $p_{T,\,b_1b_2}$ and $p_{T,\,j_1j_2}$ receive gigantic NLO corrections in the high-$p_T$ tails. The corresponding differential ${\cal K}$-factors are ${\cal K}=5$ and ${\cal K}=13$ for $p_{T,\,b_1b_2}$ and $p_{T,\,j_1j_2}$, respectively. In both cases, these higher-order effects are driven by the QCD corrections to \Bornthree and arise from the recoil of the systems against the extra jet from the real-emission part of the NLO QCD calculation, see e.g. Ref. \cite{Denner:2012yc}. Furthermore, in these phase-space regions the  \NLOfull and \NLOqcd  uncertainties increase to about $40\%$ for $p_{T,\,b_1b_2}$ and $50\%$ for $p_{T,\,j_1j_2}$.  The \NLOfull/\NLOqcd ratio increases to $14\%$ for $p_{T,\,b_1b_2}$ and it is of the order of $2\%$ for $p_{T,\,j_1j_2}$.  For $p_{T,\,b_1b_2}$, the subleading \NLOone part is rather large reaching $75\%$, \NLOfive is positive and contributes with $15\%$. The  \NLOfour contribution, containing EW Sudakov logarithms, partially cancels the latter contribution, as it is at the level of $-11\%$. In the case of $p_{T,\,j_1j_2}$ the subleading contributions are smaller with \NLOone  contributing up to $16\%$ and \NLOfive being of the order of $12\%$. An interesting effect can be observed in the case of \NLOfour, where the higher-order corrections reach $-4\%$ at around $400$ GeV and then start decreasing towards higher $p_{T,\,j_1j_2}$ values, ending at only $-1\%$. This can be explained by the fact that at such large scales also the $g\gamma$ luminosity is no longer negligible. As a consequence, the effect from the NLO QCD corrections to \Bornfour with $g\gamma$ in the initial state starts to significantly impact the total size of the higher-order effects for \NLOfour and partially compensates the contribution of the NLO  EW corrections to \Bornthree, see also e.g. Ref. \cite{Pagani:2016caq}.

For the $p_{T,\,\ell^-}$ observable we have sizeable \NLOfull corrections of the order of $+73\%$. The \NLOfull and \NLOqcd  uncertainties are similar and of the order of $15\%$. The difference between the two predictions is very small at the level of $2\%$ only. The largest subleading NLO contribution is \NLOone with $6\%$, which is partially compensated by \NLOfour with $-4\%$. The remaining NLO contributions are at $1\%$ or less.

The $H_T^{vis}$ observable is characterised by a gigantic differential $\mathcal{K}$-factor. Indeed, towards the tail of the distribution we obtain $\mathcal{K}=20$. Because the observables $H_T^{vis}$ and $H_T$ are plotted in the same range, it may seem that compared to $H_T$ this $\cal {K}$-factor is significantly higher. The only difference between these two observables is that $p_T^{miss}$ does not appear in  $H_T^{vis}$ but is part of the definition of $H_T$.  Adding $p_T^{miss}$ to $H_T$ simply shifts the entire spectrum towards higher $p_T$ values. Had we increased the plotted range for $H_T$, we would also have observed much larger higher-order QCD corrections in the tail of this distribution. We have verified that $\mathcal{K}=24$ can be already obtained for $H_T \approx 1700$ GeV. Due to the dominant \NLOthree contribution, the  \NLOfull and \NLOqcd scale uncertainties are of similar size for $H_T^{vis}$ and in the high-$p_T$ tail are of the order of $50\%$. The difference between \NLOfull and \NLOqcd is only $4\%$, which is an order of magnitude smaller than the NLO uncertainties and therefore negligible. For subleading higher-order effects, a similar impact for various contributions can be observed as in the case of $H_T$.  More precisely, \NLOone gives the largest contribution with $+64\%$ compared to \Bornthree, \NLOfive contributes $+25\%$, while \NLOfour is negative with a maximum contribution of $-5\%$ at around 1200 GeV. This latter contribution reduces to only $-1\%$ for the phase-space region just below 1500 GeV. This means that the same effect that has been observed for the $p_{T,\,j_1j_2}$ differential cross-section distribution is also demonstrated here.
\begin{figure}[t!]
    \begin{center}
	\includegraphics[width=0.49\textwidth]{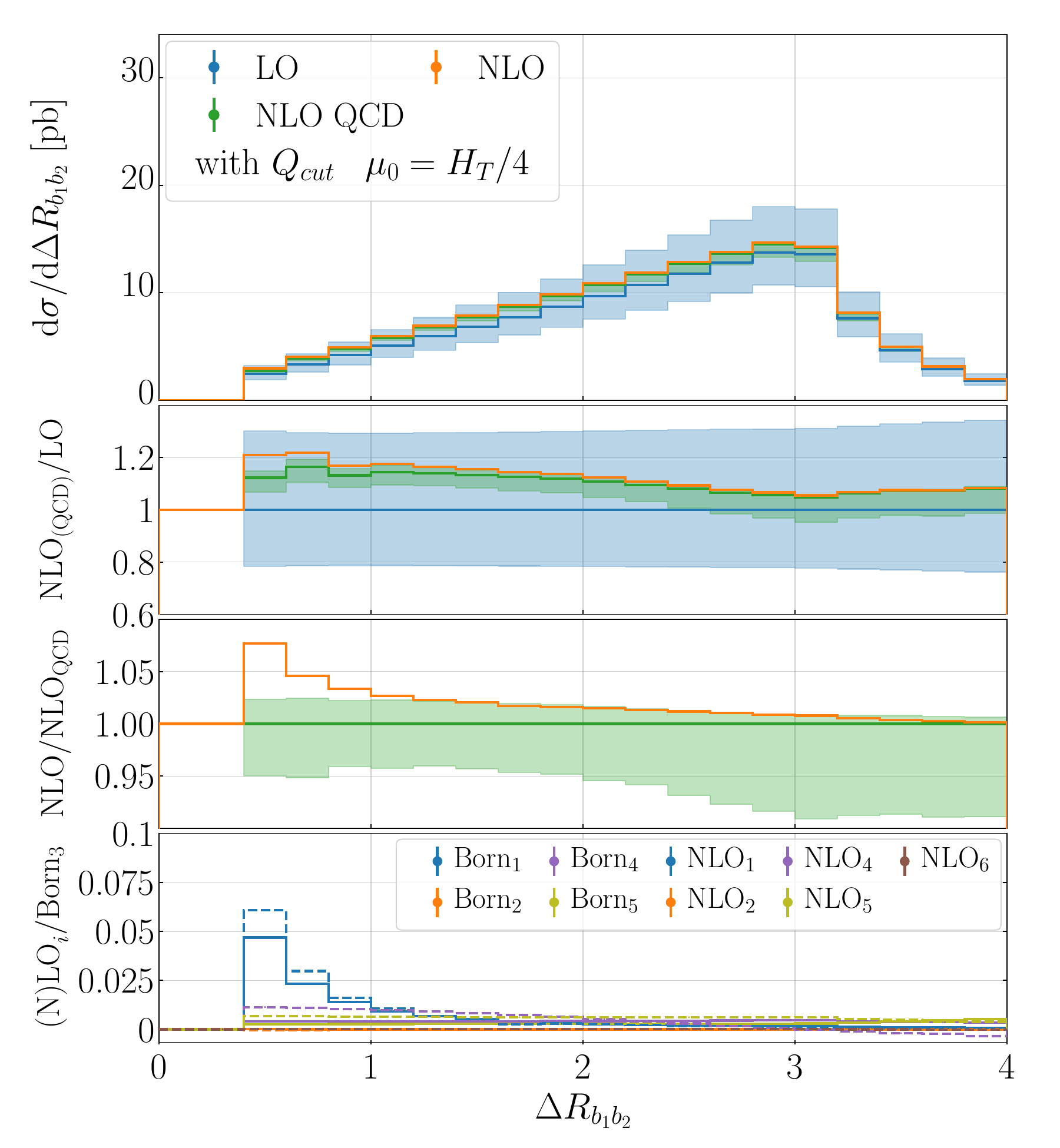}
	\includegraphics[width=0.49\textwidth]{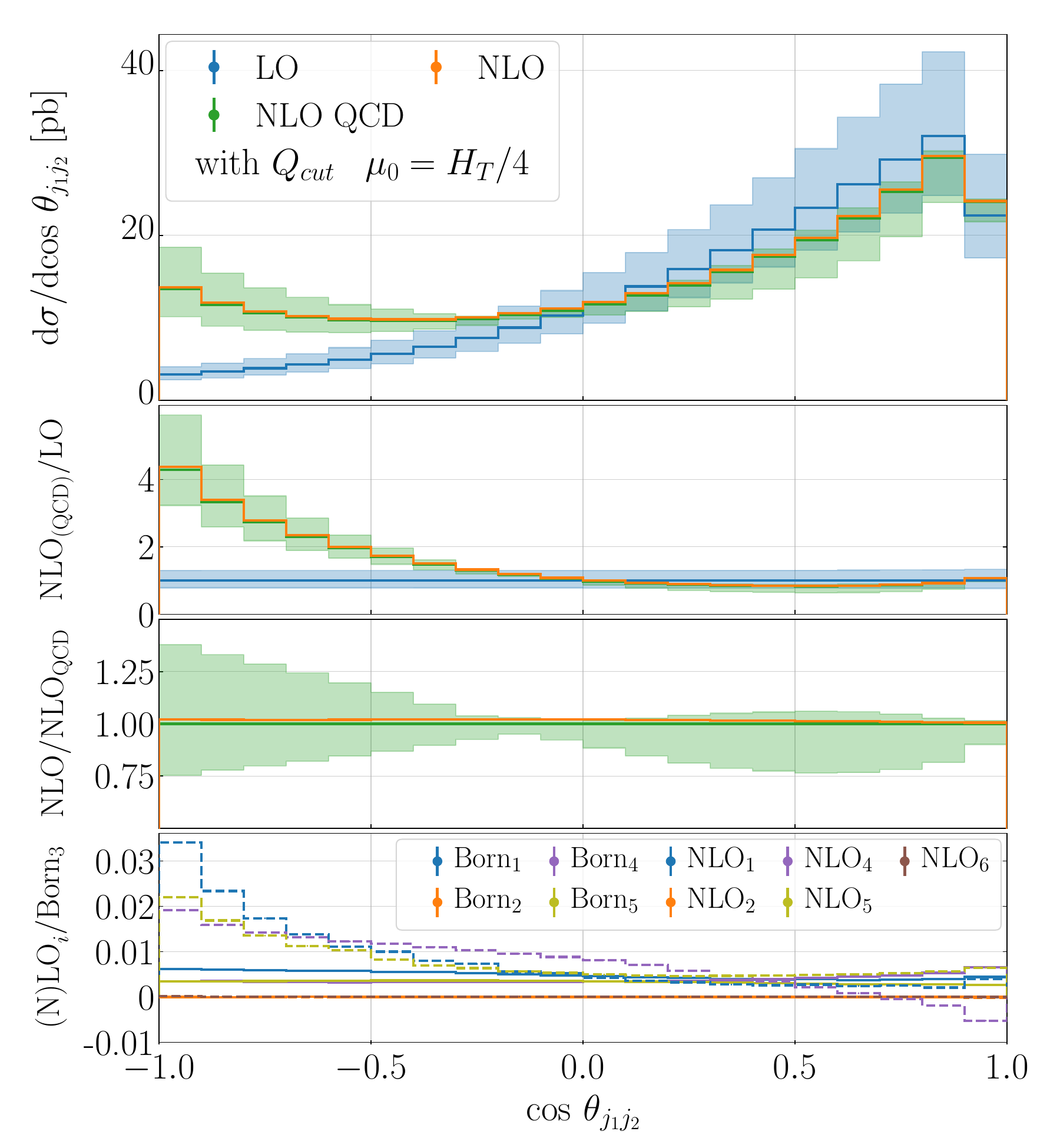}
    \end{center}
    \caption{\label{fig-ttsemi:nloangular} \it Same as Figure \ref{fig-ttsemi:nlo1}, but for  $\Delta R_{b_1b_2}$ and $\cos\theta_{j_1j_2}$.}
\end{figure}

In the next step, we investigate the impact of subleading higher-order effects on dimensionless (angular) cross-section distributions. As an example in Figure \ref{fig-ttsemi:nloangular} we show the $\Delta R_{b_1b_2}$ separation defined in the rapidity $(\Delta y_{b_1b_2})$ and azimuthal angle ($\Delta \phi_{b_1b_2}$) plane, and the cosine of the opening angle between the hardest and second hardest light jets, given by $\cos\theta_{j_1j_2} = \hat{p}_{j_1}\cdot \hat{p}_{j_2}$. In the case of $\Delta R_{b_1b_2}$, close to the minimum separation distance set by the jet algorithm, i.e. for  $\Delta R_{b_1b_2} \approx 0.4$,  we can see enhanced effects due to the QCD background, reaching $5\%$ and $6\%$ for  \Bornone and \NLOone, respectively. Consequently, in this part of the distribution there is a larger difference between the \NLOfull and \NLOqcd predictions, which is of the order of $7\%$. This effect decreases rapidly in the phase-space regions where the $b$-jets are well separated, i.e. when the probability that they originate from the top-quark decays rather than from the  $g\to b\bar{b}$ splittings increases.  All other subleading contributions are negligible, below $1\%$. For the $\cos\theta_{j_1j_2}$ observable, the largest subleading contribution comes again from \NLOone, reaching $3\%$ for $\cos\theta_{j_1j_2} \approx -1$. The latter phase-space regions correspond to a back-to-back configuration for the two light jets. At  NLO, such configurations are no longer suppressed, resulting in the ${\cal K}$-factor equal to ${\cal K}=4$. The \NLOfour and \NLOfive contributions are at the level of $2\%$.  All other subleading contributions are negligible.  We obtained similar results for other angular differential cross-section distributions that we have examined.
\begin{figure}[t!]
    \begin{center}
    \includegraphics[width=0.49\textwidth]{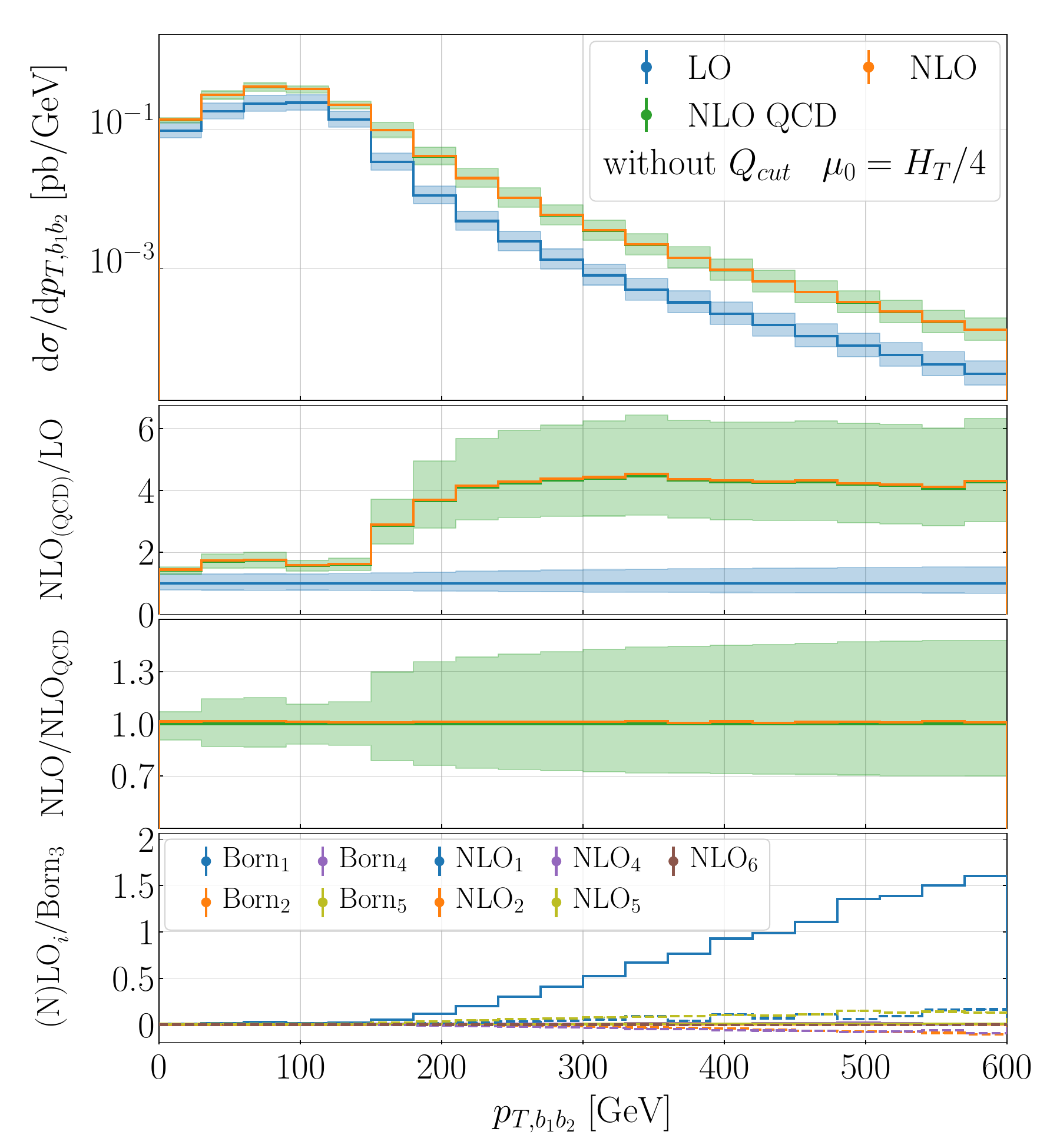}
	\includegraphics[width=0.49\textwidth]{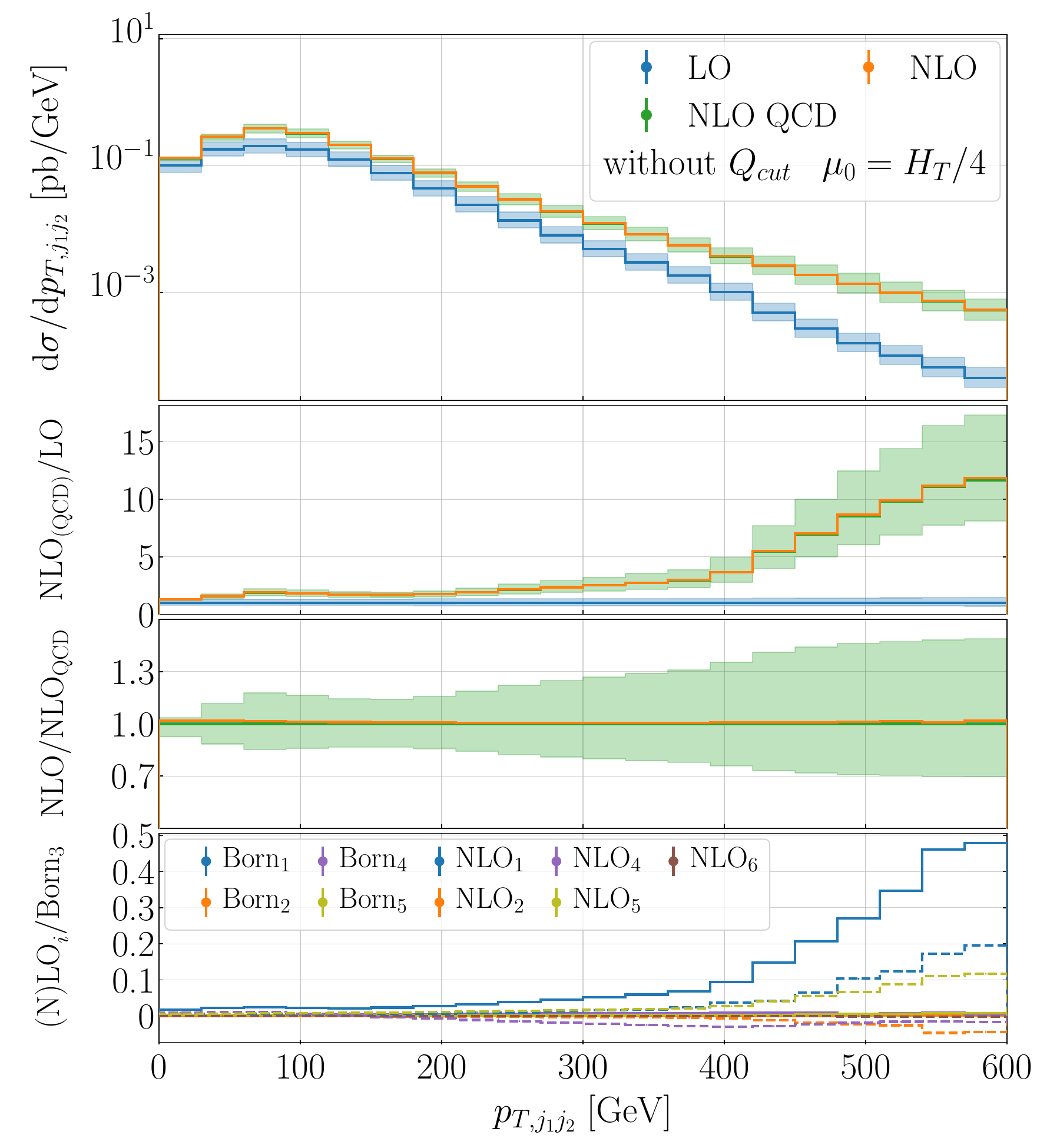}
    \includegraphics[width=0.49\textwidth]{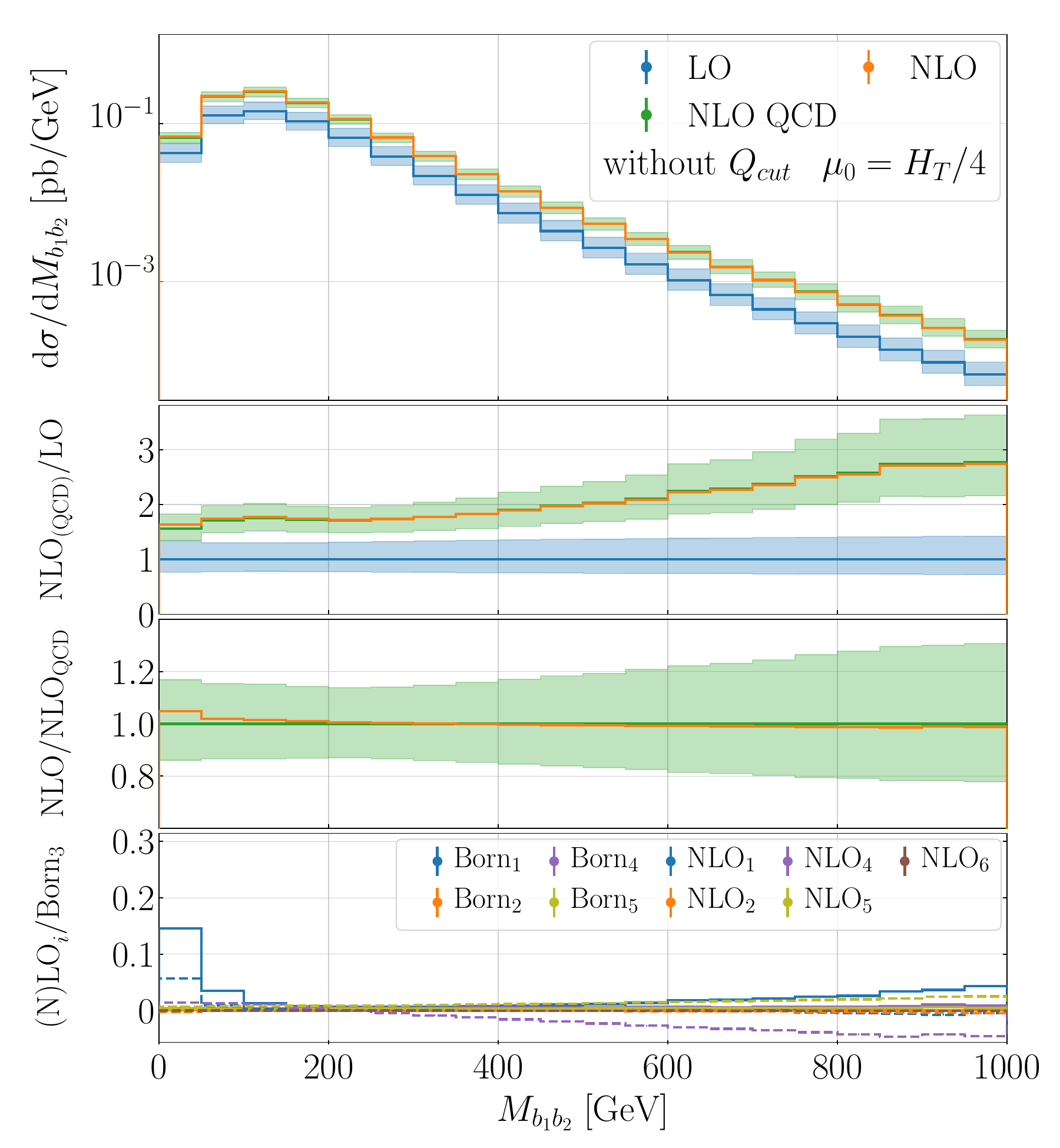}
    \includegraphics[width=0.49\textwidth]{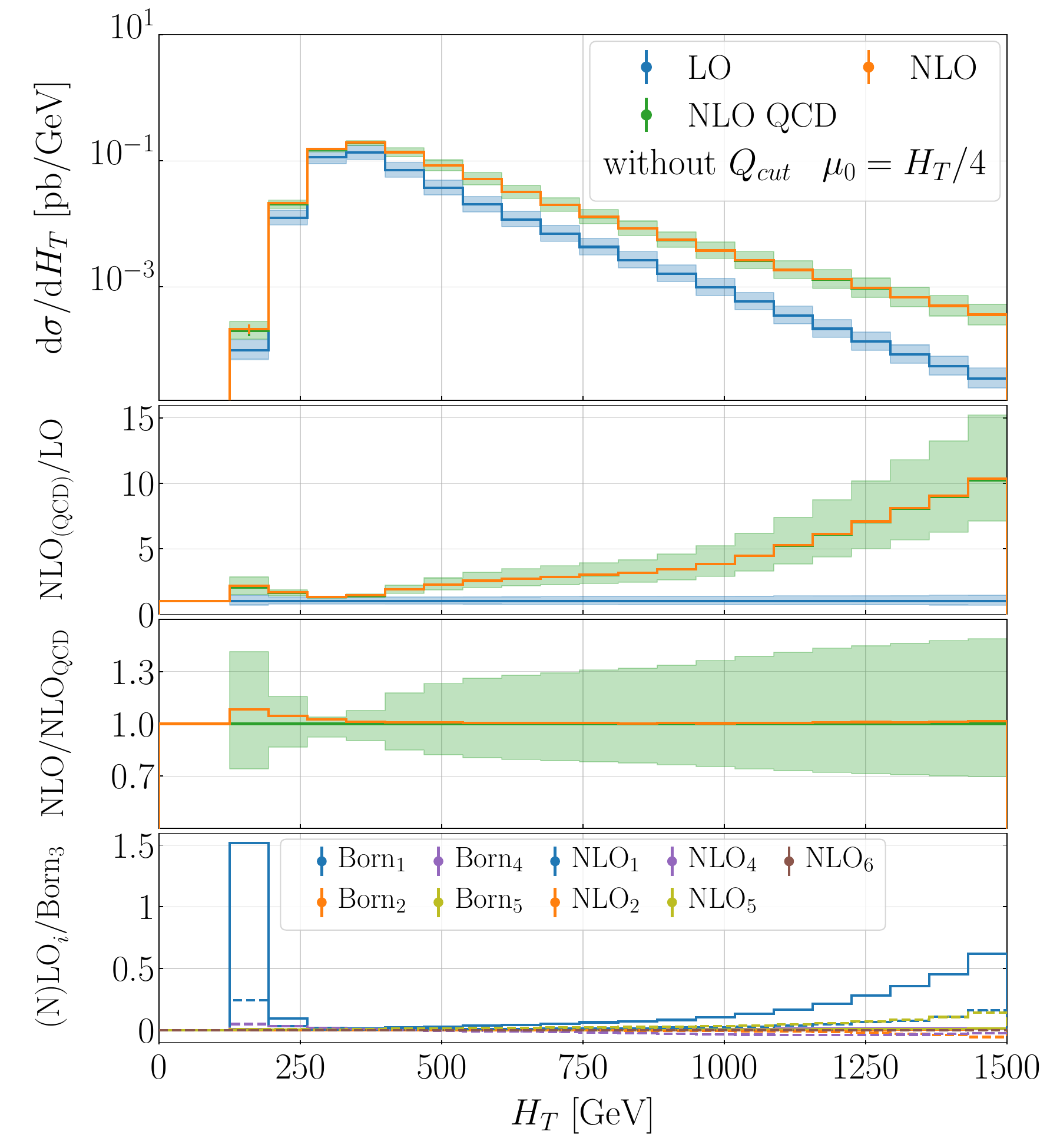}
    \end{center}
    \caption{\label{fig-ttsemi:nlo_noqcut} \it Same as Figure \ref{fig-ttsemi:nlo1}, but without the $|M_{jj}-m_W|< {\cal Q}_{cut}$ condition and for $p_{T,\,b_1b_2}$, $p_{T,\,j_1j_2}$, $M_{b_1b_2}$ and $H_T$.}
\end{figure}

As we explained earlier, the $|M_{jj}-m_W|< {\cal Q}_{cut}$ restriction improves the perturbative stability of the complete NLO calculations. We could clearly observe this in the case of the integrated cross-section results. In the following, we would like to investigate how some of the observables, which we have already examined before, change when we do not apply this condition. In Figure \ref{fig-ttsemi:nlo_noqcut} we show again the following observables $p_{T,\,b_1b_2}$, $p_{T,\,j_1j_2}$, $M_{b_1b_2}$ and $H_T$, but this time without  the ${\cal Q}_{cut}$ cut. For $p_{T,\,b_1b_2}$ the \NLOfull corrections in the tails of the distribution become smaller, reaching about $+330\%$. This can be attributed to the less restricted phase space for the \LOfull contribution, which allows a harder $p_{T,\,b_1b_2}$ spectrum before the introduction of NLO radiation, resulting  in the slightly reduced differential $\mathcal{K}$-factor. The \NLOfull  uncertainties  decrease to $36\%$ from $41\%$, while the \NLOqcd uncertainties increase to $48\%$ from $41\%$. Regarding the subleading contributions, \Bornone is less restricted without the $|M_{jj}-m_W|< {\cal Q}_{cut}$ condition, taking values comparable to \Bornthree, which are up to $+160\%$. The \NLOone part contributes $+16\%$ compared to $+75\%$ with the cut, while \NLOtwo is enhanced to $-11\%$ in the high-$p_T$ tail. The remaining two contributions \NLOfour and \NLOfive are still present, but slightly reduced to $-9\%$ and $+14\%$, respectively, while \NLOsix still contributes below $1\%$. Therefore, there are no large accidental cancellations between different subleading higher-order contributions. 

For $p_{T,\,j_1j_2}$ we can observe a reduction in the size of the  ${\cal K}$-factor. Indeed, we now obtain ${\cal K}=12$ instead of ${\cal K}=13$. The \NLOfull and \NLOqcd uncertainties remain approximately at the same level as before.  The same holds for the difference between the two results. We can notice that \Bornone now makes up almost half of \Bornthree. The  \NLOone, \NLOfour and \NLOfive subleading contributions remain at the same level. 

In the case of $M_{b_1b_2}$ we notice larger \NLOfull corrections in the tail of the spectrum, which reach $+175\%$ compared to $+40\%$ when the ${\cal Q}_{cut}$ cut has been used. This is also reflected in the magnitude of the \NLOfull and \NLOqcd uncertainties, which both triple in size and amount $30\%$. The difference between \NLOfull and \NLOqcd is negligible. The subleading contributions are mostly unaffected by the ${\cal Q}_{cut}$ cut. Indeed, only the QCD background in \NLOone increases at the beginning of the spectrum to about $+4\%$ from $+1\%$.

Finally, for $H_T$ we observe an increase in the size of the  \NLOfull corrections  to $+940\%$ as compared to $+750\%$ with the cut. The \NLOfull and \NLOqcd uncertainties are similar as before and up to $50\%$. The QCD background in \Bornone is enhanced close to the kinematical threshold for this observable, while in the tail of the distribution, \Bornone increases its contribution to about $+60\%$. In the case of the subleading higher-order effects,  \NLOone stays at the same level as before, whereas \NLOtwo, which did not contribute at all before, is now of the order of $-6\%$ in the 
high-$p_T$ tail. The electroweak Sudakov logarithms in \NLOfour are reduced to $-2\%$, while the absence of the ${\cal Q}_{cut}$ cut enhances \NLOfive to $+14\%$ from $+8\%$. 

The overall impact of the $|M_{jj}-m_W|< {\cal Q}_{cut}$ restriction, with ${\cal Q}_{cut} =20$ GeV, on the differential cross-section distributions depends on the observable under study and the phase-space region that is analysed. Without the ${\cal Q}_{cut}$ cut the QCD background in \Bornone is enhanced and becomes comparable in size to \Bornthree, especially in the 
high-$p_T$ tail. Other subleading higher-order contributions are not significantly affected by this cut. In particular, the qualitative effects of the electroweak Sudakov logarithms remain essentially unchanged.

\section{Alternative scale setting}
\label{sec:scale}

%
\begin{table}[t!]
    \centering
    \renewcommand{\arraystretch}{1.2}
    \begin{tabular}{ll@{\hskip 10mm}l@{\hskip 10mm}l@{\hskip 10mm}}
        \hline
        \noalign{\smallskip}
         &&$\sigma_{i}$ [pb] & Ratio to ${\rm Born}_3$  \\
        \noalign{\smallskip}\midrule[0.5mm]\noalign{\smallskip}
        \Bornone&\asLOone& $ 0.1270(2) $ & $ 0.43\% $ \\
        \Borntwo&\asLOtwo& $ 0.00032(2) $ & $ 0.00\% $ \\
        \Bornthree&\asLOthree& $ 29.824(2) $ & $ 100.00\% $ \\
        \Bornfour&\asLOfour& $ 0.1302(1) $ & $ 0.44\% $ \\
        \Bornfive&\asLOfive& $ 0.10086(4) $ & $ 0.34\% $ \\
        \noalign{\smallskip}\hline\noalign{\smallskip}
        \NLOone&\asNLOone& $ +\,0.152(2) $ & $ +\,0.51\% $\\
        \NLOtwo&\asNLOtwo& $ -\,0.0016(2) $ & $ -\,0.01\% $\\
        \NLOthree&\asNLOthree& $ +\,0.48(1) $ & $ +\,1.61\% $\\
        \NLOfour&\asNLOfour& $ +\,0.1074(2) $ & $ +\,0.36\% $\\
        \NLOfive&\asNLOfive& $ +\,0.1782(4) $ & $ +\,0.60\% $\\
        \NLOsix&\asNLOsix& $ +\,0.00242(1) $ & $ +\,0.01\% $\\
        \noalign{\smallskip}\hline\noalign{\smallskip}
        \LOfull&& $ 25.991(1)^{+30.3\%}_{-21.8\%} $ & $ 0.87 $ \\
        \Bornfull&& $ 30.182(2)^{+30.3\%}_{-21.8\%} $ & $ 1.01 $ \\
        \NLOqcd&& $ 30.66(1)^{+1.6\%}_{-4.8\%} $ & $ 1.03 $ \\
        \NLOfull&& $ 31.10(1)^{+1.7\%}_{-5.0\%} $ & $ 1.04 $ \\
        \noalign{\smallskip}\hline\noalign{\smallskip}
    \end{tabular}
    \caption{\it Same as in Table \ref{tab:integratedqcut}, but  for the following scale setting $\mu_0 = E_T/4$.}
    \label{tab:integratedqcut_et4}
\end{table}
\begin{figure}[t!]
    \begin{center}
	\includegraphics[width=0.49\textwidth]{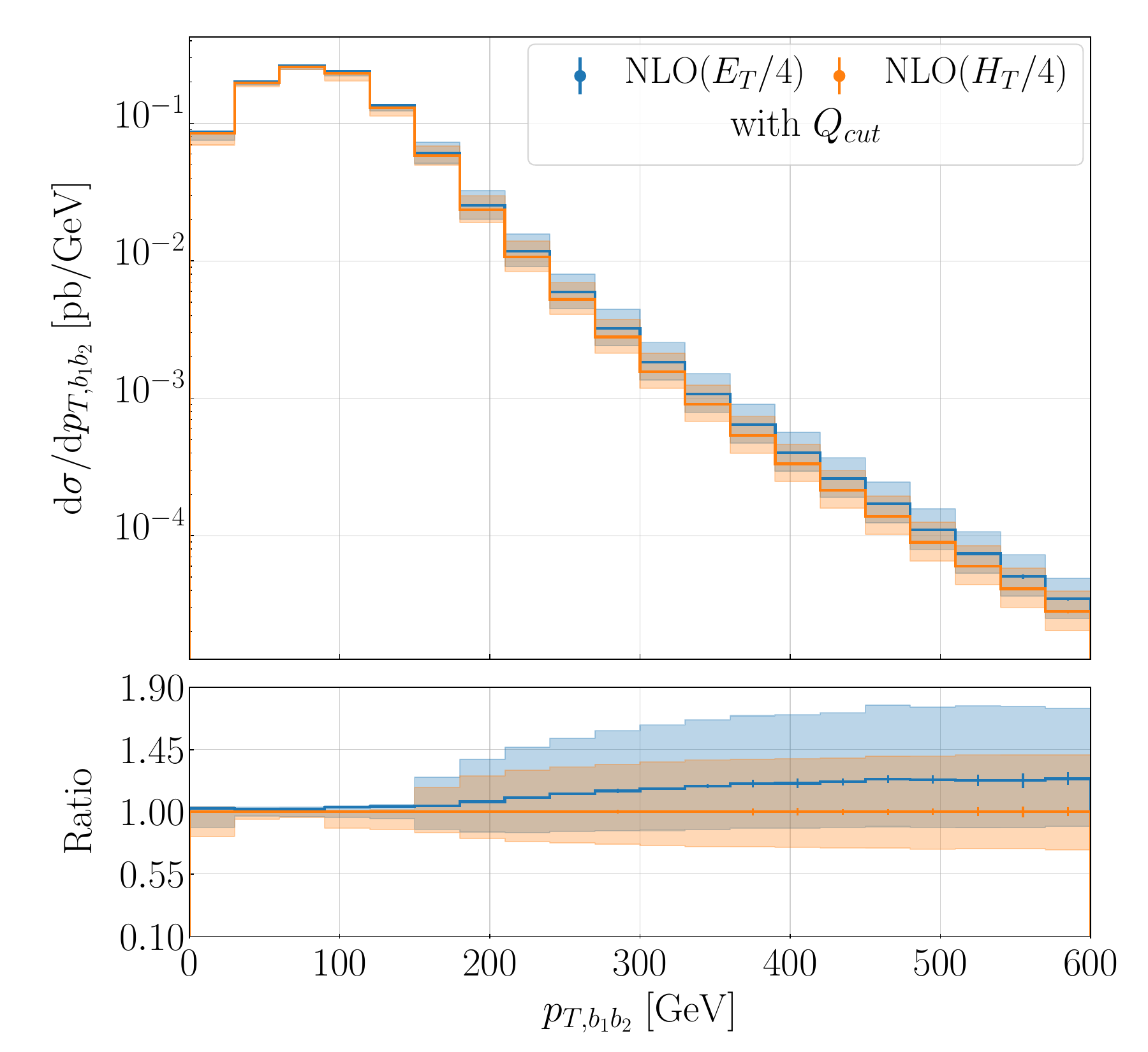}
	\includegraphics[width=0.49\textwidth]{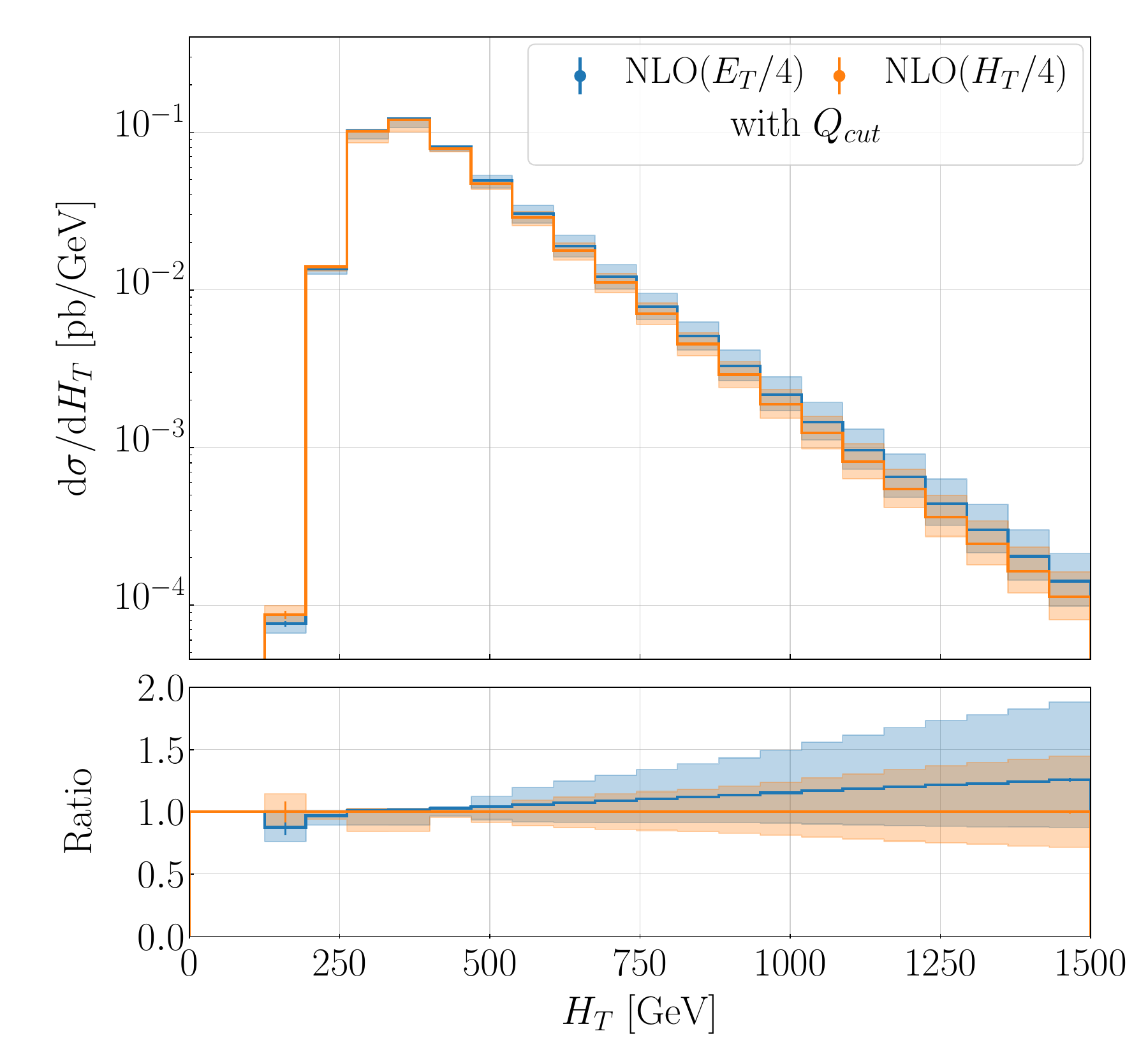}
	\includegraphics[width=0.49\textwidth]{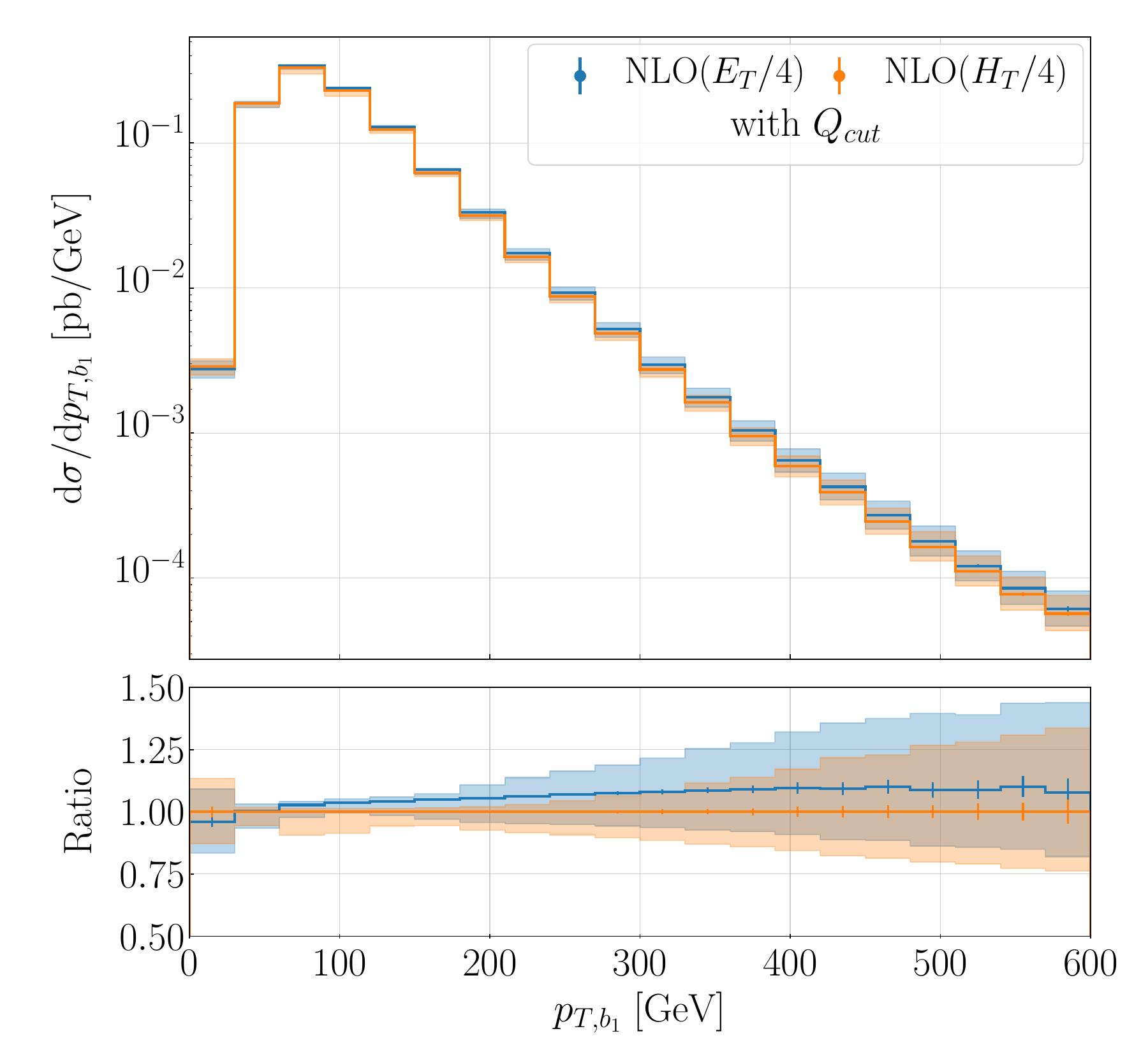}
	\includegraphics[width=0.49\textwidth]{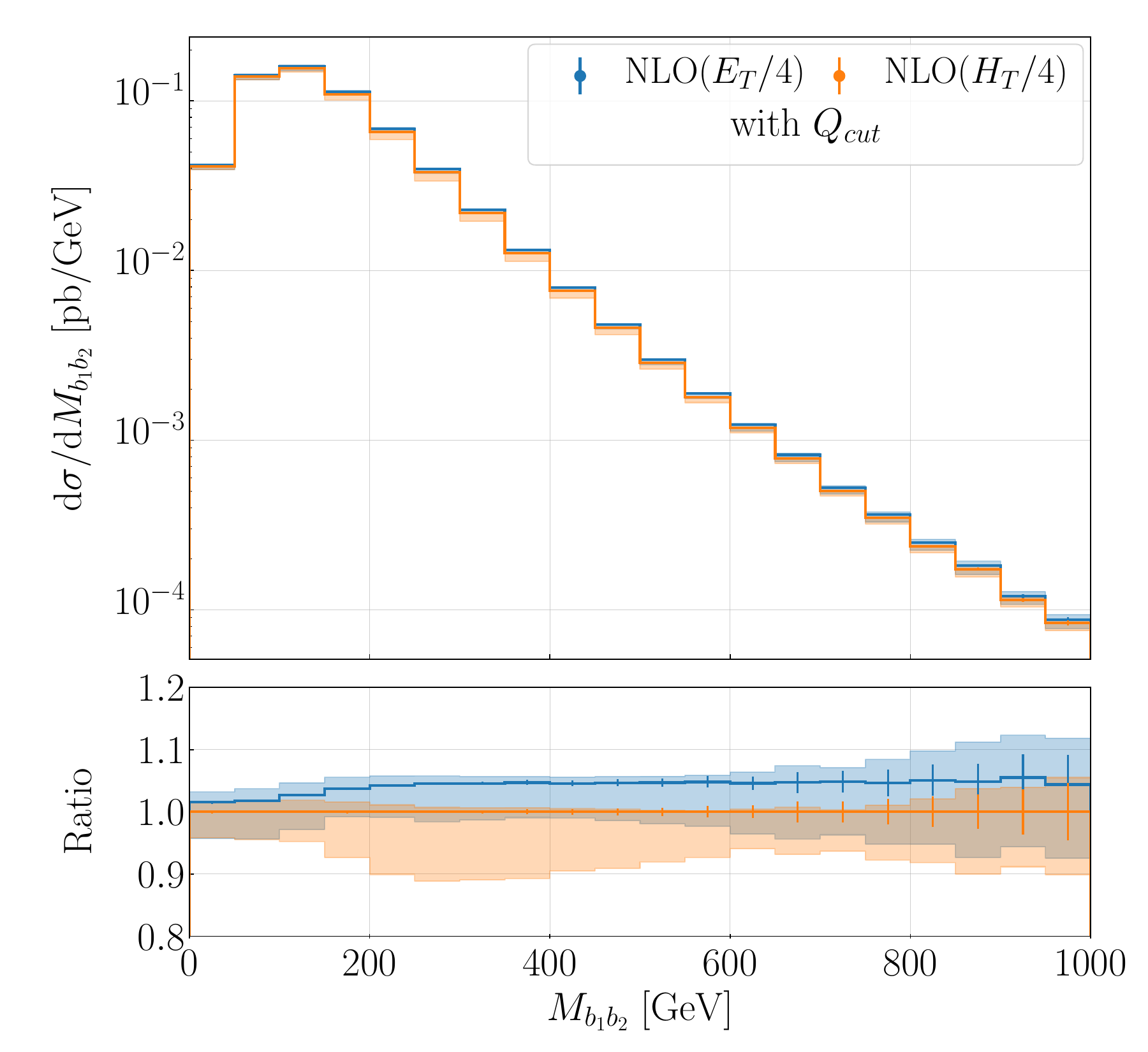}
    \end{center}
    \caption{\label{fig-ttsemi:nlo_scales} \it Differential cross-section distributions for the $pp \to \ell^-\bar{\nu}_\ell\,j_b j_b jj +X$ process at the LHC with $\sqrt{s}=13.6$ TeV as functions of  $p_{T,\,b_1b_2}$, $H_T$, $p_{T,\,b_1}$ and $M_{b_1b_2}$. Results are given for  $\mu_0 = H_T/4$ and $\mu_0 = E_T/4$ and the NLO NNPDF3.1luxQED PDF set with $|M_{jj}-m_W|< {\cal Q}_{cut}$, where  ${\cal Q}_{cut}=20$ GeV. The upper panels show the absolute \NLOfull predictions for the two scale choices together with their corresponding uncertainty bands. The second panels display their ratios.}
\end{figure}
\begin{figure}[t!]
    \begin{center}
    \includegraphics[width=0.49\textwidth]{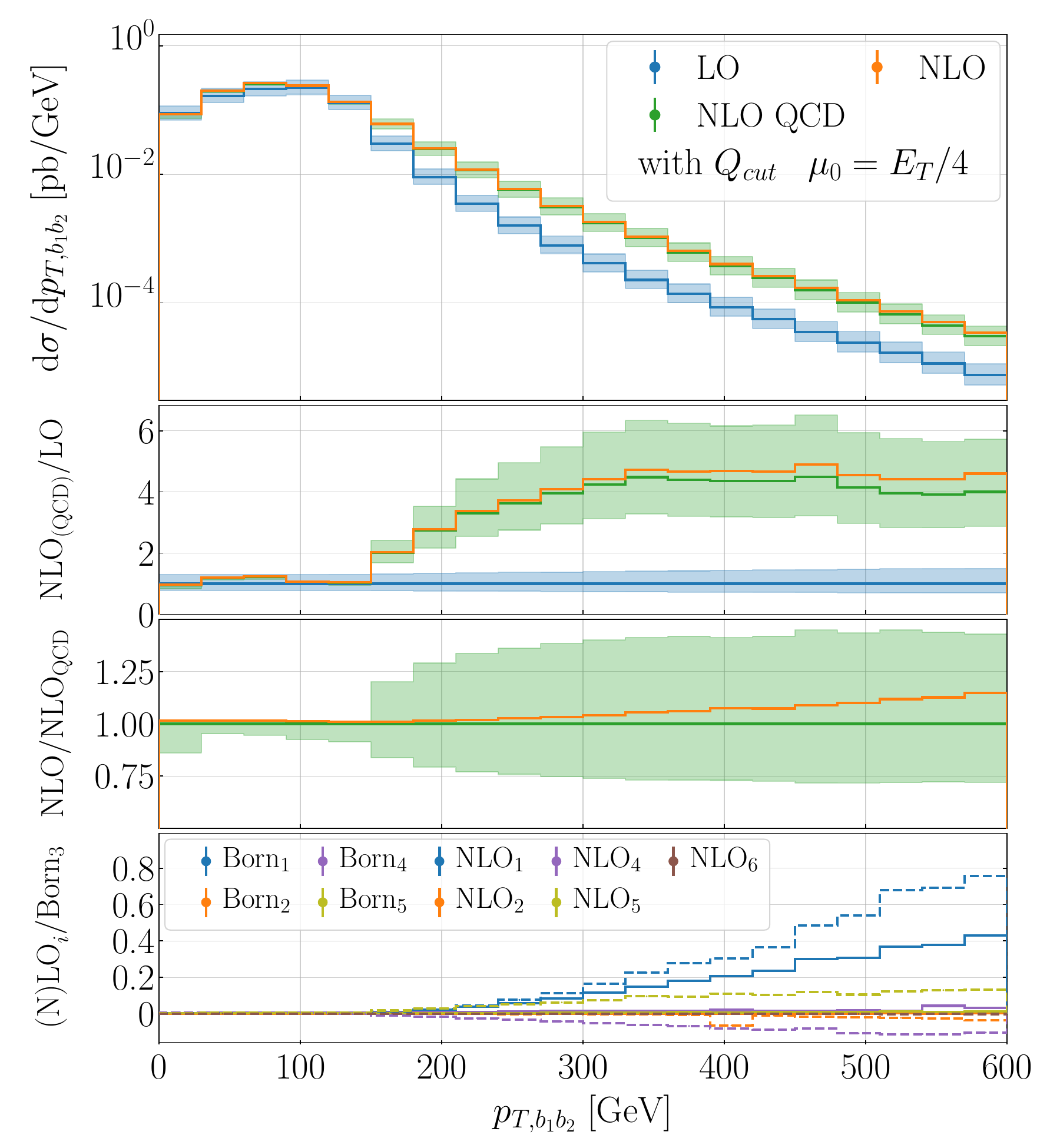}
	\includegraphics[width=0.49\textwidth]{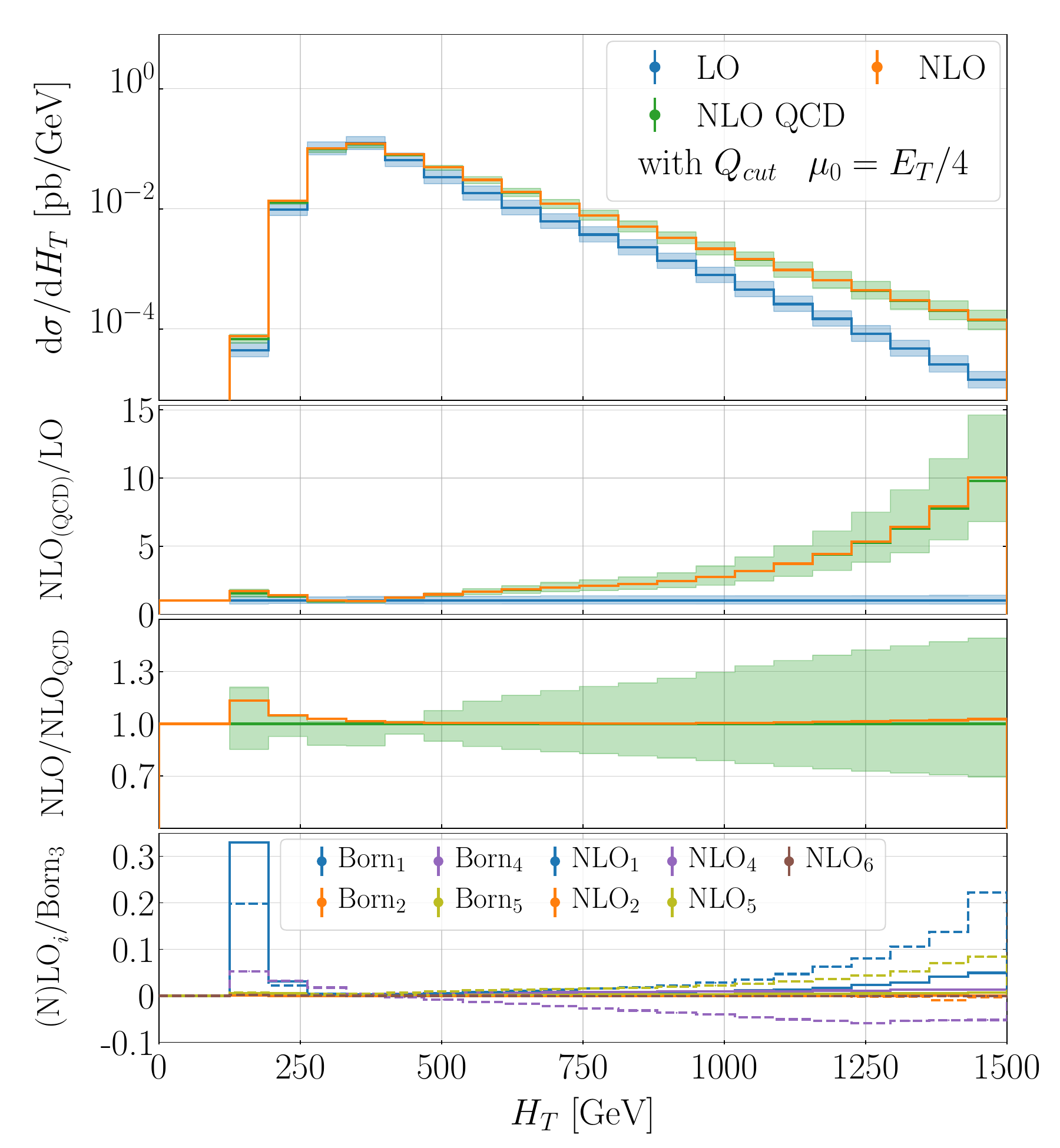}
	\includegraphics[width=0.49\textwidth]{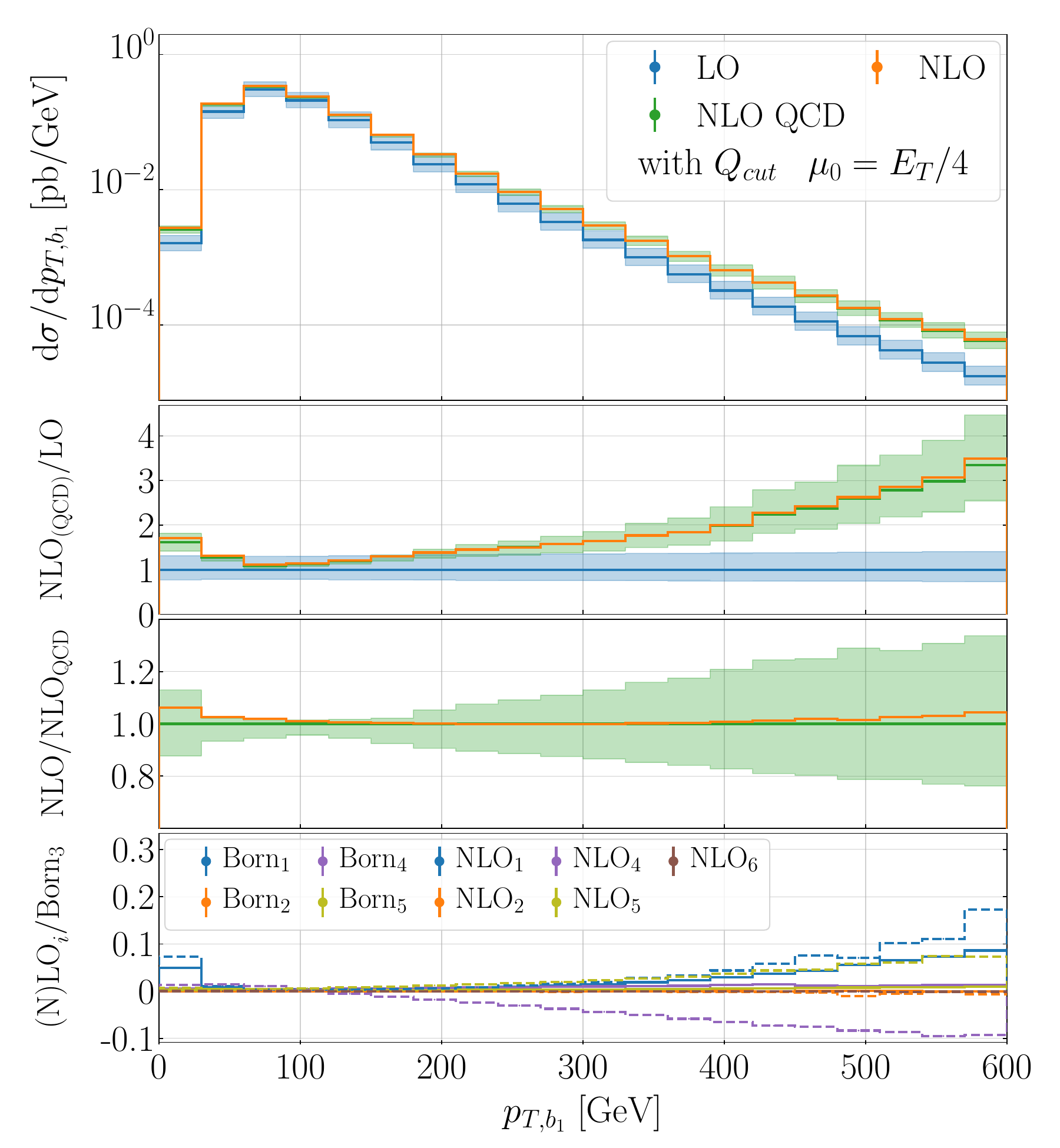}
    \includegraphics[width=0.49\textwidth]{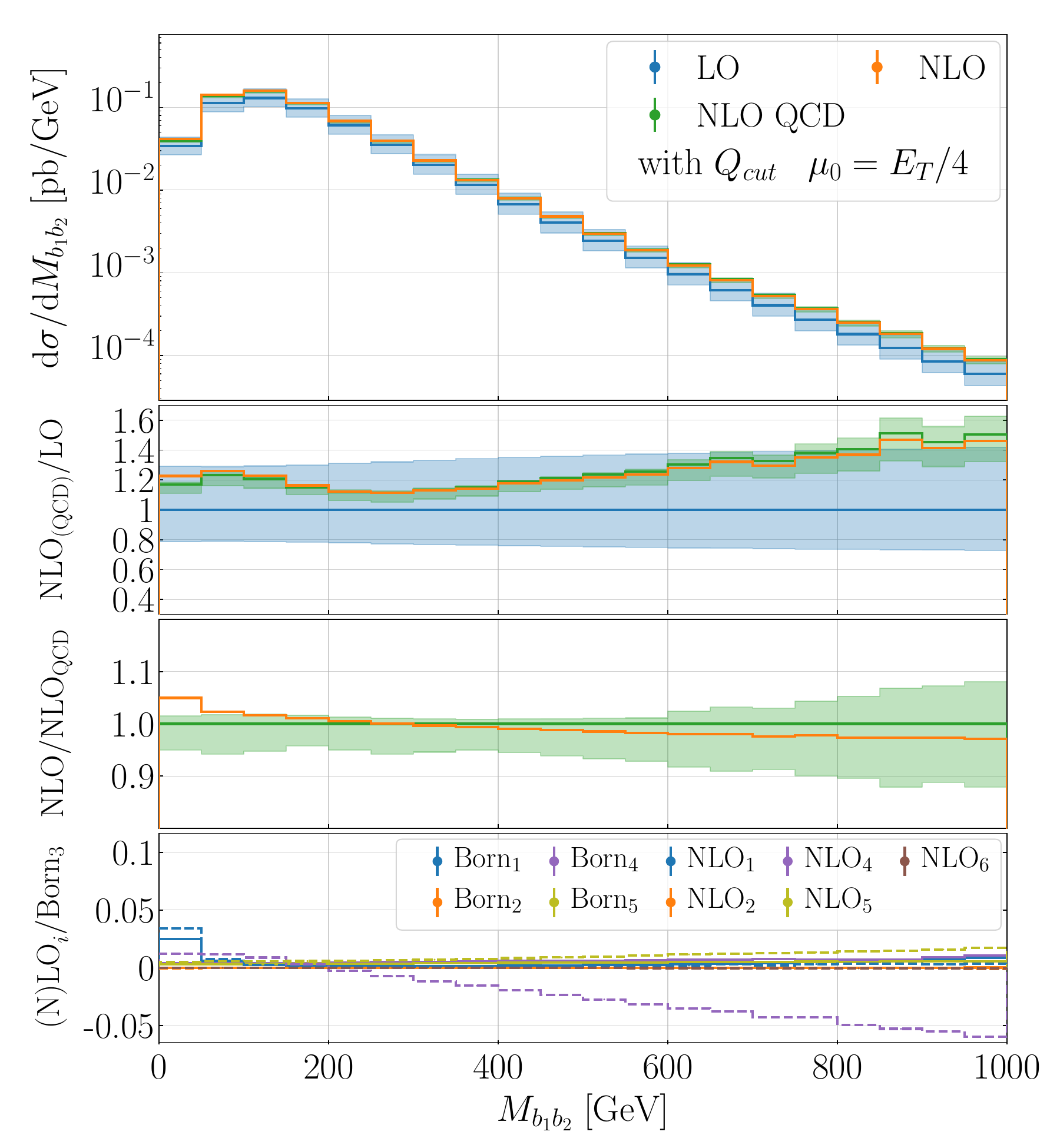}
    \end{center}
    \caption{\label{fig-ttsemi:nlo_et4} \it Differential cross-section distributions for the $pp \to \ell^-\bar{\nu}_\ell\,j_b j_b jj +X$ process at the LHC with $\sqrt{s}=13.6$ TeV as  functions of $p_{T,\,b_1b_2}$, $H_T$, $p_{T,\,b_1}$ and $M_{b_1b_2}$. Results are given for $\mu_0 = E_T/4$ and the NLO NNPDF3.1luxQED PDF set with $|M_{jj}-m_W|< {\cal Q}_{cut}$, where  ${\cal Q}_{cut}=20$ GeV. The upper panels show the absolute \LOfull, \NLOqcd and \NLOfull predictions together with the \LOfull and \NLOqcd uncertainty bands. The second panels present the ratios of \NLOqcd and \NLOfull to \LOfull,  while the ratios of \NLOfull to \NLOqcd are displayed in the third panels. The last panels depict ${\rm Born}_i$ (solid lines) and ${\rm NLO}_i$ (dashed lines) normalised to the dominant \Bornthree contribution. }
\end{figure}

Our calculations, like any fixed-order calculations, have a residual dependence on the renormalization and factorization scale, resulting from the truncation of the perturbation expansion. Consequently, the value and shape of various observables depend on $\mu_R$ and $\mu_F$ that are given as input parameters. For any given process, it is possible to define multiple scale settings in several ways. For example, one could consider the mass of a heavy particle entering the process, or even the typical momentum transfer or the total transverse energy of the process. However, it is impossible to determine which scale is truly good unless one is able to perform a similar calculation one order higher in $\alpha_s$. Since it is obviously impossible to perform complete NNLO calculations at present, the only thing left is to investigate different scale choices and estimate the dependence of our results on them. To check the robustness of our calculations, in addition to the default scale choice $\mu_0 = H_T/4$, we also consider an alternative scale setting given by $\mu_0 = E_T/4$. In  the first step, we provide the complete LO and NLO results for $\mu_0 = E_T/4$ and with the $|M_{jj}-m_W|< 20$ GeV condition at the integrated cross-section level. These results  are  shown in Table \ref{tab:integratedqcut_et4}.

The differences between the complete \LOfull and \NLOfull results for the two scale settings are very small, of the order of only a few percent, i.e. $-5\%$ for \LOfull and $+3\%$ for \NLOfull. Therefore, they are well within their corresponding theoretical uncertainties, which are at the level of $30\%$ and $5\%$ for \LOfull and \NLOfull, respectively. However, these shifts in absolute values affect the size of higher-order corrections, which increase for $\mu_0 = E_T/4$ to ${\cal K}_{\text{QCD}}=1.18$ and ${\cal K}=1.20$. The relative size of the $\text{Born}_i$ contributions remains the same. A similar situation can be observed for all the subleading $\text{NLO}_i$ contributions. The only significant change in magnitude and sign can be seen for the dominant \NLOthree contribution, where $\sigma_{\text{NLO}_3}(\mu_0=H_T/4) = -1.90$ pb $(6\%)$ is replaced by $\sigma_{\text{NLO}_3}(\mu_0=E_T/4) =+0.48$ pb $(1.6\%)$.

To check if a similar level of consistency can be observed at the differential cross-section level, in Figure \ref{fig-ttsemi:nlo_scales} we present, as an example, the complete NLO predictions for $p_{T,\,b_1b_2}$, $H_T$, $M_{b_1b_2}$ and $p_{T,\,b_1}$ for both scale settings. The ratios between these two predictions and their theoretical errors are also plotted. Overall, there is very good agreement among all the observables within the estimated theoretical uncertainties. In detail, for $p_{T,\,b_1b_2}$ the differences grow with $p_ T$ up to $24\%$ in the tail of the distribution, but they are within the estimated error which is of the order of $40\%$ in these phase-space regions. A similar pattern is repeated for the remaining observables. In each case, harder spectra are predicted with $\mu_0=E_T/4$. They increase by $26\%$, $10\%$ and $5\%$ for $H_T$, $p_{T,\,b_1}$ and $M_{b_1b_2}$, respectively. We can draw similar conclusions for the other dimensionful observables we have examined. Furthermore, we have also compared various dimensionless observables and found no significant differences in the shapes of these distributions. We therefore conclude that both dynamical scale choices  are valid, physically well motivated scale settings that, in the case of dimensionful observables, may show differences of up to $\mathcal{O}(30\%)$ in the high-$p_T$ regions due to the NLO QCD corrections. However, these differences are well within the estimated theoretical uncertainties.

Finally, we would like to investigate whether changing the scale setting has a significant impact on the subleading LO and NLO contributions. To this end, in Figure \ref{fig-ttsemi:nlo_et4} we display the same observables as before, i.e. $p_{T,\,b_1b_2}$, $H_T$, $p_{T,\,b_1}$ and $M_{b_1b_2}$, but this time we also show their LO and NLO subleading contributions. For large values of $p_{T,\,b_1b_2}$, we obtain smaller $\mathcal{K}$-factor for the $\mu_0 = E_T/4$ scale setting. It is reduced to ${\cal K}=4.6$ from  ${\cal K}=5.3$. The only significant change in subleading contribution can be found for \Bornone that increases by about $10\%$. For $H_T$, the ratio of \NLOfull to \LOfull changes from ${\cal K}=8.5$ to ${\cal K}=10$. Also in this case, a small increase from $+15\%$ to $+22\%$ can be seen for \NLOone. In the case of  $p_{T,\,b_1}$ and $M_{b_1b_2}$ any differences we observe are only at the level of a few percent for all the subleading contributions. To summarise this part, the overall picture does not change across all the observables we analysed. Changing the scale setting from $\mu_0=H_T/4$ to $\mu_0=E_T/4$ allowed us to ensure that our higher-order predictions are indeed under very good theoretical control.

\section{Summary}
\label{sec:conclusions}

In this paper we presented the first complete, NLO-accurate predictions for the production and decay of the $t\bar{t}$ pair in the \lepjet decay channel. The complete set of LO contributions and NLO corrections of EW and QCD origin is included across all partonic channels. The calculation is based on full matrix elements, computed with all resonant and non-resonant contributions, complete spin correlations, interference effects and the finite-width effects for the unstable $t$, $W$, $Z$ and $H$ particles. We provided the complete \LOfull and \NLOfull results  for the  $pp \to \ell^- \bar{\nu}_{\ell}\, j_b j_b j j + X$ process at the LHC Run-III energy of $\sqrt{s}=13.6$ TeV. On the technical side, we extended the \textsc{Helac-Dipoles} framework to ensure IR safety for the complete-NLO result in the presence of democratic jet clustering with light jets and photons.  Specifically, we incorporated the cut on the photon energy fraction directly at the level of the integrated subtraction terms together with parton-to-photon fragmentation functions, and added the photon-splitting and photon-to-jet conversion subtraction terms needed to render the complete NLO calculation finite. The implementation is incorporated in both the Catani–Seymour and Nagy–Soper subtraction schemes. Having two independent subtraction schemes allowed us to check the correctness of the changes introduced by comparing the two results.

At the integrated fiducial cross-section level we found that all subleading contributions are negligible compared to the size of the \NLOqcd uncertainties. Furthermore, the resonance-enhancing requirement on the invariant mass of the two light jets, $|M_{jj}-m_W|<{\cal Q}_{\rm cut}$, played a key role in improving the perturbative convergence of the full NLO computation. By introducing this cut, the inclusive $\cal{K}$-factor is lowered from ${\cal{K}}=1.75$ to ${\cal{K}}=1.11$, and the residual scale uncertainties are reduced from about $14\%$ to $6\%$. Without the cut, the difference between \NLOfull and \NLOqcd  increases and the QCD background, driven by \Bornone, becomes more prominent.

At the differential cross-section level we observed the impact of subleading terms primarily in the high-$p_T$ tails of dimensionful observables. In these phase-space regions one-loop electroweak Sudakov logarithms from \NLOfour yielded negative corrections that could reach the $10\%$ level. In addition, the observables that are dominated by additional QCD radiation showed characteristic patterns. For example, the transverse momentum of the two-$b$-jet system, $p_{T,\,b_1b_2}$, showed differences of up to $14\%$ between \NLOfull and \NLOqcd due to significant effects resulting  from the \NLOone background contribution. This should be compared to the case of $p_{T,\,b_2}$, where the difference between \NLOfull and \NLOqcd is $-5\%$, which is of the order of the theoretical uncertainties, reflecting the suppression of hard QCD emissions for this observable. For the transverse momentum of the dijet system, $p_{T,\,j_1j_2}$, we noted  sizeable contributions from the photon-initiated partonic channels. In the high-$p_T$ tail, the contribution from the $g\gamma$ partonic channels outweighed   the negative Sudakov EW effects from \NLOfour. In addition, we observed non-trivial cancellations between different contributions, most notably between \NLOfour and \NLOfive across many observables, which further motivates the use of the complete NLO  predictions whenever a precision at the few-percent level is required. In contrast, the dimensionless distributions are only mildly affected by the subleading contributions, which is consistent with our conclusions obtained at the integrated cross-section level.

To check the reliability of our complete NLO calculations, in addition to the default scale setting, we also considered  an alternative scale choice. A different dynamical scale choice induced no significant differences at the integrated fiducial cross-section level. For differential cross-section distributions, the largest difference between the two scale settings we observed was $30\%$ in the tails of the dimensionful observables. However, even this difference was well within the estimated scale uncertainties. The subleading contributions remained largely unchanged when this alternative scale setting was selected.

Our results demonstrated that the complete NLO predictions with full off-shell effects included for the $pp\to t\bar{t}+X$ process in the \lepjet decay channel are feasible.  Applying the $|M_{jj}-m_W|<{\cal Q}_{\rm cut}$ requirement stabilised the perturbative convergence of the results and reduced theory uncertainties, while retaining the sensitivity to genuine EW effects and photon-initiated partonic channels in the high-$p_T$ phase-space regions. 

The strategy adopted here for the IR safety is general and easily applicable to other top-quark associated processes involving photons and light jets. In particular, it opens the way to compute complete NLO corrections with full off-shell effects included  for other related processes such as $pp\to t\bar{t}H$, $pp\to t\bar{t}Z$ and $pp\to t\bar{t}W^\pm$, where the $t\bar{t}$ pair decays in the \lepjet channel.

\acknowledgments{

This work was supported by the Deutsche Forschungsgemeinschaft (DFG) under grant 396021762 $-$ TRR 257: {\it P3H - Particle Physics Phenomenology after the Higgs Discovery} and
grant 400140256 - GRK 2497: {\it The Physics of the Heaviest Particles at the LHC}. 

Support by a grant of the Bundesministerium f\"ur Forschung, Technologie und Raumfahrt (BMFTR) is additionally acknowledged.

The authors gratefully acknowledge the computing time provided to them at the NHR Center
NHR4CES at RWTH Aachen University (project number {\tt p0020216}). This is funded by the Federal
Ministry of Education and Research, and the state governments participating on the basis of the
resolutions of the GWK for national high performance computing at universities.}


\bibliographystyle{JHEP}
\end{document}